\documentclass[physics,dissertation,amsmath,amssymb,nochapterblankpages]{puthesis}
\usepackage{multicol}
\usepackage{graphicx,subfigure}
\usepackage{psfrag}
\title{Theoretical study of interacting electrons in \\one dimension \\ground
states and experimental signatures}

\author{Trinanjan Datta}{Datta, Trinanjan}
\campus{West Lafayette}
\pudegree{Doctor of Philosophy}{Ph.D.}{August}{2007}

\majorprof{Dr. Erica W. Carlson}

\newcommand{\be}{\begin{equation}}
\newcommand{\ee}{\end{equation}}

\begin{document}
\volume

%
%
%
%
%

\begin{dedication}
{\bf \emph{Dedicated to my parents Subhra and Utpal Datta}}
\end{dedication}

  \begin{acknowledgments}
  I would like to begin by acknowledging my thesis advisor Prof. Erica W.
  Carlson for offering me an opportunity to pursue a doctoral degree
  under her guidance and for suggesting the thesis problem. Her constant mentoring and support, throughout
  my doctoral work, has helped me to organize my thought process, sharpen my research skills,
  and improve my scientific writing ability. Thanks to Prof. Carlson for having faith in my abilities.

  The careful mentoring of Prof. Gabriele F. Giuliani, during my initial years as a graduate student provided me the chance to
  build a strong foundation in condensed matter physics for later years. My special thanks goes to Prof. Sherwin Love who has taught
  me materials ranging from quantum mechanics to quantum field theory and whose insights in physics have strengthened my knowledge of
  the subject. Thanks to Prof. Jiangping Hu and to Prof. Yuli Lyanda Geller for the numerous fruitful discussions and meetings I have had with them
  on condensed matter physics. Dr. Matthew Grayson deserves my sincere gratitude for his guidance on the experimental details of
  the aluminum arsenide quantum wire project. Thanks to Prof. Ramdas for willing to serve in my committee and to Prof. Hirsch for
  making the department an exciting place to work in.

  I wish to acknowledge the financial support I received from the Purdue research foundation and the Bilsland
  dissertation fellowship for completing this doctoral work.

  I wish to thank my friends Dr. George Simion, Dr. Fhokrul Md. Islam, and my office mate Zoltan Gecse for
  many useful and illuminating discussions that I have had with them about physics in
  general. I also wish to thank Dr. Robert Grantham for his great friendship over the years.

  Finally, I would like to acknowledge the love and patience of my wife, Joya, without whose support this journey would not have been possible.
\end{acknowledgments}


\tableofcontents


\listoffigures

\begin{symbols}
  AlAs & Aluminum Arsenide\cr
  GaAs & Gallium Arsenide\cr
  Si   & Silicon\cr
  Ge   & Germanium\cr
  1D   & One dimensional\cr
  QWR & Quantum wire\cr
  $T$ &Temperature\cr
  $L$ & Length of the quantum wire\cr
  $w$ & Width of the quantum wire\cr
  $d$ & Distance of the quantum wire from the back gate\cr
  $r$ &Ratio of spin to charge velocity\cr
  $E_{F}$ & Fermi energy\cr
  $v_{F}$ & Fermi velocity\cr
  $k_{F}$ & Fermi wavevector\cr
  $k^{A\pm}_{F}$ & Fermi points of band $A$ in the aluminum arsenide bandstructure\cr
  $k^{B\pm}_{F}$ & Fermi points of band $B$ in the aluminum arsenide bandstructure\cr
  $k^{o}_{F}$    & Magnitude of the Fermi wavevector measured from the bottom of each band in the aluminum arsenide bandstructure\cr
  $m_{e}$ & Free electron mass\cr
  $m^{*}$ & Effective mass of electron\cr
  $n$ & Band index\cr
  $s$ & Spin index\cr
  $k_{U}$ &Umklapp vector\cr
  $v_{\rho}$& Charge velocity\cr
  $v_{\sigma}$& Spin velocity\cr
  $v^{\pm}_{\rho}$& Charge velocity in symmetric and anti-symmetric basis\cr
  $v^{\pm}_{\sigma}$& Spin velocity in symmetric and anti-symmetric basis\cr
  $K_{\rho}$& Charge Luttinger parameter\cr
  $K_{\sigma}$& Spin Luttinger parameter\cr
  $K^{\pm}_{\rho}$& Charge Luttinger parameter in symmetric and anti-symmetric basis\cr
  $K^{\pm}_{\sigma}$& Spin Luttinger parameter in symmetric and anti-symmetric basis\cr
  $\gamma_{\rho}$& Charge interaction strength\cr
  $\phi_{\rho}$ &Bosonic charge field\cr
  $\partial_{x}\theta_{\rho}$ &Conjugate momentum of $\phi_{\rho}$\cr
  $\phi_{\sigma}$ &Bosonic spin field\cr
  $\partial_{x}\theta_{\sigma}$ &Conjugate momentum of $\phi_{\sigma}$\cr
  $\alpha$ & Short distance cutoff in the bosonization theory\cr
  ${R}^{\dag}_{ns}$ &Right moving fermion creation operator in band $n$ with spin $s$\cr
  ${L}^{\dag}_{ns}$ &Left moving fermion creation operator in band $n$ with spin $s$\cr
  ${\eta}_{Rns}$ &Klein factor for the right moving fermion in band $n$ with
  spin $s$\cr
  ${\eta}_{Lns}$ &Klein factor for the left moving fermion in band $n$ with
  spin $s$\cr
  $\Lambda$ &Length scale in the renormalization group scheme\cr
  $CDW$ &Charge density wave\cr
  $SDW$ &Spin density wave\cr
  $SS$ &Singlet superconductivity \cr
  $TS$ &Triplet superconductivity\cr
  $MDC$&Momentum distribution curve\cr
  $EDC$ &Energy distribution curve\cr
  $A^{<}(k,\omega)$ &Single hole spectral function
  at momentum $k$ and energy $\omega$\cr
  $E_{\rm kink}$ &Kink energy\cr
  $FFLO$ &Fulde-Ferrell-Larkin-Ovchinnikov\cr
\end{symbols}




\begin{abstract}
This dissertation focuses on a theoretical study of interacting
electrons in one dimension. The research elucidates the ground state
(zero temperature) electronic phase diagram of an aluminum arsenide
quantum wire which is an example of an interacting one dimensional
electron liquid. Using one dimensional field theoretic methods
involving abelian bosonization and the renormalization group we show
the existence of a spin gapped quantum wire with electronic ground
states such as charge density wave and singlet superconductivity.
The superconducting state arises due to the unique umklapp
interaction present in the aluminum arsenide quantum wire
bandstructure discussed in this dissertation. It is characterized by
Cooper pairs carrying a finite pairing momentum. This is a
realization of the Fulde-Ferrell-Larkin-Ovchinnikov state which is
known to lead to inhomogeneous superconductivity. The dissertation
also presents a theoretical analysis of the finite temperature
single hole spectral function of the one dimensional electron liquid
with gapless spin and charge modes (Luttinger liquid). The hole
spectral function is measured in angle resolved photoemission
spectroscopy experiments. The results \emph{predict} a kink in the
effective electronic dispersion of the Luttinger liquid. A
systematic study of the temperature and interaction dependence of
the kink provides an alternative way to detect spin-charge
separation in one dimensional systems where the peak due to the spin
part of the spectral function is suppressed.


\end{abstract}

%
%
%

\chapter{INTRODUCTION \label{introduction}}

Electronic systems in which the kinetic energy is treated as the
starting point with the Coulomb interaction as a perturbation can be
described by a gas of weakly interacting quasiparticles. The quantum
numbers of the quasiparticle are similar to the non-interacting
particle they are derived from. The quasiparticle parameters, such
as mass and charge, are redefined to their effective values due to
interactions. The principal effects of the mutual electron-electron
interaction are assumed to be adequately captured in these effective
parameters. This is the Landau Fermi liquid paradigm
\cite{Landau,Abrikosov,Pines}. Although it has been a cornerstone of
solid state physics for over fifty years, there is increasing
experimental \cite{Stewart,Chang,Shen} and theoretical evidence
\cite{Devreese,Fradkin,Sachdev,concepts} for its inadequacy in
systems where strong correlations dominate. The band structure limit
of nearly free electrons \cite{Mermin} is not an appropriate
starting point and one should approach the problem with the
interactions considered on an equal footing with the kinetic energy
of the system\cite{Auerbach,Fradkin,Nagaosa1,Nagaosa2,Fazekas}. With
this in mind we define \emph{strongly correlated} systems as being
those in which interactions have a profound effect on the ground
state and the low-lying excitations. Examples include transition
metal oxides\cite{Fazekas}, heavy fermion compounds\cite{Radousky,
Hewson,Mohn,Levy}, quantum Hall systems
\cite{Ando,Klitzing,Gossard,Chang,Hu}, high-temperature
superconductors, \cite{concepts,Zhang} and one dimensional (1D)
electronic systems \cite{Giamarchi,Gogolin,Miranda,Haldane} among
others.

One dimensional electronic systems are inherently correlated. Due to
the reduced phase space in 1D these systems behave in a way which is
radically different from their higher dimensional counterparts, two
and three dimensional electronic systems\cite{Mermin,Landau}. The
usual concept of an electron-like elementary excitation gives away
to a more exotic class of fractionalized excitations referred to as
the spinon and the holon. These collective spin and charge modes,
respectively, are the stable elementary excitations of the
interacting 1D electron liquid and they propagate with different
velocities, a phenomenon referred to as spin-charge
separation\cite{Giamarchi}. In a spirit similar to the Landau Fermi
liquid theory, the paradigm for describing the 1D systems where the
stable elementary spin and charge excitation fields have not
acquired an expectation value (\emph{i.e.} gapless) is called
Luttinger liquid\cite{Giamarchi,Gogolin,Devreese} - a terminology
coined by Haldane\cite{Haldane}.

Interacting 1D electrons in the Luttinger liquid phase are
characterized by spin-charge
separation\cite{tomonaga1950,luttinger1960,mattis-lieb1965},
suppression of the density of states near the Fermi level, and a
power law behavior in the correlation functions. Experimental
evidence for Luttinger liquid behavior has been reported in many 1D
systems, via, {\em e.g.}, a suppression of the density of states
near the Fermi level in ropes of carbon
nanotubes\cite{bockrath-smalley1997} or power law behavior in the
conductance vs. temperature in  edge states of the fractional
quantum Hall effect\cite{milliken-webb1996,chang1996} and carbon
nanotubes\cite{bockrath-smalley1999}. Direct evidence of spin-charge
separation is evident in the measured single hole spectral function
of the Mott-Hubbard insulator SrCuO$_2$\cite{zx-2006}.

A useful quantity to detect spin-charge separation is the finite
temperature single hole spectral function\cite{Orgad} relevant in
angle resolved photoemission spectroscopy (ARPES) experiments. In
the literature, theoretical analyses of this effect have mostly been
performed for the zero temperature spectral
function\cite{voit1993,meden1992} of the Luttinger liquid. Since
experiments are performed at nonzero temperatures a reliable
comparison between theory and experiment can only be made with the
finite temperature spectral functions, as described in chapter
\ref{kink}.

In nature there are materials which are
quasi-1D\cite{Devreese,Bourbonnais,Jerome}. Experimental data on
their electronic structure and theoretical analysis of these
compounds\cite{concepts,Balents,Giam_Schulz,Fabrizio,Khveshchenko,Oron,Kusmartsev,Nersesyan,Yakovenko,Arrigoni,Dagotto,
Orignac,Finkelstein,Nagaosa,kimura,kimurapair,spingap,Shelton,Riera,Noack,Poilblanc,Tsunetsugu,White,Varma,Apel,Lin,LBF,Furusaki,Wu}
supports the idea that there may be a temperature regime where the
electronic structure can be characterized as 1D. Luttinger liquid
physics or other instabilities of the interacting 1D electron liquid
are then usually assumed to provide a correct physical description
of the ground state. However there are many assumptions behind
this\cite{concepts,Chang}. Furthermore, being quasi-1D in nature
these systems do not allow theoretical ideas\cite{Fradkin,Gogolin}
and techniques which have been primarily developed for true 1D
problems to be tested. Fortunately advances in semiconductor device
fabrication technology have led experimentalists to create systems
such as quantum wires (QWR's) \cite{Yacoby,Pfeiffer,Motohisa,Kurdak}
and carbon nanotubes (CNTs) \cite{BalentsFisher,Krotov,Egger} which
are more fitting as examples of a 1D system. The aluminum arsenide
(AlAs) QWR studied in this dissertation is such an example.

This dissertation focuses on two projects. The first is concerned
with characterizing the ground state electronic phase diagram of an
AlAs QWR. Using 1D field theoretic methods involving abelian
bosonization and the renormalization group this QWR is shown to have
a spin gapped electronic phase of matter with a novel singlet
superconducting ground state arising due to the unique umklapp
interaction present in the AlAs bandstructure. The singlet state has
Cooper pairs which carry a finite pairing momentum. This
dissertation also presents a theoretical analysis of the finite
temperature single hole Luttinger liquid spectral function measured
in the ARPES experiments. The results \emph{predict} a kink in the
effective electronic dispersion of the finite temperature Luttinger
liquid. Being a finite temperature effect all previous analyses
which focused on the zero temperature spectral function had failed
to capture this feature in the electronic dispersion. The kink
analysis provides a way to detect spin-charge separation in 1D
systems where the spin peak is muted due to repulsive interactions.

The dissertation is organized as follows. Chapter \ref{introduction}
provides a general introduction to 1D strongly correlated systems.
Chapter \ref{SCES} presents the reasons for studying the systems and
the main motivation behind the projects. Chapter \ref{theory}
introduces the interacting 1D electron liquid and, provides a brief
overview of bosonization, the renormalization group, and the finite
temperature spectral function of the Luttinger liquid. Chapter
\ref{QWR} focuses on the model Hamiltonian used to study the AlAs
QWR, the interactions present, the RG approach applied to the
system, and finally a discussion of the resulting phase diagram of
the AlAs QWR. Chapter \ref{kink} presents the work on the finite
temperature single hole spectral function of the Luttinger liquid.
These chapters are followed by a list of references and appendices
which detail the calculations.

%
%
%

\chapter{GROUND STATES AND EXPERIMENTAL SIGNATURES\label{SCES}}
One of the challenging and motivating aspect of 1D strongly correlated physics is to determine,
characterize, and explain the nature of electronic phases of matter
arising from an interplay of strong electronic correlations and low
dimensionality. Experimental signatures of these systems are equally
intriguing\cite{datta-2007}.

In general, strongly correlated
systems\cite{Fradkin,Auerbach} can support a rich variety of novel
electronic phases of matter like high-temperature superconductivity
in ceramic layered copper-oxide materials \cite{concepts} and
quantum Hall states \cite{Hu} in the two dimensional electron gas.
Some of the exotic ground states also appear in magnetic systems
where the electronic spins can order on long length-scales and give
rise to low-energy magnetic excitations called spin-waves
\cite{Mattis,Auerbach,Fradkin}. The spin degrees of freedom could
also be correlated only on short length-scales and have gapped
(confined) excitations. This is the ``spin liquid" phase \cite{Lee}.
In most cases variation of the external parameters such as
temperature, pressure or chemical doping can help tune transitions
from one novel ground state to another\cite{Sachdev}. Novel quantum
phase transitions can also be realized in artificially engineered 1D
nanostructures such as QWR's where the electrons move along one
direction but their transverse motions are quantum mechanically
confined.

Experimental probes capable of detecting Luttinger
liquid physics include ARPES\cite{Shen}, tunneling
measurements\cite{auslaender-halperin2005}, neutron
scattering\cite{tranquada-in-schrieffer}, and conductance measurements
\cite{Yacoby,Pfeiffer}. ARPES and neutron scattering experiments are
primarily used for the quasi-1D materials. The tunneling and conductance
measurements are employed for the QWR and carbon nanotube systems.

Laboratory fabricated 1D systems, such as a QWR,
present to us a unique challenge of studying a genre of correlated
electron device in which there is a theoretically controlled way of
incorporating strong electron correlations. From a broader
perspective because the device parameters in a QWR are
experimentally tunable, they offer a genuine opportunity to study
transitions between ground states in 1D, for example,
charge-density-wave (CDW) to singlet superconductivity (SS). The AlAs QWR studied in this dissertation
encourages the CDW and the SS electronic phases. In general, QWR systems allow to confirm 1D theories in a way that
is not possible in bulk three dimensional materials or quasi-1D
materials (since bulk materials are never really 1D)\cite{Devreese,concepts}.
There is a tremendous potential to control these physical systems we
study as an aid to test theoretical predictions, and perhaps even
paving the way towards designing new composite materials.

Spin-charge separation in quasi-1D systems can be probed
using ARPES measurements which measure the single hole spectral
function. The effect can be hard to detect experimentally due to
interactions, finite temperature and experimental resolution. This dissertation presents a
theoretical analysis of the finite temperature single hole Luttinger
liquid spectral function\cite{Orgad} which predicts a kink in the
effective electronic dispersion. This unique signature, a result of
finite temperature and interactions had been overlooked previously.

\section{Novel electronic ground states in a quantum wire}
A QWR is an excellent realization of a 1D system
\cite{Yacoby,Pfeiffer,Motohisa,Kurdak,Picciotto, Stormer} on which
controlled experiments and theoretical calculations \cite{Stone,
Safi, Ponomarenko, Maslov, Yaroslav} can be performed. 1D combined
with strong interactions cause the QWR's to display markedly
non-Fermi liquid behavior
\cite{Yacoby,Pfeiffer,Maslov,Ponomarenko,Safi,Stone}.

AlAs is a heavy mass system with degenerate valleys and
anisotropic mass. By exploiting the valley degeneracy in AlAs, a
single QWR has recently been fabricated with two degenerate
nonoverlapping bands separated in momentum-space by half an umklapp
vector ($k_{U}$) \cite{Moser,JM}, as illustrated in
Fig.~\ref{subfig:bandstructure}, using the cleaved edge overgrowth
technique. The arrangement of the bands in momentum-space allows for
the possibility of multiple Fermi points (more than the usual case
of two from a single band) at the Fermi
energy (see Fig.~\ref{subfig:dispersion}). The presence of multiple Fermi
points implies multiple charge and spin channels, causing a rich
phase diagram. Multiple Fermi points have been a recurring
experimental and theoretical theme in recent years within the
context of quasi-one dimensional systems \cite{Devreese,Bourbonnais,
Jerome,JeromeSchulz,Balents,Giam_Schulz,Schulz,Fabrizio,Khveshchenko,Oron,
Kusmartsev,Nersesyan,Yakovenko,
Arrigoni,Dagotto,Orignac,Finkelstein,
Nagaosa,kimurapair,spingap,Shelton,Riera,Noack,Poilblanc,Tsunetsugu,White,
Varma,Apel,Lin,LBF,Furusaki}. In the context of these 1D AlAs QWR
systems they are exciting because of the potential for
experimentally accessible new ground states. The multiple Fermi
points in this system are present even at the lowest densities. They
have a new class of interactions, the everpresent umklapp interaction (see Fig.
\ref{subfig:umklappcooper}), which has the possibility
to favor exotic electronic phases of matter; as described in chapter
\ref{QWR}. Specifically, a SS state with
Cooper pairs carrying a finite pairing momentum is encouraged. Such a state is a realization of the Fulde-Ferrell-Larkin-Ovchinnikov
(FFLO) state, but in the present context for an interacting 1D
system. This is known to lead to inhomogeneous superconductivity\cite{casalbuoni:263}.

One dimension is characterized by strong quantum fluctuations. This prevents long-range order from developing in these system at finite
temperatures. However, in order to realize a true long-range order and a phase transition at a finite temperature one can couple
a set of 1D systems to dimensionally crossover to a two dimensional system\cite{carlson:3422}. For \emph{e.g.}, in the present context an array
of QWR's could be fabricated to investigate dimensional crossover from a 1D to a two dimensional system.

\section{Experimental signature of spin-charge separation}

The interacting 1D electron liquid is characterized by spin-charge separation where the stable elementary excitations, the holon and
the spinon, propagate with different velocities. This is expected to give rise to separate spin and charge peaks in the single hole spectral
function of the Luttinger liquid. Until recently a clear experimental detection of these two peaks has proven difficult,
since the combined effects of interactions, thermal broadening, and finite experimental resolution can suppress the spin peak for
repulsive interactions. The measured peak in the effective electronic dispersion propagates with a combination of the spin and charge velocity at
low energies but with only the charge velocity at high energies. This change in velocity gives rise to a kink in the effective electronic dispersion.
A systematic study of the temperature and interaction dependence of this kink has been performed in this dissertation. This is an useful experimental
signature especially when the two peaks are not directly visible in the Luttinger liquid spectral function to confirm the spin-charge separated
nature of the material under investigation.

Angle resolved photoemission spectroscopy experiments on SrCuO$_2$\cite{zx-2006} have measured spin-charge separation by detecting the
spin and charge dispersions separately in the single hole spectral function. Previous attempts to confirm spin-charge
separation through detection of separately dispersing spin and charge peaks with ARPES\cite{claessen1995,segovia1999} have been
overturned\cite{bluebronze-not,Si-Au-not}, or lack independent verification of the spin and charge energy scales\cite{TTF-TCNQ}. Indirect experimental
evidence of spin-charge separation in 1D systems also exists. For example, the tunneling measurements via real-space imaging of Friedel oscillations using scanning tunneling
microscopy on single-walled carbon nanotubes\cite{lee-eggert2004} and momentum- and energy- resolved tunneling between two coupled
QWRs\cite{auslaender-halperin2005}.

Theoretical analysis of the finite temperature single hole spectral function\cite{datta-2007} indicates that within
Luttinger liquid theory, the spinon branch is suppressed compared to the holon branch for repulsive interactions. This presents a
difficulty in directly confirming spin and charge dispersions through measurements proportional to the single particle spectral function. Nevertheless
spin-charge separation can be detected via the systematic temperature dependence of a kink in the electronic dispersion, even in cases where the spin
peak is not directly resolvable, as described in chapter \ref{kink}.

%
%

\chapter{{INTERACTING ONE DIMENSIONAL ELECTRON LIQUID}{\label{theory}}}

\section{Luttinger liquid paradigm and Bosonization}
\begin{figure}[h]
\centering
{\includegraphics[width=2in]{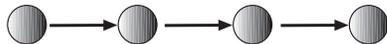}}
\caption{In a 1D interacting system an individual electron cannot
move without pushing all the neighboring electrons. As a result only
\emph{collective} excitations can exist
\cite{Giamarchi}.}\label{collective}
\end{figure}

\begin{figure}[t]
\centering
  \subfigure[Two and three dimensions]
  {\label{subfig:eholeconthigh}{\psfrag{w}{$\omega$}\psfrag{2kf}{$2k_{F}$}\psfrag{0}{$0$}\psfrag{q}{$q$}
  \includegraphics[width=2in]{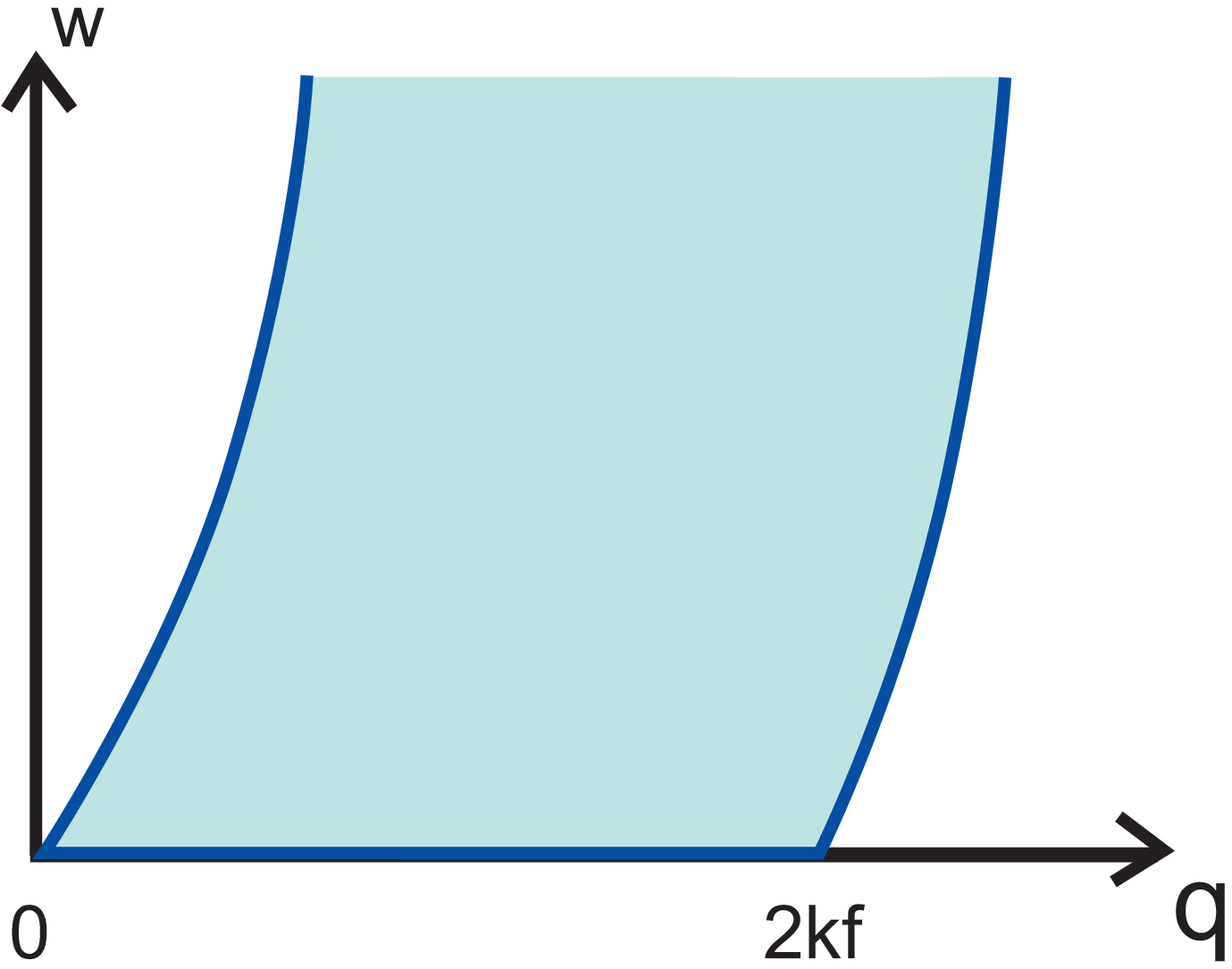}}}
  \subfigure[One dimension]{\label{subfig:eholecont1D}{\psfrag{w}{$\omega$}\psfrag{0}{$0$}\psfrag{v}{$v_{F}q$}
  \psfrag{2kf}{$2k_{F}$}\psfrag{q}{$q$}
  \includegraphics[width=2in]{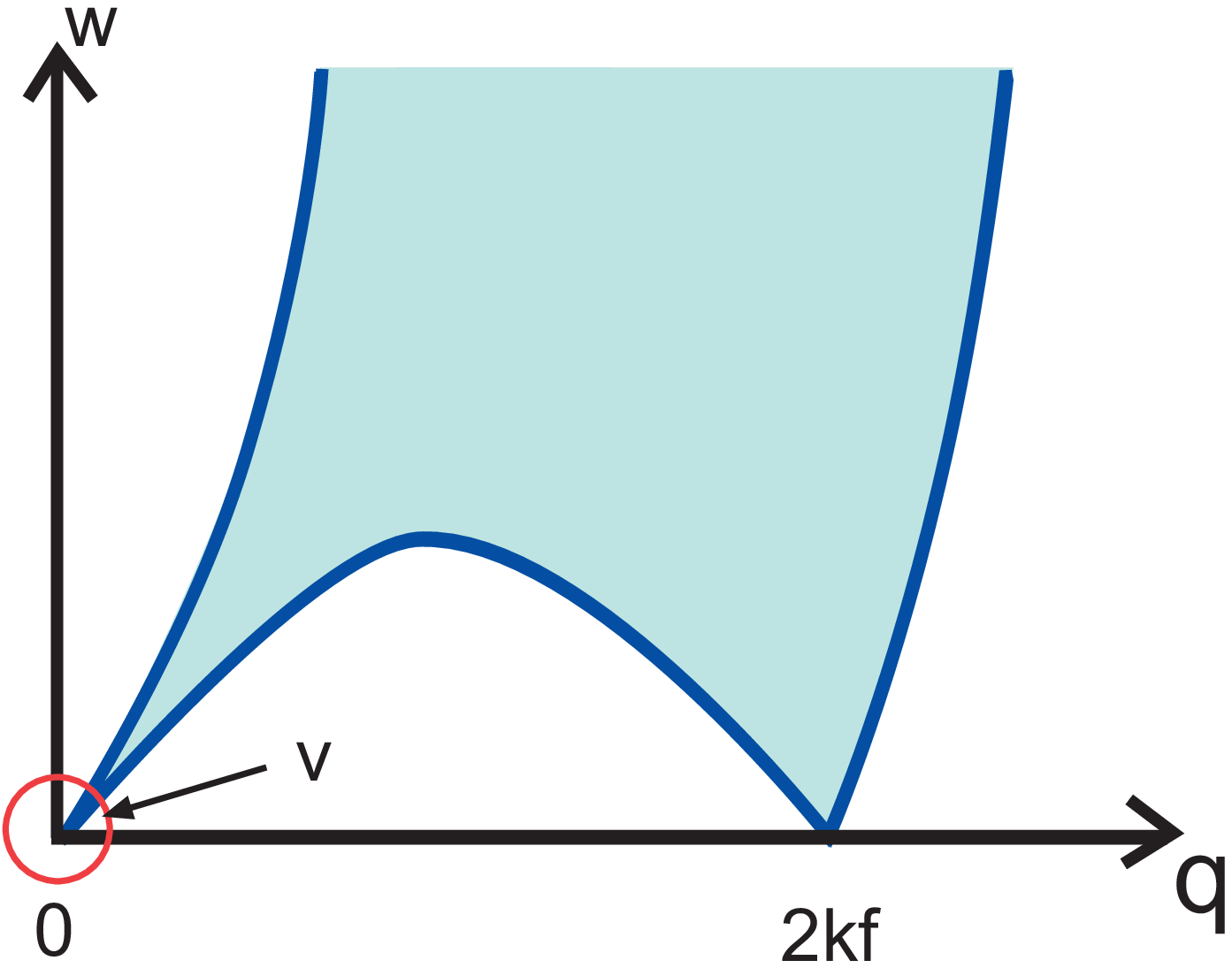}}}
\caption{Particle-hole spectrum. The momentum of the particle-hole
excitations are denoted by $q$ and their energy by $\omega$. The
Fermi velocity is given by $v_{F}$. (a) Due to the large volume of
the available phase space in two and three dimensions, a
particle-hole pair, for $q<2k_{F}$, can have a continuum of energies
extending from zero. Interactions cannot form coherently propagating
particle-hole pairs. (b) Contrary to higher dimensions, in one
dimension due to the reduced phase space the only allowed low-energy
excitations are for the two Fermi points, $q=0$ and $q=2k_{F}$.
Particle-hole excitations now have both a well defined energy and
momentum for $\omega \rightarrow 0$ and $q \rightarrow 0$. A
coherently propagating particle-hole pair can now form with a result
that collective bosonic excitations are stable \cite{Giamarchi}.}
\label{fig:eholecontinuum}
\end{figure}

Interactions have drastic effects in 1D compared to higher
dimensions. As a consequence of strong electron-electron
interactions the familiar concept of an electron-like quasiparticle
has no meaning. Due to the reduced phase space individual motion of
an electron is impossible (see Fig.~\ref{collective}), and all the
stable elementary excitations are collective (see
Fig.~\ref{fig:eholecontinuum}).
A single fermionic excitation appears to split into a collective
excitation carrying charge (holon) and another collective excitation
carrying spin (spinon). The electron is said to have
`fractionalized'. This is spin-charge separation where the
collective charge and spin modes have in general different
velocities. The minimal quantum numbers of the gapless modes are
charge, spin, and (crystal) momentum. Here the ``spin-modes" have
spin 1/2 and charge 0 (spinon), and ``charge modes" have spin 0 and
charge $e$ (holon). This is the interacting 1D electron gas where
the bosonic quasiparticles are the key to solving our 1D problem. In
this context it is important to note that the Luttinger liquid is a
particular phase of the 1D electron gas where all the charge and
spin modes are gapless \cite{Giamarchi}. The notion of a Luttinger
liquid implies that all gapless 1D electronic systems share these
properties at low energies \cite{Haldane}.

The particle-hole spectrum (see Fig.~\ref{fig:eholecontinuum}) holds
the key to understanding the nature of the stable elementary
excitations in 1D. In the low energy $(\omega \rightarrow 0)$ and
low momentum limit $(q \rightarrow 0)$, particle-hole excitations in
1D form stable collective bosonic excitations which become the basis
for solving the 1D models. As shown in Fig.~\ref{fig:eholecontinuum}
a particle-hole pair in two and three dimensions can have a
continuum of energies extending from zero for $q<2k_{F}$. Any
electron-hole pair which tries to propagate coherently decays
immediately into the electron-hole continuum. However in 1D, due to
the Pauli exclusion principle there is a volume of excluded phase
space (see Fig.~\ref{fig:eholecontinuum}). The only allowed
low-energy excitations are for the two Fermi points, $q=0$ and
$q=2k_{F}$. For low energy and low momentum the particle-hole
excitations have both a well defined energy and momentum. A
coherently propagating particle-hole pair can now form with a result
that collective bosonic excitations are stable \cite{Giamarchi}.

\begin{figure}[t]
\centering
  \subfigure[Dispersion]{\label{subfig:dispersion}{\psfrag{k}{$k$}\psfrag{E}{$E$}\psfrag{ef}{$E_{F}$}\psfrag{mkf}{$-k_{F}$}
  \psfrag{kf}{$k_{F}$}
  \includegraphics[width=2in]{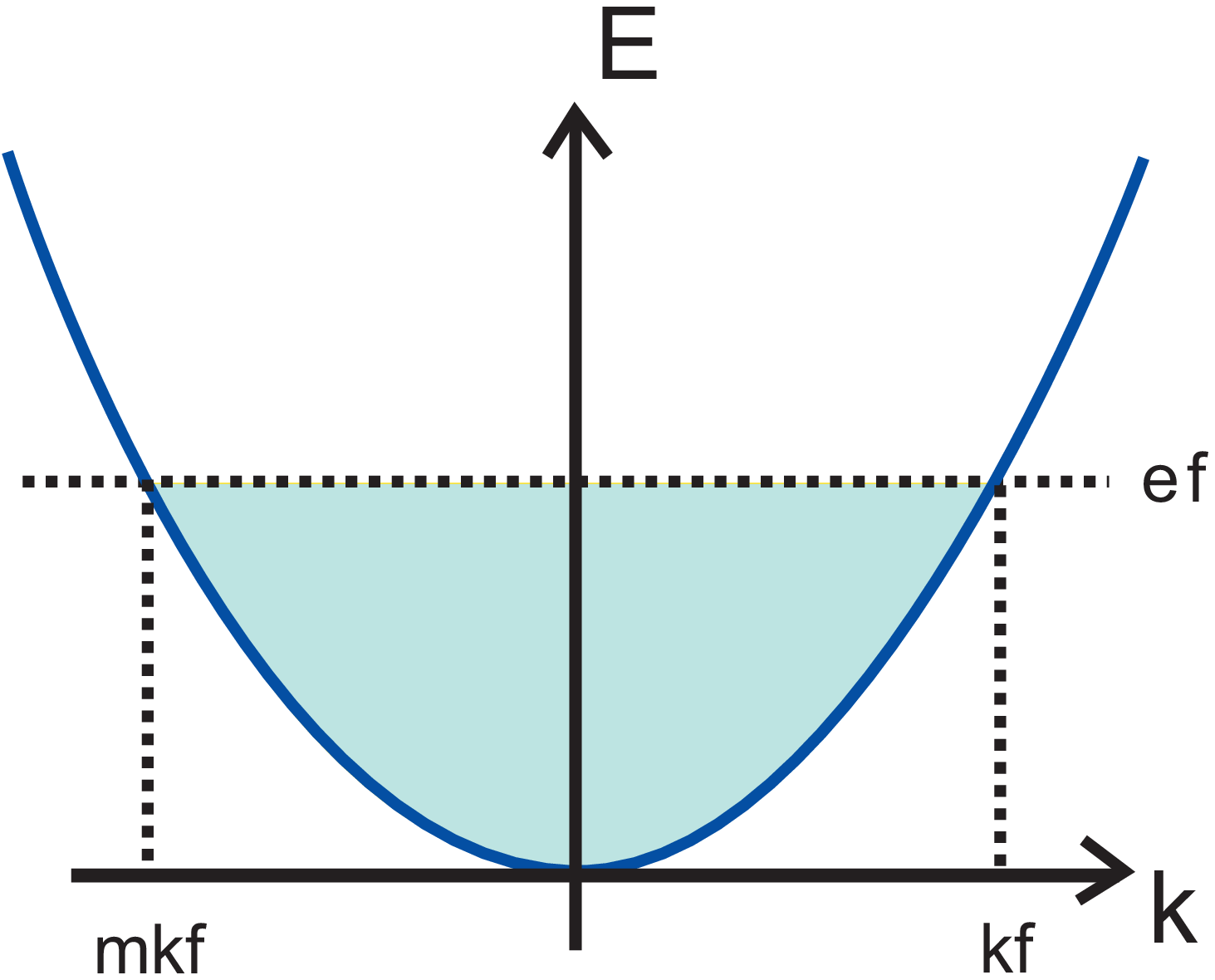}}}
  \subfigure[Linearized dispersion]{\label{subfig:lineardispersion}{\psfrag{E}{$E$}\psfrag{k}{$k$}\psfrag{R}{$R$}
  \psfrag{L}{$L$}\includegraphics[width=2in]{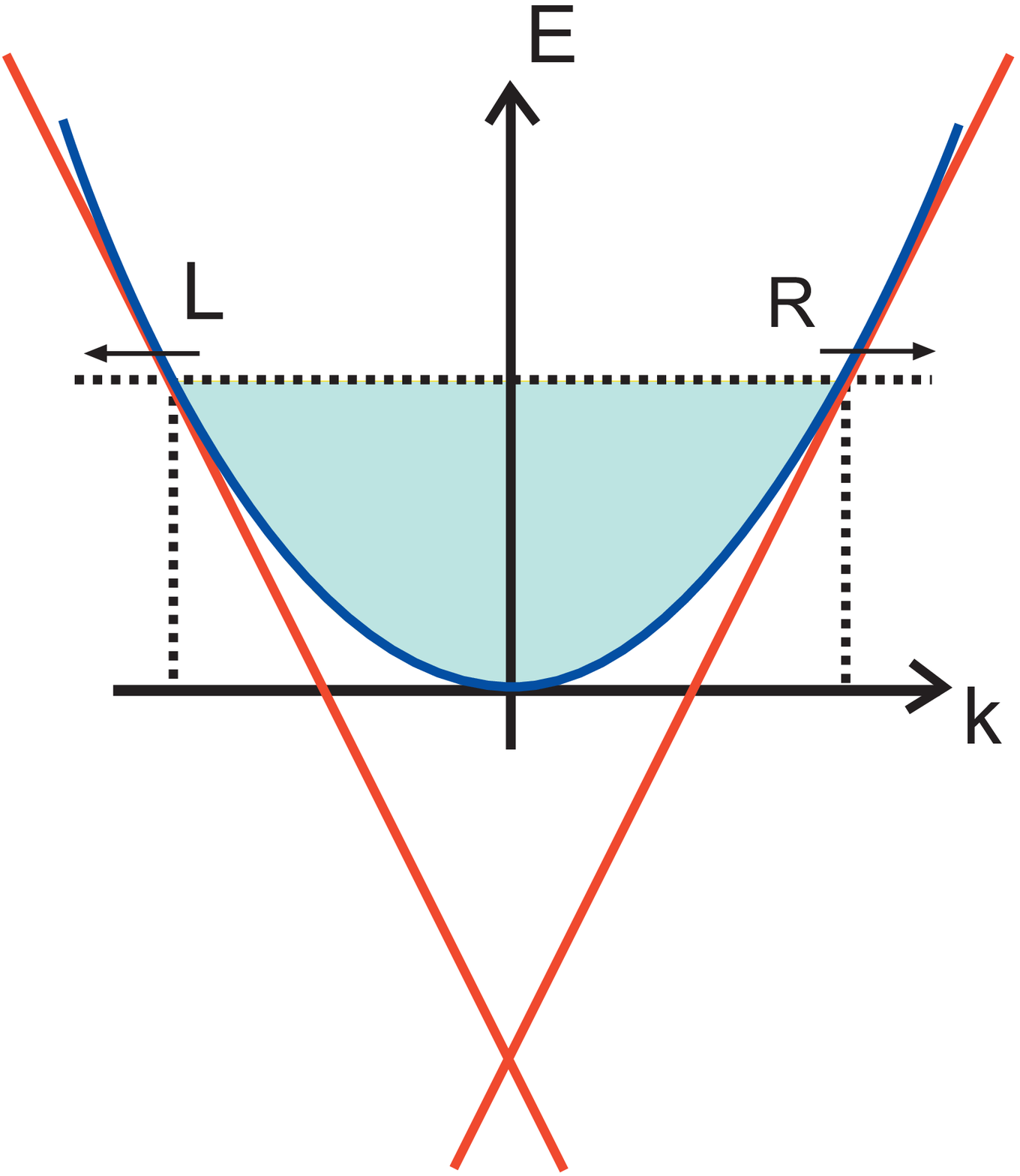}}}
\caption{The original model of fermions with band curvature (a) is
replaced by a model of fermions with a linear spectrum (b). This
causes to introduce two species of fermions (right ($R$) and left
($L$) going fermions). Here $E_{F}$ is the Fermi energy and $k_{F}$
the Fermi wavevector \cite{Giamarchi,Gogolin,Fradkin}.
\label{fig:linearization}}
\end{figure}

The continuum model of a 1D interacting electron gas consists of
approximating it by a pair of linearly dispersing branches of right-
and left- moving, $R$ and $L$, respectively, spin-half fermions
constructed around the right and left Fermi points respectively as
shown in Fig.~\ref{fig:linearization}. This approximation captures
the essential physics in the limit of low energy and long wavelength
where the only important processes involve the fermionic excitations
in the vicinity of the Fermi points. Now we use the basic idea of
\emph{bosonization} where we associate with the right- and left-
moving fermionic fields a corresponding bosonic field.
The crucial physical ingredient involves in recognizing that in 1D
the stable elementary excitations in the limit of low energy and
momentum are the collective charge and spin modes. The
\emph{bosonization identity} \cite{Miranda,Giamarchi,Fradkin} is
\begin{equation}
\psi_{\xi,s}\equiv \frac{\eta_{\xi,s}}{\sqrt{2\pi
\alpha}}\exp[-i\Phi_{\xi,s}(x)]\label{fermionicfields}
\end{equation}
which expresses the fermionic fields in terms of self dual fields
$\Phi_{\xi,s}(x)$ obeying
\begin{equation}
[\Phi_{\xi,s}(x),\Phi_{\xi,s^{'}}(x)]=-i\pi
\delta_{\xi,\xi^{'}}\delta_{s,s^{'}}sign(x^{'}-x)
\end{equation}
with $\xi=+1$ for right moving fields and $\xi=-1$ for left moving
fields. The spin index $s=\{\uparrow,\downarrow$\}. The Klein
factors $\eta_{\xi,s}$ are responsible for reproducing the correct
anticommutation relations between different Fermionic species and
$\alpha$ is the short distance cutoff that is taken to zero at the
end of the calculation. The fields $\Phi_{\xi,s}(x)$ are in turn
combinations of the bosonic fields $\phi_{\rho}$ (charge) and
$\phi_{\sigma}$ (spin) and their conjugate momenta
$\partial_{x}\theta_{\rho}$ and $\partial_{x}\theta_{\sigma}$. It is
expressed as
\begin{equation}
\Phi_{\xi,s}=\sqrt{\frac{\pi}{2}}[(\theta_{\rho} - \xi\phi_{\rho})+
s(\theta_{\sigma} - \xi\phi_{\sigma})]
\end{equation}
where $\phi_{\nu}=(\phi_{\uparrow}\pm\phi_{\downarrow})/ \sqrt{2}$,
$\theta_{\nu}=(\theta_{\uparrow}\pm
 \theta_{\downarrow})/ \sqrt{2}$, and $\nu = \rho,\sigma$ (charge and spin modes)
correspond to the $\pm$ combinations respectively. The bosonic
fields satisfy the commutation relation
[$\phi_{\nu}(x),\partial_{y}\theta_{\nu^{'}}(y)$]=
i$\pi\delta(x-y)\delta_{\nu,\nu^{'}}$. With the above identification
we can cast the original fermionic Hamiltonian in the equivalent
general bosonic form

\begin{equation}
H=\frac{1}{2}\int dx \sum_{\nu=\rho,\sigma} v_{\nu}
 \bigg[K_{\nu}(\partial_{x}\theta_{\nu})^2 +
 \frac{(\partial_{x}\phi_{\nu})^2}{K_{\nu}}\bigg] +
 H_{int}
\end{equation}
where $H_{int}$ are the bosonized interactions which may couple spin
and charge as described in chapter \ref{QWR},
Eq.~\ref{interactionsI} or they may be completely decoupled. In the
later case spin-charge separation can be formally defined as a
statement where the Hamiltonian can be expressed as a sum of two
pieces involving only charge or spin fields in the absence of
interactions mixing spin and charge modes in the bosonic theory. The
phenomenon of spin-charge separation holds for the spinfull problem
even in the presence of forward and back-scattering fermionic
interactions\cite{Giamarchi,Morandi}.

Physically $\phi_{\rho}$ and $\phi_{\sigma}$ are the phases of the
charge density wave (CDW) and spin density wave (SDW) fluctuations,
and $\theta_{\rho}$ is the superconducting phase. The parameter
$K_{\nu}$, a measure of the electron-electron interaction strength
in the theory, is referred to as the Luttinger parameter, see
appendix \ref{sec:luttparam}. For $K=1$, it refers to a
non-interacting theory. The repulsive and attractive regimes are
given by $K<1$ and $K>1$ for repulsive and attractive interactions
respectively. Furthermore, $K_{\sigma}=1$ for systems in which there
are no explicit spin symmetry breaking fields or spontaneous
breakdown of spin-rotation invariance. The velocities for the charge
and spin modes are given by $v_{\nu}$ where $\nu=\rho,\sigma$.

It is a salient feature of the 1D electron gas that all the
properties of such systems, including fermionic correlation
functions, can be expressed in terms of the bosonic fields (apart
from the Klein factors) \cite{Voit,Senechal,Morandi}. Any 1D problem
involving only the forward scattering interactions can be solved
\emph{exactly} by using the boson representation in which it is
non-interacting \cite{Giamarchi,Gogolin,Miranda}. Furthermore it is
advantageous that when spin-charge separation holds the Hamiltonian
is separable, and so wavefunctions, and therefore correlation
functions, factor.
\begin{figure}[b]
\centering {\psfrag{la}{$\Lambda$}\psfrag{lb}{$\Lambda/b$}
  \includegraphics[width=3.5in]{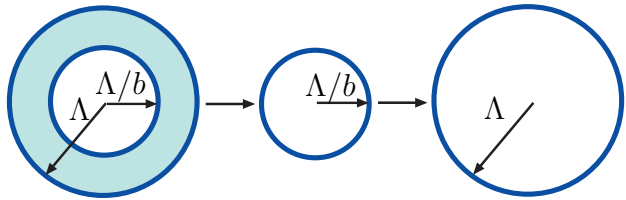}}
\caption{The momentum shell renormalization group process involves
decomposing the field into fast and slow moving fourier modes. All
the fields with wavevector $q$ lying in the momentum shell
$\Lambda/b < q <\Lambda$, with the scale factor $b$, are then
integrated out leaving a reduced volume of radius $\lambda/b$. Next,
the wavevectors are rescaled by $q^{'}= bq$. As a result we recover
the original volume in the momentum space
\cite{Goldenfeld,Ma,Wilson,Bellac,Justin}.
\label{fig:renormalization}}
\end{figure}

The effects of some perturbations on the low energy properties of
Luttinger liquids can be studied using the renormalization group
idea. In considering the 1D AlAs QWR model, as described in chapter
\ref{QWR}, where such a situation may arise we employ this scheme to
study the phase diagram. We focus on the long distance physics that
can be precisely derived from the effective bosonized field theory.
The coupling constants which appear in the problem are then
effective parameters and implicitly include much of the high energy
physics. A weak coupling perturbative renormalization group
treatment
\cite{Goldenfeld,Ma,Giamarchi,Gogolin,Lubensky,Balents,Arrigoni,spingap,Bellac,Justin}
of all the interactions is then employed to reveal the low energy,
long wavelength physics. The procedure involves in thinning the
degrees of freedom followed by a rescaling of length scales. The
physics of the problem is then studied by investigating the
dependence of the coupling constants on the length scale. The
momentum shell renormalization group procedure carried out in this
dissertation is described in Fig.~\ref{fig:renormalization}.

\section{Finite temperature spectral functions in Luttinger liquid}
The finite temperature single hole correlation function,
$G^{<}_{\xi}(x,t;T)$, is defined as
\begin{equation}
G^{<}_{\xi}(x,t;T)=\langle\psi^{\dag}_{\xi,s}(x,t)\psi_{\xi,s}(0,0)\rangle
\end{equation}
where $\xi=\pm$ for the right and left moving fermionic fields
$\psi_{\xi,s}$ (refer Eq \ref{fermionicfields}) respectively. The
spin index $s=\{\uparrow,\downarrow\}$. The spatial and temporal
coordinates are denoted by $x$ and $t$. The temperature is denoted
by $T$. Recently, explicit analytic expressions for the above finite
temperature correlation function in the Luttinger liquid have been
obtained under various conditions\cite{Orgad}.

The spectral function, $A^<(k,\omega)$, relevant for the ARPES
experiments is obtained from the above by Fourier transforming. In
the spin-rotationally invariant case, the finite-temperature single
hole spectral function\cite{Orgad} may be written in terms of the
scaled variables $\tilde{k}=\frac{v_{\sigma}k}{\pi T}$ and
$\tilde{\omega}=\frac{\omega}{\pi T}$ with the Boltzmann constant
$k_{B}=1$
\begin{eqnarray}
\lefteqn{A^{<}(\tilde{k},\tilde{\omega})\propto
\int^{\infty}_{-\infty}
dq~h_{\frac{1}{2}}(\tilde{k}-2rq)\times }\nonumber \\
&h_{\gamma_{\rho}+\frac{1}{2}}\bigg[{\tilde{\omega}-\tilde{k} \over
2}+(1+r)q\bigg] h_{\gamma_{\rho}}\bigg[{\tilde{\omega}-\tilde{k}
\over 2}-(1-r)q\bigg]
\end{eqnarray}
where $k$ is the momentum measured with respect to the Fermi
wavevector $k_{F}$, $\omega$ the energy relative to Fermi energy
$E_{F}$ and $r=v_{\sigma}/v_{\rho}$ is the ratio between the spin
velocity and the charge velocity and $h_{\gamma}$ is related to the
beta function,
\begin{equation}
h_{\gamma}(k)=\Re e\bigg[(2i)^{\gamma}B\bigg(\frac{\gamma -ik}{2},
1- \gamma \bigg)\bigg]~.
\end{equation}
The charge interaction strength $\gamma_\rho$ is related to the
charge Luttinger parameter $K_{\rho}$ by $\gamma_{\rho} =
\frac{1}{8}(K_{\rho}+K_{\rho}^{-1}-2)$, {\em i.e.} $\gamma_\rho = 0$
in the noninteracting case, and $\gamma_\rho$ increases with
increasing interaction strength. Because of spin rotation
invariance, we use $K_\sigma = 1$ and $\gamma_\sigma = 0$. The
scaled form of the spectral function arises from the critical nature
of the Luttinger liquid model. The corresponding zero temperature
expressions are also documented in the
literature\cite{voit1993,meden1992,Giamarchi}.

%
%

\chapter{PHASE DIAGRAM OF THE ALUMINUM ARSENIDE QUANTUM WIRE\label{QWR}}
\section{Introduction}
\begin{figure}[h]
\centering{\psfrag{E}{E}\psfrag{k}{k}\includegraphics[width=2in]{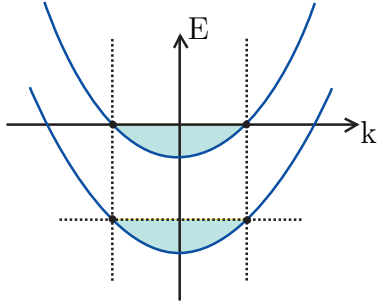}}
\caption{Bandstructure of a quantum wire where there is only one
band of lowest energy. When the first two bands are filled,
interactions between the four Fermi points lead to gaps. The
possibility of density reorganization is shown, in which it becomes
energeticaly favorable for the first two subbands to match their
Fermi momenta. This costs the kinetic energy of moving the densities
away from the noninteracting values, but gains the CDW gap energy
\cite{Starykh}. The notations $E$ and $k$ refer to energy and
momentum respectively.\label{fig:1}}
\end{figure}

In the usual realization of a QWR, transverse quantization leads to
a succession of nested energy bands. While much of the attention has
been focused on elucidating the theoretical properties of a single
band QWR
\cite{Starykh,Stone,Safi,Ponomarenko,Maslov,Apel,Kane,Fisher}, there
is a theoretical and practical urgency to focus on the novel
phenomena which may arise in these systems when more than one energy
band is involved.
\begin{figure}[h]
\centering{\includegraphics[width= 4.0in]{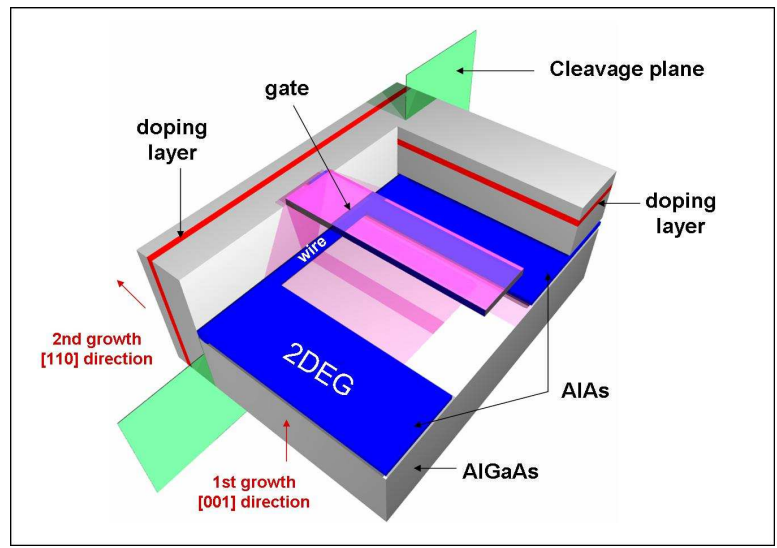}
\caption{Aluminum arsenide quantum wire fabricated using the cleaved
edge overgrowth technique. The notation 2DEG refers to the two
dimensional electron gas which couples to the quantum wire from
either side. \emph{Picture courtesy of Dr. M. Rother, Ph.D Thesis
(2000), Technische Universitaet Muenchen,
Germany.}}\label{fig:CEOQWR}}
\end{figure}
A theoretical attempt has been made in this direction by Starykh
\emph{et.al.} \cite{Starykh} where they examined and demonstrated
the possibility of gapped phases in a QWR focusing on the nested
bandstructure arrangement (see Fig.~\ref{fig:1}) which arises due to
the quantum confinement. According to their theoretical proposal
when the electronic density in such a system is tuned so that the
lowest two successive energy levels are occupied, there are gapped
phases possible, for e.g., an interband CDW, with anti-correlated
charge density waves in each band, and a ``Cooper phase" at zero
pairing momenta with strong singlet superconducting fluctuations.
However due to the possibility of density reorganization (see
Fig~\ref{fig:1}) a mechanism in which it becomes energetically
favorable for the two lowest subbands to match their densities, the
interband CDW is the most likely state. The Cooper phase may exist
just as the second band becomes occupied, when the difference in
Fermi momenta is the largest \cite{Starykh}.

In this regard one of the questions which could be posed is the
following: Is there a possibility of a robust superconducting phase
with a spin gap in a multiple Fermi point QWR? To answer the
question we consider the theoretical treatment of an AlAs QWR which
has been recently fabricated by Moser \emph{et.al.} \cite{JM,Moser}
using the cleaved edge overgrowth technique \cite{Yacoby} (see
Fig.~\ref{fig:CEOQWR}). Our calculations indicate that in the clean
limit (\emph{i.e.} no disorder) this QWR has the possibility to
realize a spin gap with a stable SS phase with finite pairing
momentum. The finite pairing of the Cooper pairs is an indication of
a Fulde-Ferrell-Larkin-Ovchinnikov (FFLO) state which is known to
lead to inhomogeneous superconductivity\cite{casalbuoni:263}.

AlAs is a heavy mass semiconductor with three degenerate valleys in
the first Brillouin zone and anisotropic mass \cite{shayegan-2006}.
The arrangement of the bands in momentum-space makes it an ideal
candidate for a system where multiple Fermi points are present even
at the lowest densities. There are two degenerate nonoverlapping
bands separated in momentum-space by half an umklapp vector
($k_{U}$), as illustrated in Fig.~\ref{fig:2}. For low densities
with only the lowest bands of transverse quantization occupied, the
density-reorganizing interband CDW instability discussed above is
forbidden. In addition, since the two (degenerate) band minima are
connected by half an umklapp vector, there is a class of umklapp
excitations unique to this bandstructure which exist {\em at all
densities} (see Fig. \ref{subfig:umklappcooper}) and which have the
possibility to favor novel electronic phases.

In this chapter, we use abelian bosonization and the weak coupling
renormalization group scheme to theoretically characterize this QWR.
We find that for repulsive interactions, the wire can be tuned so
that the ground state has a spin gap which favors either a divergent
CDW correlation function or a divergent singlet superconducting
correlation function with finite pairing momenta for the Cooper
pairs. Even though the original problem contains a repulsive
electron-electron interaction, there is a possibility of generating
an effective attraction between the electrons in the course of
renormalization of the electron-electron interaction. Such a
repulsion induced superconductivity is a novel realization in itself
and the fact that a mesoscopic system such as the AlAs QWR discussed
in this dissertation can allow for its physical realization is a
great source of excitement. The final phase has one gapless (total)
charge mode with no gapless spin modes. In the literature this is
also referred to as the C1S0 phase
\cite{Varma,Balents,spingap,Starykh}.

The outline of this chapter is as follows. In section~\ref{sec:alas}
we describe the AlAs bandstructure. In section~\ref{sec:model} we
state the model Hamiltonian used to describe the AlAs QWR and
classify the low energy long wavelength fermionic interaction
processes which are important. In section~\ref{sec:bosonization} we
bosonize the Hamiltonian in a symmetric and an anti-symmetric basis
of the bosonic fields constructed from the two bands $A$ and $B$
(see Fig.~\ref{subfig:bandstructure}). In section~\ref{sec:RG} we
derive the renormalization group equations and discuss the
electronic phase diagram. In section~\ref{sec:conclusion} we state
the conclusions.

\section{Quantum wire bandstructure\label{sec:alas}}
In bulk AlAs, conduction-band minima (or valleys) occur at the six
equivalent X-points of the Brillouin zone. The constant energy
surface consists of six half ellipsoids (three full ellipsoids in
the first Brillouin zone), with their major axes along one of the
$\langle 100 \rangle$ directions. These valleys are highly
anisotropic with an anisotropic effective mass of 1.1$m_{e}$ in the
longitudinal direction and 0.19$m_{e}$ in the transverse direction,
where $m_{e}$ is the bare mass of the electron. Since a large
effective mass leads to a reduced kinetic energy, many-body effects
related to the Coulomb potential are expected to be enhanced in AlAs
compared to the light-mass, 0.067$m_{e}$, GaAs system.

In the AlAs QWR fabricated by the cleaved edge overgrowth process
\cite{Yacoby,JM}, the initial growth direction is along [001] with
the AlAs layer flanked by GaAs and AlGaAs on either side. This
creates the two dimensional electron gas. To prepare the QWR the
heterostructure is then cleaved in the perpendicular direction
specified by [110] (the cleavage plane). The 1D channel thus formed
at the edge of the cleaved plane together with the tantulum gate
deposited on top of the two dimensional electron gas then helps to
define the length of the QWR.

The bandstructure of the QWR can be estimated, approximately, by
taking slices through the bandstructure of the corresponding two or
three dimensional material. For special systems such as CNTs, which
have cylindrical boundary conditions one can just take slices with
various transverse momenta which is a good quantum number in these
systems. However, most other physical systems do not have such
elegant boundary conditions, and in general there will be boundary
reflections that mix states of different transverse momenta. The
effect of this is that the dispersion of the lowest lying band,
rather than corresponding to a slice along a particular constant
transverse momentum, corresponds instead to a slice that tracks the
minimum of the two or three dimensional dispersion relation.
\begin{figure}[p]
\centering \subfigure[Two valleys, four Fermi
points]{\label{subfig:alasbandstruct}
  {\psfrag{x}{X-valley}\psfrag{y}{Y-valley}\psfrag{g}{$\Gamma$}\psfrag{kx}{$k_{x}$}\psfrag{ky}{$k_{y}$}
  \includegraphics[width=2.5in]{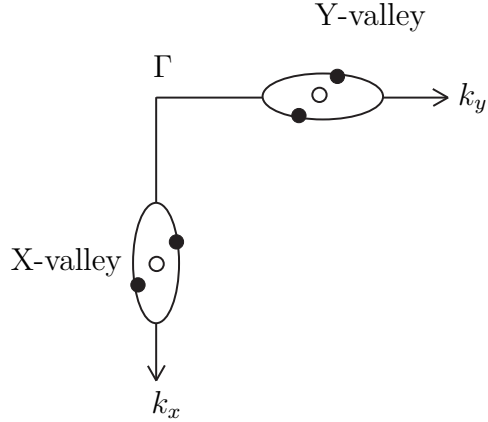}}}
  \subfigure[Quantum wire bandstructure
  wire]{\label{subfig:bandstructure}{\psfrag{k}{k}\psfrag{E}{E}\psfrag{umk}{$\frac{k_{U}}{2}$}\psfrag{A}{A}\psfrag{B}{B}
  \includegraphics[width=3.5in]{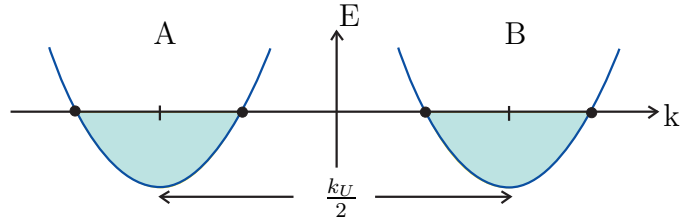}}}
\caption{(a)The aluminum arsenide quantum wire bandstructure
considered here has the four Fermi points indicated in the diagram.
The two ellipses refer to the degenerate X and Y valleys
\cite{Shkolnikov} situated at the X-point (indicated by the open
circle) of the Brillouin zone. Note the energy minimum for aluminum
arsenide is not at the $\Gamma$-point. The distance between the two
open circles is half an umklapp vector. (b)In our calculation the
quantum wire bandstructure is modelled with the two degenerate bands
labeled A and B referring to the degenerate X and Y valleys in an
aluminum arsenide bandstructure. Even at the lowest densities, there
are two degenerate subbands. In addition, the band minima are
separated by half an umklapp vector, $k_{U}/2$, giving rise to new
``umklapp'' interactions which are present at all fillings. The
Fermi points, represented by black dots on the figure, are at
$k^{A\pm}_{F}= -\frac{k_{U}}{4}\pm k^{o}_{F}$ and $k^{B\pm}_{F}=
\frac{k_{U}}{4}\pm k^{o}_{F}$ where $k_U$ is the umklapp vector, and
$k_{F}^{o}$ is the magnitude of the Fermi wavevector measured from
the bottom of each band. The notations $E$ and $k$ refer to energy
and momentum respectively. \label{fig:2}}
\end{figure}

For the cleaved edge AlAs QWR when one performs this analysis
\cite{yloh} we obtain the bandstructure arrangement as shown in
Fig.~\ref{subfig:bandstructure}. In the lowest lying band the
transverse momentum has no role to play. The bandstructure consists
of two degenerate bands $A$ and $B$, as shown in
Fig.~\ref{subfig:bandstructure}, referring to the degenerate X and Y
valleys in the AlAs bandstructure (see
Fig.~\ref{subfig:alasbandstruct}). Even at the lowest densities,
there are two degenerate subbands. In addition, the band minima are
separated by half an umklapp vector, $k_{U}/2$, giving rise to
umklapp interactions which are present at all fillings and is not
related to the commensurability of the electron gas with the
underlying lattice. The four Fermi points, represented by black dots
in Fig.~\ref{subfig:alasbandstruct} and
Fig.~\ref{subfig:bandstructure}, are at $k^{A\pm}_{F}=
-\frac{k_{U}}{4}\pm k^{o}_{F}$ and $k^{B\pm}_{F}= \frac{k_{U}}{4}\pm
k^{o}_{F}$ where $k_U$ is the umklapp vector, and $k_{F}^{o}$ is the
magnitude of the Fermi wavevector measured from the bottom of each
band.

We also note that the band structure of AlAs electrons is similar to
Si, except that in Si there are six ellipsoids centered around six
equivalent points along the $\Delta$-lines of the Brillouin zone,
while in AlAs we have three (six half) ellipsoids at the $X$-points.
Typically these valleys are denoted by the directions of their major
axes: $X, Y,$ and $Z$ for the [100], [010], and [001] valleys,
respectively. For our purposes a crucial difference between AlAs and
Si is the manner in which the valleys are occupied in a quantum well
from which eventually the QWRs are fabricated. When the electrons
are confined along the [001] direction in a (001) Si-MOSFET or a
Si/Si-Ge heterostructure, the two $Z$ valleys, with their major axes
pointing out of plane, are occupied because the larger mass of
electrons along the confinement direction lowers their energy. In
AlAs quantum wells grown on a (001) GaAs substrate, however, the $Z$
valley is occupied only if the well thickness is less than
approximately 5 $nm$ \cite{Gunawan}. For larger well thicknesses, a
biaxial compression of the AlAs layer, induced by the lattice
mismatch between AlAs and GaAs, causes the $X$ and $Y$ valleys with
their major axes lying in the plane to be occupied
\cite{shayegan-2006}. For the AlAs QWR fabricated by Moser
\emph{et.al.} \cite{JM} using the cleaved edge overgrowth technique
(see Fig.~\ref{fig:2}), the well thickness is 15 $nm$.
As a result the $X$ and $Y$ valleys are lower in energy and provide
the two degenerate conduction bands which can be occupied by
electrons. For Si such a possibility is precluded as stated above.

There are also crucial differences between the AlAs QWR and the CNT
bandstructure. In the CNTs the two bands, around the Dirac points,
between which the electron-electron scattering processes take place
are not separated by half an umklapp vector as in the AlAs QWR
\cite{BalentsFisher}. Furthermore, the umklapp interactions which
are generated in the CNT systems are present only at half-fillings
and not just at any electronic density. At half-filling, a metallic
CNT maps to the Hubbard model also at half-filling. While the
Hubbard model has umklapps at half-filling, these do not correspond
to umklapps in the original CNT. Rather, they are merely extra
interactions which are only allowed by symmetry at the Dirac points,
i.e. at half-filling. Also, a metallic tube is unlike our QWR, in
that the pseudospin which is equivalent to the sublattice quantum
number prevents an electron from backscattering from one branch to
another around the same Brillouin zone points. A doped
semiconducting tube could perhaps do this. We in fact do include
backscattering within the same subband which is strictly forbidden
in a metallic CNT.

Furthermore, empirical evidence \cite{Gunawan,Shkolnikov} suggests
that in an AlAs quantum well the spin degeneracy is not lifted in
the absence of magnetic field leading us to conclude that spin-orbit
coupling effects can be safely ignored in the theoretical
formulation of the present problem.

\section{The Fermionic Hamiltonian\label{sec:model}}
In a QWR electrons are quantum mechanically confined to move along
one direction with their motion in the remaining transverse
directions confined via a potential $V_{conf}(\vec{r_{\perp}})$
where $\vec{r}_{\perp}=(y,z)$ denotes the transverse coordinates of
quantization. Electron-electron interactions within the wire are
described by $U(\vec{r}$) which is purely repulsive. The Hamiltonian
is a sum of two independent terms in the transverse and longitudinal
directions with the result that the wavefunction (and therefore the
correlation functions) can be decomposed as a product of $
\phi(\vec{r}_{\perp})$ and $\psi_{s}(x)$ where
$\phi(\vec{r}_{\perp})$ is the orthogonal wavefunction of transverse
quantization of the two degenerate bands ($X$ and $Y$ valleys) and
$\psi_{s}(x)$ the longitudinal part. In order to describe the
physics along the longitudinal direction we now promote the
wavefunction, $\psi_{s}(x)$, to the level of a field operator (for a
field theoretic description) responsible for creating and
annihilating the electrons taking part in the various scattering
processes. With this in mind the second quantized Hamiltonian
suitable for our purposes of study is
\begin{eqnarray}\label{eq:hamiltonian}
H = \sum_{s} \int d^{3}r \Psi^{\dag}_{s}(\vec{r})
\left(-\frac{1}{2m}\vec{\nabla}^{2}_{r} -\mu +
V_{conf}(\vec{r}_{\perp})
\right )\Psi_{s}(\vec{r}) \nonumber \\
+ \frac{1}{2}\sum_{s,s^{'}} \int d^{3}r
d^{3}r^{'}U(\vec{r}-\vec{r^{'}})
\Psi^{\dag}_{s}(\vec{r})\Psi^{\dag}_{s^{'}}(\vec{r}^{'})\Psi_{s^{'}}(\vec{r}^{'})\Psi_{s}(\vec{r})
\end{eqnarray}
where $\Psi_{s}(\vec{r})= \phi(\vec{r}_{\perp})\psi_{s}(x) $ is now
the field operator for an electron species of spin
$s=\{\uparrow,\downarrow$\}, and $\mu$ is the chemical potential in
the leads. Because the low energy, long wavelength excitations occur
around the vicinity of the Fermi points (see
Fig.~\ref{subfig:bandstructure}) a further decomposition is possible
with $\Psi_{s}(\vec{r})= \phi(\vec{r}_{\perp})(\psi_{As}(x) +
\psi_{Bs}(x))$. The coordinate $x$ is in the long direction of the
wire. The longitudinal part of the field can be naturally expanded
in terms of the right- and left- moving excitations, $R_{ns}(x)$ and
${L}_{ns}(x)$, respectively, residing around the Fermi points of the
two bands (indicated by the black dots in
Fig.~\ref{subfig:bandstructure}) with $\psi_{As}(x)=
R_{As}(x)e^{ik^{A+}_{F}x} + L_{As}(x)e^{ik^{A-}_{F}x}$ and
$\psi_{Bs}(x)= R_{Bs}(x)e^{ik^{B+}_{F}x} +
L_{Bs}(x)e^{ik^{B-}_{F}x}$. The band index is $n=A,B$ and the Fermi
momenta are defined by $ k^{A\pm}_{F}=-\frac{k_{U}}{4}\pm k^{o}_{F}$
and $k^{B\pm}_{F}= \frac{k_{U}}{4}\pm k^{o}_{F}$ where $k_U$ is the
umklapp vector, and $k_{F}^{o}$ is the magnitude of the Fermi
wavevector measured from the bottom of each band, as shown in
Fig.~\ref{subfig:bandstructure}.

\begin{figure}[p]
\centering
  \subfigure[Intraband forward scattering ($U^{F}_{intra}$)]{\psfrag{E}{E}\psfrag{k}{k}\label{subfig:intraforward}
  \includegraphics[width=3.5in]{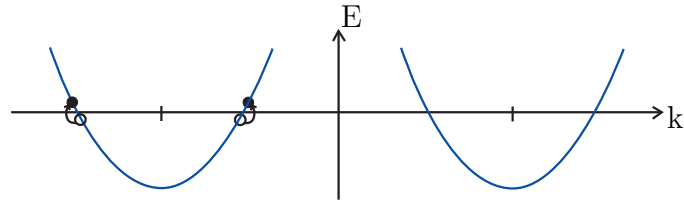}}
  \subfigure[Intraband backscattering ($U^{B}_{intra}$)]{\psfrag{E}{E}\psfrag{k}{k}\label{subfig:intrabackscattering}
  \includegraphics[width=3.5in]{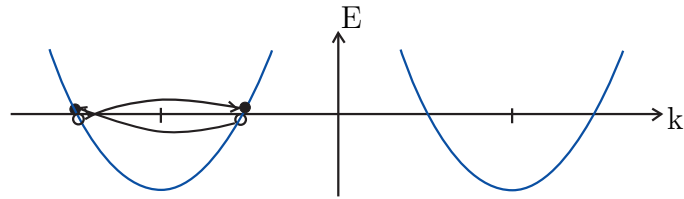}}
  \caption{{Representative low energy long wavelength \emph{intraband electron-electron
scattering processes}. In each figure $E$ and $k$ refer to energy
and momentum respectively. The open and filled dots represent
electrons which are taking part in the scattering processes with the
arrows depicting the direction of scattering. The Fermi points are
at $k^{A\pm}_{F}= -\frac{k_{U}}{4}\pm k^{o}_{F}$ and $k^{B\pm}_{F}=
\frac{k_{U}}{4}\pm k^{o}_{F}$ (as described in Fig.
\ref{subfig:bandstructure}) where $k_U$ is the umklapp vector, and
$k_{F}^{o}$ is the magnitude of the Fermi wavevector measured from
the bottom of each band.}
  \label{fig:3}}
\end{figure}

The interaction part of the Hamiltonian reduces to a sum of two
types of fermionic processes. The first type describes the
interaction of electrons within the same band, $n=A,B$, and contains
the forward $(U^{F}_{intra})$ and the backward scattering
$(U^{B}_{intra})$ processes. These are referred to as the intraband
interaction processes. The second type describes electron-electron
scattering processes involving both bands, classified as interband
interactions. The relevant interband interaction terms in the
fermionic language include forward (U$^{F}$), backward $(U^{B}_{d},
U^{B}_{x}$ and $U^{B}_{inter})$, and inter-valley umklapp $(U_{um}$)
scattering. The Fermionic interaction terms of the problem are
expressed below via the right- and left- moving excitations,
$R_{ns}(x)$ and $L_{ns}(x)$. In the interaction terms stated below
the band index is $n=A,B$, the spin index is
$s=\{\uparrow,\downarrow \}$, $k_U$ is the umklapp vector and
$k_{F}^{o}$ is the magnitude of the Fermi wavevector measured from
the bottom of each band as shown in Fig.~\ref{subfig:bandstructure}.
The notation $\xi_{n}=\delta_{nA}- \delta_{nB}$. The matrix element
$M$, which is the interaction kernel, is given by the expression
\begin{equation}\label{eqn:intxnkernel}
M(x-x^{'})=\int
U(\vec{r}-\vec{r}^{'})\phi^{2}(\vec{r}_{\perp})\phi^{2}(\vec{r}^{'}_{\perp})d\vec{r}_{\perp}d\vec{r}^{'}_{\perp}.
\end{equation}
The Fermionic interaction terms are as follows
\\ \\
{\bf Intraband interaction} (see Fig.~\ref{fig:3})
\\ \\
The intraband forward and backward scattering terms are shown in
Fig. \ref{fig:3} and their expressions stated in
Eqs.~\ref{eqn:intraforward} and ~\ref{eqn:intrabackscattering}. The
intraband forward scattering interaction term, $U^{F}_{intra}$, is
\begin{eqnarray}
U^{F}_{intra}=\frac{1}{2}\sum_{n=A,B}\int dx dx^{'}
M(x-x^{'})\sum_{s,s^{'}}\bigg[R^{\dag}_{ns}(x)R_{ns}(x)
+L^{\dag}_{ns}(x)L_{ns}(x)\bigg]\nonumber\\
\times\bigg[R^{\dag}_{ns^{'}}(x^{'})R_{ns^{'}}(x^{'})+L^{\dag}_{ns^{'}}(x^{'})L_{ns^{'}}(x^{'})\bigg]\label{eqn:intraforward}\end{eqnarray}

and the intraband backscattering term, $U^{B}_{intra}$ is
\begin{eqnarray}
U^{B}_{intra}=\frac{1}{2} \sum_{n=A,B}\int
dxdx^{'}M(x-x^{'})\sum_{s,s^{'}}
\bigg[R^{\dag}_{ns}(x)L^{\dag}_{ns^{'}}(x^{'})R_{ns^{'}}(x^{'})L_{ns}(x)
e^{-2ik^{o}_{F}(x-x^{'})}\nonumber\\+L^{\dag}_{ns}(x)R^{\dag}_{ns^{'}}(x^{'})L_{ns^{'}}(x^{'})R_{ns}(x)
e^{2ik^{o}_{F}(x-x^{'})}\bigg]\label{eqn:intrabackscattering}\end{eqnarray}

\begin{figure}[htp]
\centering
  \subfigure[Forward scattering ($U^{F}$)]{\psfrag{E}{E}\psfrag{k}{k}\label{subfig:interforward}\includegraphics[width=3.5in]{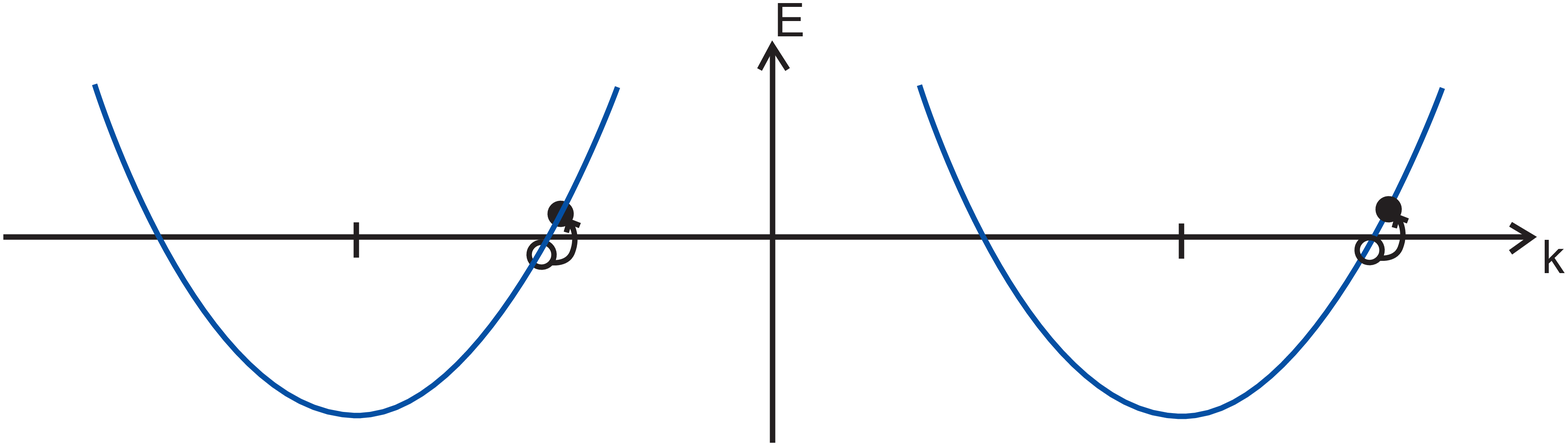}}
  \subfigure[Direct backscattering ($U^{B}_{d}$)]{\psfrag{E}{E}\psfrag{k}{k}
  \label{subfig:directbackscattering}\includegraphics[width=3.5in]{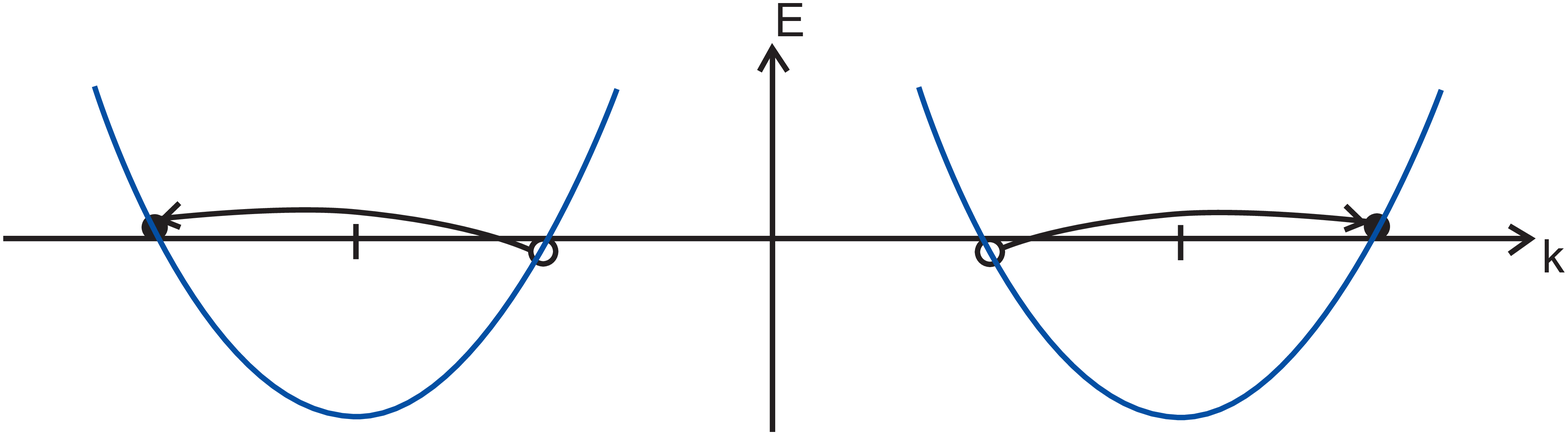}}
  \subfigure[Exchange backscattering ($U^{B}_{x}$)]{\label{subfig:xbackscattering}{\psfrag{E}{E}\psfrag{k}{k}
  \includegraphics[width=3.5in]{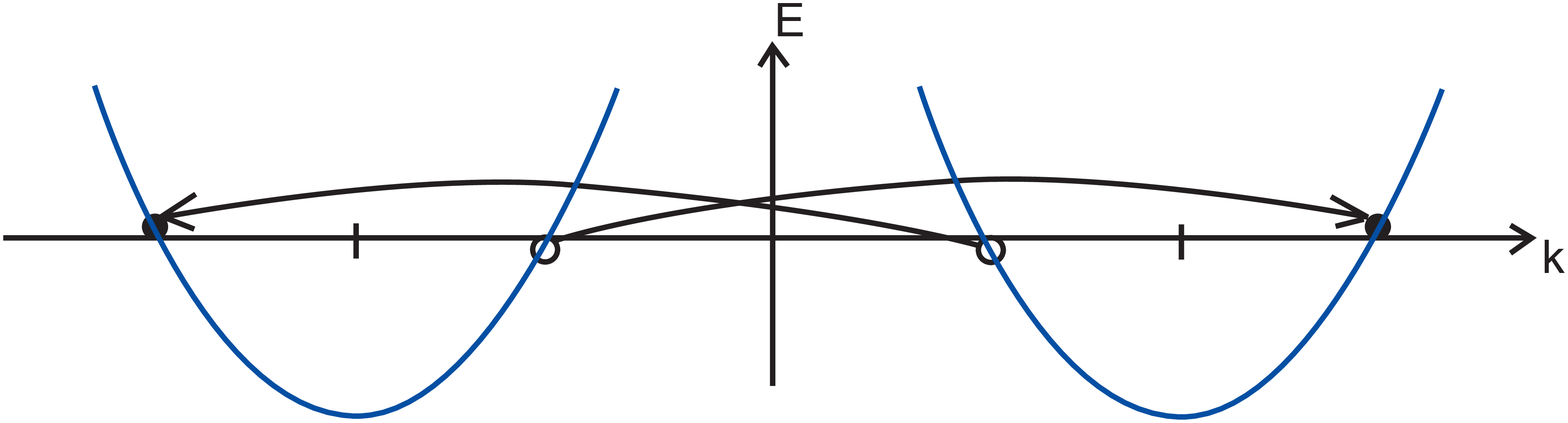}}}
\subfigure[Interband backscattering
($U^{B}_{inter}$)]{\psfrag{k}{k}\psfrag{E}{E}\label{subfig:interbackscattering}\includegraphics[width=3.5in]{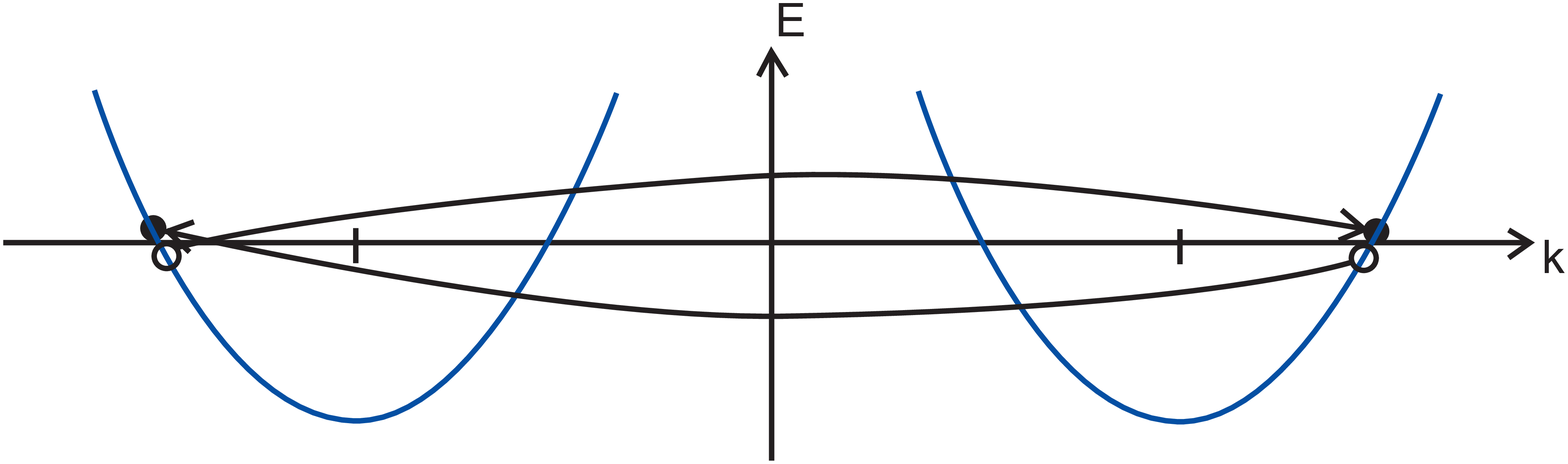}}
 \subfigure[Inter-valley umklapp scattering
 ($U_{um}$)]{\psfrag{E}{E}\psfrag{k}{k}\label{subfig:umklappcooper}\includegraphics[width=3.5in]{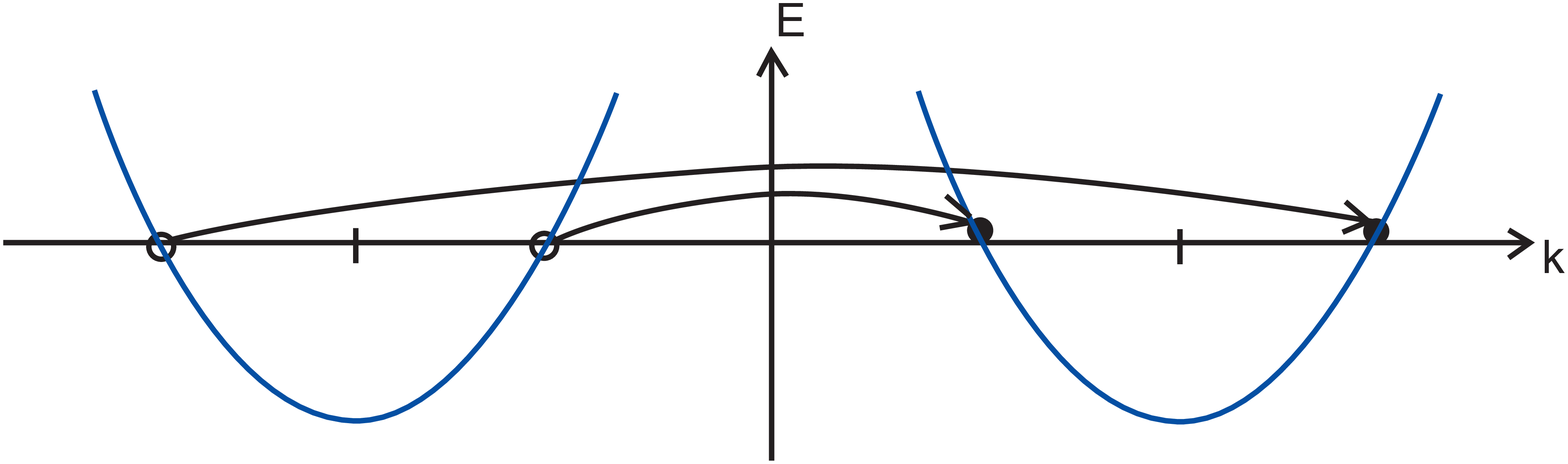}}
\caption{{Representative low energy long wavelength \emph{interband
electron-electron scattering processes}. In each figure $E$ and $k$
refer to energy and momentum respectively. The open and filled dots
represent electrons which are taking part in the scattering
processes with the arrows depicting the direction of scattering. The
Fermi points are at $k^{A\pm}_{F}= -\frac{k_{U}}{4}\pm k^{o}_{F}$
and $k^{B\pm}_{F}= \frac{k_{U}}{4}\pm k^{o}_{F}$ (as described in
Fig. \ref{subfig:bandstructure}) where $k_U$ is the umklapp vector,
and $k_{F}^{o}$ is the magnitude of the Fermi wavevector measured
from the bottom of each band. The inter-valley umklapp scattering
(Fig. \ref{subfig:umklappcooper}) shown above is unique to this
aluminum arsenide quantum wire bandstructure arrangement.}
\label{fig:4}}
\end{figure}

The interband scattering terms are shown in Fig. \ref{fig:4} and
their expressions stated in Eqs.~\ref{eqn:forwardscattering} --
~\ref{eqn:umklapcooper}.

{\flushleft \bf Interband interaction} (see Fig.~\ref{fig:4})\\
\\The forward scattering term, $U^{F}$, is
\begin{eqnarray}
U^{F}=\frac{1}{2}\sum_{n\neq m}\int dx dx^{'}
M(x-x^{'})\sum_{s,s^{'}}\bigg[R^{\dag}_{ns}(x)R_{ns}(x)
+L^{\dag}_{ns}(x)L_{ns}(x)\bigg]\nonumber\\\times\bigg[R^{\dag}_{ms^{'}}(x^{'})R_{ms^{'}}(x^{'})+L^{\dag}_{ms^{'}}(x^{'})L_{ms^{'}}(x^{'})\bigg]
\label{eqn:forwardscattering}
\end{eqnarray}
The direct backscattering term, $U^{B}_{d}$, is
\begin{eqnarray}
U^{B}_{d}=\frac{1}{2} \sum_{n\neq m}\int
dxdx^{'}M(x-x^{'})\sum_{s,s^{'}}
\bigg[R^{\dag}_{ns}(x)L_{ns}(x)L^{\dag}_{ms^{'}}(x^{'})R_{ms^{'}}(x^{'})
e^{-2ik^{o}_{F}(x-x^{'})}\nonumber
\\+L^{\dag}_{ns}(x)R_{ns}(x)R^{\dag}_{ms^{'}}(x^{'})L_{ms^{'}}(x^{'})
e^{2ik^{o}_{F}(x-x^{'})}\bigg]\nonumber\\\label{eqn:directbackscattering}
\end{eqnarray}
The exchange backscattering term, $U^{B}_{x}$, is
\begin{eqnarray}
U^{B}_{x}=-\frac{1}{2}\sum_{n\neq m}\int dx
dx^{'}M(x-x^{'})e^{i\frac{k_{U}}{2}(x-x^{'})\xi_{n}}\sum_{s,s^{'}}
\bigg[R^{\dag}_{ns}(x)L_{ns^{'}}(x^{'})L^{\dag}_{ms^{'}}(x^{'})R_{ms}(x)
\nonumber
\\+L^{\dag}_{ns}(x)R_{ns^{'}}(x^{'})R^{\dag}_{ms^{'}}(x^{'})L_{ms}(x)
\bigg]\nonumber\\\label{eqn:exchangebackscattering}
\end{eqnarray}
The interband backscattering term, $U^{B}_{inter}$, is
\begin{eqnarray}
U^{B}_{inter}=-\frac{1}{2}\sum_{n\neq m}\int dx dx^{'}
M(x-x^{'})e^{i\frac{k_{U}}{2}(x-x^{'})\xi_{n}}\sum_{s,s^{'}}
\bigg[R^{\dag}_{ns}(x)R_{ns^{'}}(x^{'})R^{\dag}_{ms^{'}}(x^{'})R_{ms}(x)
\nonumber\\+
L^{\dag}_{ns}(x)L_{ns^{'}}(x^{'})L^{\dag}_{ms^{'}}(x^{'})L_{ms}(x)
+R^{\dag}_{ns}(x)R_{ns^{'}}(x^{'})L^{\dag}_{ms^{'}}(x^{'})L_{ms}(x)e^{-i2k^{o}_{F}(x-x^{'})}
\nonumber
\\+L^{\dag}_{ns}(x)L_{ns^{'}}(x^{'})R^{\dag}_{ms^{'}}(x^{'})R_{ms}(x)
e^{i2k^{o}_{F}(x-x^{'})}\bigg]\nonumber
\\\label{eqn:interbandbackscattering}\end{eqnarray} and finally the
inter-valley umklapp scattering term, $U_{um}$, is
\begin{eqnarray}
&U_{um}=\frac{1}{2}\sum_{n\neq m}\int dx
dx^{'}M(x-x^{'})e^{i\frac{k_{U}}{2}(x-x^{'})\xi_{n}}\sum_{s,s^{'}}
\bigg[R^{\dag}_{ns}(x)L^{\dag}_{ns^{'}}(x^{'})e^{-ik^{o}_{F}(x-x^{'})}\nonumber
\\+&L^{\dag}_{ns}(x)R^{\dag}_{ns^{'}}(x^{'})
e^{ik^{o}_{F}(x-x^{'})}\bigg]
\bigg[R_{ms^{'}}(x^{'})L_{ms}(x)e^{-ik^{o}_{F}(x-x^{'})}+L_{ms^{'}}(x^{'})R_{ms}(x)e^{ik^{o}_{F}(x-x^{'})}\bigg]
\nonumber\\\label{eqn:umklapcooper}
\end{eqnarray}
The interband forward scattering term is shown in
Fig.~\ref{subfig:interforward}. The backscattering term has three
types - direct, exchange and interband. We define direct
backscattering, U$^{B}_{d}$, as an event where backscattering
processes occur in each band separately (see
Fig.~\ref{subfig:directbackscattering}). Exchange backscattering,
U$^{B}_{x}$, has the same initial and final states as direct
backscattering, but the electrons switch bands (see
Fig.~\ref{subfig:xbackscattering}). Interband backscattering,
$U^{B}_{inter}$, is backscattering between two electrons in
different bands (see Fig.~\ref{subfig:interbackscattering}). The
inter-valley umklapp scattering is unique to this bandstructure (see
Fig.~\ref{subfig:umklappcooper}). It is a scattering process where
two electrons starting, for \emph{e.g.}, in band A with total
momentum -$k_{U}$/2 scatter into band B, with total momentum
$k_{U}$/2. The momentum difference between the initial and final
states is an umklapp vector which can be exchanged with the lattice.
This is an umklapp process which is \emph{present at all densities}
and is not related to the commensurability of the electron gas with
the underlying lattice. It is unique to the AlAs QWR bandstructure
and as we shall see later will play an important role in classifying
the phase diagram.
\section{Bosonizing the Hamiltonian\label{sec:bosonization}}
The low energy properties of the interacting 1D electron gas can be
conveniently described within the framework of the bosonization
technique \cite{Miranda,Giamarchi,Gogolin,Devreese}. Within this
approach one can associate with the right- and left- moving,
${R}_{ns}(x)$ and ${L}_{ns}(x)$, respectively, fermionic field
operators a combination of bosonic fields
$\phi_{n\nu}=(\phi_{n\uparrow}\pm\phi_{n\downarrow})/ \sqrt{2}$ and
$\theta_{n\nu}=(\theta_{n\uparrow}\pm\theta_{n\downarrow})/
\sqrt{2}$, where $\nu = \rho,\sigma$ (the charge and spin modes)
correspond to the $\pm$ combination, $s=\{\uparrow,\downarrow$\} is
the spin index and $n=A,B$ is the band index. The bosonic fields
satisfy the commutation relation
[$\phi_{n\nu}(x),\partial_{x^{'}}\theta_{n^{'}\nu^{'}}(x^{'})$]=i$\pi\delta(x-x^{'})\delta_{\nu,\nu^{'}}\delta_{n,n^{'}}$
with $\hbar$ set equal to one. We then have
\begin{equation}
{R}_{ns}(x) = \frac{\eta_{Rns}}{\sqrt{2\pi\alpha}}e^{i\sqrt{\pi
\over 2} [\theta_{n\rho}(x)- \phi_{n\rho}(x) +s(\theta_{n\sigma}(x)
- \phi_{n\sigma}(x))]}
\end{equation}and,
\begin{equation}
{L}_{ns}(x) = \frac{\eta_{Lns}}{\sqrt{2\pi\alpha}}e^{i\sqrt{\pi
\over 2} [\theta_{n\rho}(x)+ \phi_{n\rho}(x) +
s(\theta_{n\sigma}(x)+ \phi_{n\sigma}(x))]}
\end{equation}
where $\alpha$ is the short distance cutoff, $\eta_{Rns}$ and
$F_{Lns}$ are the Klein factors for the right- and left- moving
fields of band $n$ with species of spin $s$. They are required to
preserve the anti-commutation relations of the fermionic fields. The
convenient field variables for the Hamiltonian in our problem will
be a linear combination of the boson fields constructed out of the
two bands. We define the transformation to a symmetric and an
anti-symmetric basis as $\phi^{\pm}_{\nu}
=\frac{1}{\sqrt{2}}(\phi_{A\nu} \pm \phi_{B\nu})$ and
$\theta^{\pm}_{\nu} = \frac{1}{\sqrt{2}}(\theta_{A\nu} \pm
\theta_{B\nu})$. Upon bosonization and subsequent transformation the
parts of the Hamiltonian corresponding to free motion produce the
harmonic terms in the symmetric and the anti-symmetric bosonic
fields $(\phi^{\pm}_{\nu}$ and $\theta^{\pm}_{\nu})$ only. The
intraband interactions and the interband interactions, however,
generate both cosine interaction terms and harmonic terms.
Furthermore while bosonizing we also used the fact that the
exponential $e^{i k_{U}x}=1$. This is true because the electrons in
the QWR move in an underlying lattice and their spatial coordinate
$x=ma_{x}$, where $a_{x}$ is the lattice spacing in the long
direction of the wire and $m$ is an integer. When multiplied with
the umklapp vector, $k_{U}=2\pi/a_{x}$, and exponentiated the result
is $\exp[-2m\pi]=1$.

The Hamiltonian, $H$, can be written in the following canonical form
\begin{equation}\label{eq:bosonizedhamiltonian}
H=\frac{1}{2}\sum_{\mu=\pm}\sum_{\nu=\rho,\sigma}\int dR
\bigg[v^{\mu}_{\nu}K^{\mu}_{\nu}(\partial_{R}\theta^{\mu}_{\nu})^{2}
              + \frac{v^{\mu}_{\nu}}{K^{\mu}_{\nu}}(\partial_{R}\phi^{\mu}_{\nu})^{2}\bigg]+H_{int}
\end{equation}
where the bosonic interaction terms, $H_{int}$, are
 \begin{eqnarray}
 \label{interactionsI}
 H_{int}=\frac{t_{1}}{2\pi^{2}\alpha^{2}}\int dR \cos[\sqrt{4\pi}\phi^{+}_{\sigma}]\cos[\sqrt{4\pi}\phi^{-}_{\sigma}]\nonumber\\
 +\frac{t_{2}}{2\pi^{2}\alpha^{2}}\int dR \cos[\sqrt{4\pi}\phi^{-}_{\rho}]\cos[\sqrt{4\pi}\phi^{+}_{\sigma}]\nonumber\\
 +\frac{t_{3}}{2\pi^{2}\alpha^{2}}\int dR \cos[\sqrt{4\pi}\phi^{-}_{\rho}]\cos[\sqrt{4\pi}\phi^{-}_{\sigma}]\nonumber\\
 +\frac{t_{4}}{2\pi^{2}\alpha^{2}}\int dR \cos[\sqrt{4\pi}\phi^{-}_{\rho}]\cos[\sqrt{4\pi}\theta^{-}_{\sigma}]\nonumber\\
 +\frac{t_{5}}{2\pi^{2}\alpha^{2}}\int dR \cos[\sqrt{4\pi}\theta^{-}_{\sigma}]\cos[\sqrt{4\pi}\phi^{-}_{\sigma}]\nonumber\\
 +\frac{t_{6}}{2\pi^{2}\alpha^{2}}\int dR \cos[\sqrt{4\pi}\theta^{-}_{\sigma}]\cos[\sqrt{4\pi}\phi^{+}_{\sigma}]\nonumber\\
 +\frac{t_{7}}{2\pi^{2}\alpha^{2}}\int dR \sin[\sqrt{4\pi}\theta^{-}_{\sigma}]\sin[\sqrt{4\pi}\phi^{+}_{\sigma}]\nonumber \\
 +\frac{t_{8}}{2\pi^{2}\alpha^{2}}\int dR \cos[\sqrt{4\pi}\theta^{-}_{\rho}]\cos[\sqrt{4\pi}\theta^{-}_{\sigma}]\nonumber\\
 +\frac{t_{9}}{2\pi^{2}\alpha^{2}}\int dR  \cos[\sqrt{4\pi}\theta^{-}_{\rho}]\cos[\sqrt{4\pi}\phi^{+}_{\sigma}]\nonumber\\
 +\frac{t_{10}}{2\pi^{2}\alpha^{2}}\int dR \cos[\sqrt{4\pi}\theta^{-}_{\rho}]\cos[\sqrt{4\pi}\phi^{-}_{\sigma}]\nonumber\\
 +\frac{t_{11}}{2\pi^{2}\alpha^{2}}\int dR \cos[\sqrt{4\pi}\theta^{-}_{\rho}]\cos[\sqrt{4\pi}\phi^{-}_{\rho}]
 \end{eqnarray}
with the coupling constants defined in terms of the Coulomb matrix
element $M(a)$
\begin{eqnarray}
\label{initconditions}
 &t_{1}=t_{2}=2\int da M(a)\cos[2k^{o}_{F}a] \nonumber \\
 &t_{3}= 2\int da M(a)\bigg(\cos[2k^{o}_{F}a]-\cos\bigg[\frac{k_{U}a}{2}\bigg]\bigg) \nonumber \\
 &t_{4}= t_{5}= -t_{10}= -t_{11}= -2\int M(a)\cos\bigg[\frac{k_{U}a}{2}\bigg] \nonumber \\
 &t_{6}= -t_{9}= -2\int da M(a)\cos[2k^{o}_{F}a]\cos\bigg[\frac{k_{U}a}{2}\bigg]\nonumber \\
 &t_{7}=  -2\int da M(a)\sin[2k^{o}_{F}a]\sin \bigg[\frac{k_{U}a}{2}\bigg]\nonumber \\
 &t_{8}= 2\int da M(a)\bigg(1- \cos[2k^{o}_{F}a]\bigg)\cos \bigg[\frac{k_{U}a}{2}\bigg]
 \end{eqnarray}

In the Hamiltonian, $R=(x+x^{'})/2$  is the center-of-mass
coordinate of two electrons and $a= x -x^{'}$ their relative
coordinate in the long direction of the QWR. In the quadratic part
the bare symmetric and anti-symmetric Luttinger parameters,
$K^{\pm}_{\nu}$, quoted in appendix \ref{sec:luttparam}, can be
expressed in terms of the original Luttinger parameters $K_{\nu}$.
Furthermore the Luttinger parameters which have been derived here
starting from the Fermionic Hamiltonian, Eq.~\ref{eq:hamiltonian},
are the bare effective parameters\cite{Giamarchi}. The symmetric and
anti-symmetric velocities $v^{\pm}_{\nu}$, refer appendix
\ref{sec:luttparam}, can also be expressed in terms of the original
velocities $v_{\nu}$.

The coupling constants of the problem are denoted by $t_{i}$, where
$i=1,\dots,11$. They are related to the fourier components of the
interaction kernel, $M(a)$, with the wavevectors of the problem -
$k_{U}$ and $k_{F}^{o}$. It is instructive to observe that the
umklapp vector is independent of the electron density in the QWR
whereas $k_{F}^{o}$ is not. This allows for the interesting
possibility to tune $k_{F}^{o}$ experimentally and control the
initial conditions in the renormalization group equations itself and
possibly get oneself into various electronic phases of matter as
allowed in an interacting 1D electron gas problem. Also, the initial
conditions depend on the QWR parameters - the width ($w$), the
length $(L)$, and the distance of the QWR from the back-gate $(d)$.

The first bosonized interaction term with the coupling $t_{1}$
arises from intraband backscattering type processes (refer
Fig.~\ref{subfig:intrabackscattering}) with a coupling strength
which depends on the $2k_{F}^{o}$ cosine fourier component of the
interaction kernel $M(a)$. The second is a result of the direct
backscattering type interaction as shown in
Fig.~\ref{subfig:directbackscattering} and has a coupling strength
equal to that of an intraband backscattering process. This is
evident from the diagram of the scattering processes shown in
Figs.~\ref{subfig:intrabackscattering} and
~\ref{subfig:directbackscattering}. The third interaction term has
contributions from both the direct and exchange backscattering
processes, whereas the fourth term is exclusively generated from
exchange backscattering. The exchange backscattering process has a
strength which depends on the $k_{U}/2$ cosine fourier component of
the interaction kernel $M(a)$. The third interaction term is a
combination of both and has the $2k_{F}^{o}$ and the $k_{U}/2$
cosine fourier components. These first four interaction terms have
in their bosonized expressions the charge density wave phase
$\phi_{\rho}^{-}$  for the relative charge channel in the
antisymmetric basis together with the spin fields
$\phi_{\sigma}^{+}, \phi_{\sigma}^{-}$, and $\theta_{\sigma}^{-}$.
The next three terms $t_{5},t_{6}$, and $t_{7}$ are generated from
the interband backscattering process shown in
Fig.~\ref{subfig:interbackscattering}. They involve only the spin
fields $\phi_{\sigma}^{+}, \phi_{\sigma}^{-}$, and
$\theta_{\sigma}^{-}$. For $t_{6}$ and $t_{7}$ the coupling depends
on both the $2k_{F}^{o}$ and $k_{U}/2$ cosine fourier component of
the interaction kernel. The strength of $t_{5}$ is proportional to
$\frac{k_{U}}{2}$.

Bosonization of the inter-valley umklapp scattering fermionic
processes produce six bosonized interaction terms in total. Out of
those, two are of the form
\begin{eqnarray}
\sim
\cos[\sqrt{4\pi}\theta^{-}_{\rho}]\cos[\sqrt{4\pi}\phi^{-}_{\rho}]\cos[\sqrt{4\pi}\theta^{-}_{\sigma}]\cos[\sqrt{4\pi}\phi^{-}_{\sigma}]\nonumber
\\
\sim
\sin[\sqrt{4\pi}\theta^{-}_{\rho}]\sin[\sqrt{4\pi}\phi^{-}_{\rho}]\sin[\sqrt{4\pi}\theta^{-}_{\sigma}]\sin[\sqrt{4\pi}\phi^{-}_{\sigma}]
\end{eqnarray}
 From a purely physical standpoint these terms involve self
destructing competing dual fields. Furthermore the couplings
associated with these interactions can never grow since the
perturbations involved have a scaling dimension which is greater
than two and makes them irrelevant in the renormalization group
sense. We therefore ignore their contribution in computing the
renormalization group equations. We retain only the remaining four
interaction terms labeled as $t_{8},t_{9},t_{10}$ and $t_{11}$. All
these terms involve the superconducting phase $\theta_{\rho}^{-}$
for the relative charge channel in the anti-symmetric basis togther
with the spin fields of the problem. Their coupling strength depends
on the $2k_{F}^{o}$ and $k_{U}/2$ cosine fourier component. Finally
we note that the Hamiltonian remains quadratic in the total charge
fields $\phi_{\rho}^{+}$ and $\theta_{\rho}^{+}$.

In summary, starting from the interacting 1D Hamiltonian (see
Eq.~\ref{eq:hamiltonian}) we have classified the important low
energy long wavelength fermionic interaction processes and then
bosonized them (see Eqs.~\ref{eq:bosonizedhamiltonian} and
~\ref{interactionsI}). In the next section \ref{sec:RG} we use a
weak coupling renormalization group treatment, described earlier in
chapter~\ref{theory}, to determine which of the coupling constants
associated with the bosonized interaction terms diverge.

\section{Analyzing the renormalization group equations\label{sec:RG}}

To analyze the low energy, long wavelength behavior of the
interacting system, we employ the renormalization group approach. In
this approach the shortwavelength modes are systematically
eliminated leading to a set of coupled differential equations for
the coupling constants. For the present problem, one can derive the
appropriate renormalization group equations for the entire set of
interactions, $t_{i}$, where $i=1,\dots,11$, in perturbation theory
about the noninteracting fixed point, the quadratic Hamiltonian
Eq.~\ref{eq:bosonizedhamiltonian} using the momentum shell
renormalization group procedure described in chapter~\ref{theory}.
The renormalization group equations for these coupling constants (up
to second order, $\mathcal{O}(2)$) of the problem are quoted in
appendix \ref{sec:rg}.

We study the renormalization group equations for the regimes -
$K^{-}_{\rho}<1$ (repulsive) and $K^{-}_{\rho}>1$ (attractive) of
the bosonic theory in the relative charge channel. The
$K^{-}_{\rho}<1$  case is repulsive since it promotes the CDW phase,
$\phi^{-}_{\rho}$. The $K^{-}_{\rho}>1$ case is attractive because
it promotes the SS phase, $\theta^{-}_{\rho}$. The equations are
integrated numerically for some suitable initial conditions until
one or more couplings grow to be of order one, $\mathcal{O}(1)$. The
order one $\mathcal{O}(1)$ couplings are considered to be large and
thus pin (gap) the appropriate bosonic modes. We then replace the
gapped fields with their expectation value both in the Hamiltonian
and in the bosonized version of the correlation functions. The
correlation functions which do not vanish then help to determine the
divergent susceptibilities and the possible thermodynamic phase for
that regime.

The initial condition under which we begin the renormalization group
flow is determined by evaluating the coupling constants,
Eq.~\ref{initconditions}, for the QWR parameters width, length and
the distance from the back-gate. The phenomenological form for the
interaction kernel is $M(a) =
\frac{e^{2}}{4\pi\epsilon_{o}}\frac{e^{-(a/d)}}{\sqrt{w^{2}+
a^{2}}}$, which has the nature of a screened Coulomb potential where
$e$ is the electronic charge and $\epsilon_{o}$ the permittivity of
free space. The screening involves two length scales. The distance
of the QWR to the back-gate, $d$, and the width, $w$, of the wire.
The width $w$ provides the short-distance cutoff whereas the
distance to the back-gate $d$ is the long-distance cutoff. Using
these parameters and the expression for the coupling constants,
Eq.~\ref{initconditions}, we can make an estimate for the initial
conditions.

For the quantum wire of Moser \emph{et.al.}\cite{JM,Moser} we have a
transverse size of $w \sim 15 nm$ separated from the metallic gate
by a distance $d \sim 300 nm$. The wire length is $L \sim 1 \mu m$.
We then have $w/d \sim$ 0.05 and the ratio $L/d$ = 10/3. The
parameters used to estimate $k^{o}_{F}$ for the aluminum arsenide
bandstructure are: density of the electrons in the quantum wire
$\sim 10^{8}m^{-1}$ and the effective mass of the electron $m^{*}=
0.33m_{e}$ along the long direction where $m_{e}$ is the bare mass
of the electron \cite{JM}. Furthermore, experimental evidence
\cite{Gunawan,Shkolnikov} suggests that spin rotational invariance
is not broken in the AlAs quantum wells in the absence of a magnetic
field. The problem is then $SU(2)$ invariant and we can set
$K^{\pm}_{\sigma}=1$ to begin the renormalization group flow.

The analysis, with the above initial conditions, for the repulsive
regime in the relative charge channel, $K_{\rho}^{-}<1$, shows that
the coupling constants $t_{2}$ and $t_{3}$ diverge, in fact
$(t_{2},t_{3}) \rightarrow (\infty,\infty)$. The fields which get
gapped are the antisymmetric charge field $\phi^{-}_{\rho}$ and the
spin fields $\phi^{\pm}_{\sigma}$. Their possible acquired
expectation values are $(\langle \phi^{-}_{\rho}\rangle,\langle
\phi^{+}_{\sigma}\rangle,\langle\phi^{-}_{\sigma}\rangle)$=$(\pm
\sqrt{\pi}/2, 0, 0)$ or $(0,\pm\sqrt{\pi}/2,\pm\sqrt{\pi}/2)$. The
final phase in either case is a dominant divergent intraband (odd
combination) 2$k^{o}_{F}$-CDW. The corresponding $CDW$ correlation
function, $\hat{O}^{intra,odd}_{CDW,2k^{o}_{F}}$ (refer appendix
\ref{sec:correlation}), decays with the power law $\chi_{CDW}\sim
(1/r)^{K^{+}_{\rho}/2}$ where $K_{\rho}^{+}$ is the total charge
Luttinger parameter.

Although the original screened Coulomb interaction is repulsive, in
the course of renormalization it can be led to an effective
attractive regime, $K_{\rho}^{-}>1$. For this case the
renormalization group flows indicate that the divergent coupling
constants are $t_{9}$ and $t_{10}$ where both $(t_{9},t_{10})
\rightarrow (\infty,\infty)$. From the interactions terms we can
then deduce that the dual antisymmetric charge field
$\theta_{\rho}^{-}$ gets gapped together with the spin fields
$\phi^{\pm}_{\sigma}$. They acquire the expectation values of
$(\langle \theta^{-}_{\rho}\rangle,\langle
\phi^{+}_{\sigma}\rangle,\langle\phi^{-}_{\sigma}\rangle)$=$(\pm
\sqrt{\pi}/2, 0, 0)$ or $(0,\pm\sqrt{\pi}/2,\pm\sqrt{\pi}/2)$. These
gapped fields lead to a state with a divergent intraband SS
correlation function $\Delta^{intra,singlet}$ (appendix
\ref{sec:correlation}) with the power law $\chi_{SS}\sim
(1/r)^{1/2K^{+}_{\rho}}$.

\begin{figure}[h]
\centering
\psfrag{intra}{Intraband}\psfrag{inter}{Intraband}\psfrag{Krho1}{$K^{-}_{\rho}<1$}\psfrag{eqn}{$2k_{F}^{o}-CDW$}
\psfrag{Krho2}{$K^{-}_{\rho}>1$}\psfrag{sing}{SS}
\psfrag{kpneq}{$(k_{p}\neq 0)$}\psfrag{1}{1}
\psfrag{krho}{$K_{\rho}^{-}$}{
\includegraphics[width=3.0in]{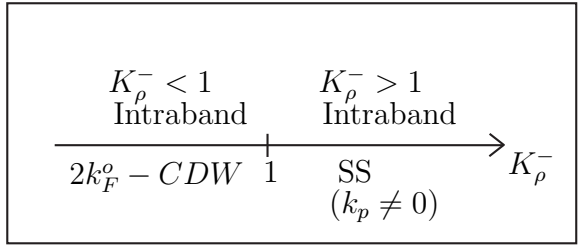}}
\caption{{\emph {Phase diagram}}: Divergent correlation functions
for the quantum wire in the presence of inter-valley umklapp
scattering for the two regimes K$^{-}_{\rho} <$ 1 (repulsive) and
K$^{-}_{\rho}>$ 1 (attractive). The notation CDW stands for charge
density wave and SS for singlet superconducitvity. In the diagram
$k_{p}$ denotes the pairing momenta of the center-of-mass of the
Cooper pairs. The finite pairing momentum is an indication of a
Fulde-Ferrell-Larkin-Ovchinnikov (FFLO) state which is encouraged in
this wire due to the presence of the umklapp interactions. Such a
state is also known to lead to inhomogeneous
superconductivity.\label{fig:6}}
\end{figure}

We also find through our analysis that although for the finite sized
wire with $L/d \sim 3.33$, the initial conditions for the
renormalization group flows differ from the true 1D limit of $L/d
\rightarrow \infty$, due to the screening of the back gate this is
not a severe effect, and the system remains in the same basin of
attraction as the infinite wire. The renormalization group results
can now be summarized in the phase diagram shown in Fig.~\ref{fig:6}

For the AlAs QWR bandstructure, due to the presence of four Fermi
points we have in general eight fields, $\phi^{\pm}_{\nu}$ and their
duals $\theta^{\pm}_{\nu}$ where $\nu=\rho,\sigma$ refer to the
charge and spin modes. For a gapless system, \emph{Luttinger
liquid}, we would have a C2S2 phase where the notation refers to the
number of gapless charge $(C)$ and spin $(S)$ modes. But in the
present problem due to the presence of interactions certain modes as
predicted by the renormalization analysis get gapped. Furthermore,
due to the absence of any bosonized interaction terms in the total
charge mode we expect the total charge field to remain gapless. As a
result we have the C1S0 phase \cite{Varma,Balents} where the
notation refers to one gapless total charge mode $(\phi^{+}_{\rho})$
with all other charge and spin modes gapped.

One of the novel aspects of this superconductivity is that it
originated from a purely repulsive screened Coulomb interaction. In
the process of renormalization, in the relative charge channel for
$K^{-}_{\rho}>1$, an attractive interaction was generated. This
resulted in a pairing of up and down spins within a band and
subsequent scattering of it from one band to another via the umklapp
Cooper scattering process. Energetically the process can be
understood as a competition between Coulomb repulsion of pairing and
the kinetic energy of the pair where the latter wins. Such a kinetic
energy driven mechanism for superconductivity has been proposed
earlier in the context of high temperature
superconductors\cite{spingap}.

Another unique aspect of this superconductivity is that the pairing
momenta of the Cooper pairs is nonzero. It is half the umklapp
vector. Being at a finite wavevector one could think of this as a
density modulated superconducting state where the superfluid density
varies from one lattice site to another with the wavevector
$\frac{k_{U}}{2}$. Such an inhomogeneous superconducting state is
known to arise in the FFLO state \cite{casalbuoni:263}.

\section{Conclusion\label{sec:conclusion}}
QWRs provide an opportunity for technological innovation. In this
context a phase diagram helps to understand what electronic phase
the QWR may predominantly find itself in since it has an important
effect on the transport properties. In this paper we have
investigated the possibility of a spin gapped AlAs QWR. Using 1D
field theoretic methods and perturbative renormalization group we
are able to conclude that the novel AlAs QWR, fabricated by Moser
\emph{et al.} \cite{Moser,JM} under investigation will have the
possibility of a spin gapped state with divergent 2$k^{o}_{F}$-CDW
or SS fluctuations. While the CDW wave phase is robust deep in the
repulsive region, there is a part of the phase diagram which
promotes a non-trivial SS with finite-momentum Cooper pairing
leading to an inhomogeneous superconducting state. The finite
pairing momentum is an indication of a FFLO state.

%
%
%
\chapter{LUTTINGER LIQUID KINK\label{kink}}
\section{Introduction}\label{sec:kinkintro}
One of the most dramatic consequences of confining electrons to one
spatial dimension is the prediction of spin-charge separation. That
is, due to many-body interactions the electron is no longer a stable
quasiparticle, but decays into separate spin and charge modes
\cite{tomonaga1950,luttinger1960,mattis-lieb1965}. A direct
experimental observation of spin-charge separation has proven
difficult although evidence for Luttinger liquid behavior has been
reported in many 1D systems, via, {\em e.g.}, a suppression of the
density of states near the Fermi level in ropes of carbon
nanotubes\cite{bockrath-smalley1997} or power law behavior in the
conductance vs. temperature in  edge states of the fractional
quantum Hall effect\cite{milliken-webb1996,chang1996} and carbon
nanotubes\cite{bockrath-smalley1999}. Until now, very limited direct
evidence for spin-charge separation has been reported.
Tunneling measurements later provided evidence for explicit
spin-charge separation in 1D systems, via real-space imaging of
Friedel oscillations using scanning tunneling microscopy on
single-walled carbon nanotubes\cite{lee-eggert2004} and momentum-
and energy-resolved tunneling between two coupled
QWRs\cite{auslaender-halperin2005}, both of which observed multiple
velocities indicative of spin-charge separation. More direct
evidence of spin-charge separation would be to measure separate spin
and charge dispersions in a single-particle spectral
function.\cite{zx-2006} Despite much effort in this area, this has
only been achieved recently in an unambiguous way in the
Mott-Hubbard insulator SrCuO$_2$.\cite{zx-2006}
Other claims of the detection of separately dispersing spin and
charge peaks with ARPES\cite{claessen1995,segovia1999} have been
overturned\cite{bluebronze-not,Si-Au-not}, or lack independent
verification of the spin and charge energy scales.\cite{TTF-TCNQ}

Part of the difficulty in directly measuring spin and charge
dispersions through measurements proportional to the single particle
spectral function is that within Luttinger liquid theory, the spinon
branch is muted compared to the holon branch. Finite temperature and
experimental resolution only compound the problem, making direct
detection of the spinon branch in, {\em e.g.}, ARPES difficult. In
this chapter, we show how spin-charge separation can nevertheless be
detected via the systematic temperature dependence of a \emph{kink}
in the effective electronic dispersion, even in cases where the spin
peak is not directly resolvable.
\begin{figure}[t]
\centering{{\psfrag{B}{Binding
Energy}\psfrag{m}{Momentum}\psfrag{vs}{$v_{\sigma}$}\psfrag{vc}{$v_{\rho}$}\includegraphics[width=4.0in]{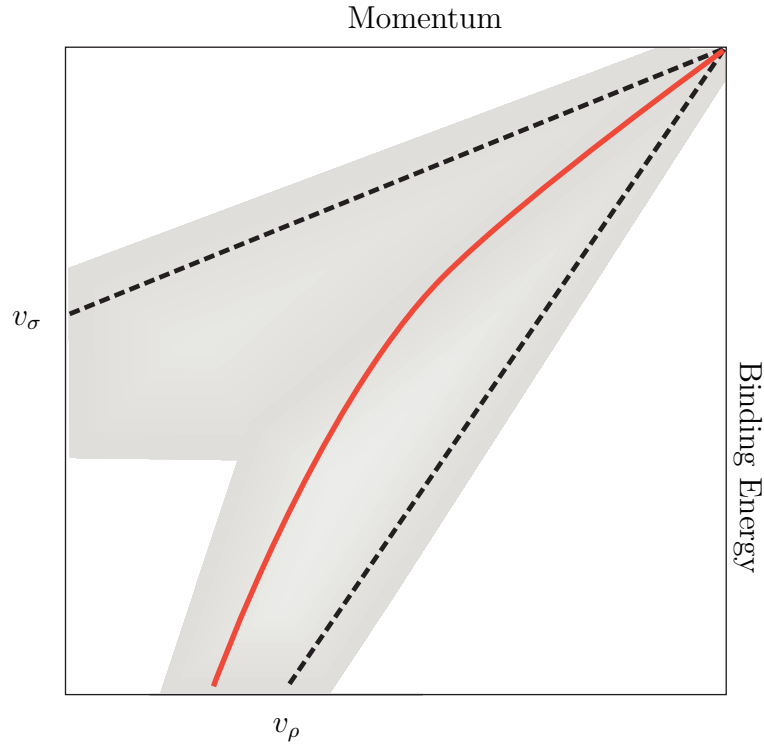}}
\caption{Schematic diagram of peak broadening due to interactions
and finite temperature in the Luttinger liquid spectral function.
The dashed lines denote the zero temperature dispersion tracking the
charge part with velocity $v_{\rho}$ and the spin part with velocity
$v_{\sigma}$. At $T\ne 0$, the peaks become thermally broadened as
indicated by the shaded regions. In this case the effective
dispersion now tracks the solid red line, so that the low energy
part tracks the sum of the two broad spin and charge peaks,
resulting in a low energy velocity $v_l$ which is between the spin
and charge velocities. Note the high energy effective dispersion is
parallel to the charge part but displaced, an effect due to finite
temperature and interactions. This results in the high energy part
extrapolating back to a value $k_{ex}\neq k_{F}$.
}\label{fig:thermalbroadening}}
\end{figure}
\section{Theory}\label{sec:kinktheory}

Although the electron is not an elementary excitation of the
Luttinger liquid because it is unstable to spin-charge separation,
an effective electronic dispersion may still be defined by the
existence of (generally broad) peaks in the spectral function. At
zero temperature in 1D, there are two sharp peaks in the electronic
spectral function, one dispersing at the velocity $v_\sigma$ of the
collective spin modes (spinons) and the other at the velocity
$v_\rho$ of the collective charge modes
(holons)\cite{meden1992,voit1993}. However, at finite temperature,
the spin and charge peaks are broadened, as shown schematically in
Fig.~\ref{fig:thermalbroadening}. At low binding energy, this causes
the two to merge into one broad peak with an effective dispersion
which lies between the spin velocity $v_\sigma$  and the charge
velocity $v_\rho$. Although the two peaks separate at higher binding
energies, interactions and temperature strongly suppress the spin
peak for repulsive interactions. As a result, the dominant (and most
easily measurable) peak will disperse with the charge velocity at
high energy. This gives rise to a kink in the effective electronic
dispersion. Since the Luttinger liquid is quantum critical, the kink
energy scales linearly with temperature, $E_{\rm kink} \propto
a(r,\gamma_\rho) T$, where $a$ is a function of the velocity ratio
$r = v_\sigma/v_\rho$ and the interaction strength $\gamma_{\rho}
=\frac{1}{8}(K_{\rho}+K_{\rho}^{-1}-2)$ where $K_\rho$ is the charge
Luttinger parameter. For $\gamma_{\rho}=0.15-0.30$ and $r=0.2-0.4$,
the range of $a$ is
$a=3.3-3.9$. The kink is stronger for lower values of $r$, but
diminishes again for strong enough interaction strength. Moreover,
the high energy linear effective dispersion extrapolates to the
Fermi energy at a wavevector $k_{ex} \ne k_F$ which is shifted from
the Fermi wavevector by an amount which scales linearly with
temperature.
Recently explicit analytic expressions for correlation functions in
the Tomonaga-Luttinger liquid at finite temperature were obtained
under various conditions\cite{Orgad}. We consider here the single
hole spectral function, $A^<(k,\omega)$, since it is directly
proportional to the intensity observed in ARPES  experiments. In the
spin-rotationally invariant case, the finite-temperature single hole
spectral function\cite{Orgad} may be written in terms of the scaled
variables $\tilde{k}=\frac{v_{s}k}{\pi T}$ and
$\tilde{\omega}=\frac{\omega}{\pi T}$
with the Boltzmann constant $k_{B}=1$,
\begin{eqnarray}
\lefteqn{A^{<}(\tilde{k},\tilde{\omega})\propto
\int^{\infty}_{-\infty}
dq~h_{\frac{1}{2}}(\tilde{k}-2rq)\times }\nonumber \\
&h_{\gamma_{c}+\frac{1}{2}}\bigg[{\tilde{\omega}-\tilde{k} \over
2}+(1+r)q\bigg] h_{\gamma_{c}}\bigg[{\tilde{\omega}-\tilde{k} \over
2}-(1-r)q\bigg]\nonumber
\\
\end{eqnarray}
where $r=v_\sigma/v_\rho$ is the ratio between the spin velocity and
the charge velocity and $h_{\gamma}$ is related to the beta function
\begin{equation}
h_{\gamma}(k)=\Re e\bigg[(2i)^{\gamma}B\bigg(\frac{\gamma -ik}{2},
1- \gamma \bigg)\bigg]
\end{equation}
The charge interaction strength $\gamma_\rho$ is related to the
charge Luttinger parameter $K_{\rho}$ by $\gamma_{\rho} =
\frac{1}{8}(K_{\rho}+K_{\rho}^{-1}-2)$, {\em i.e.} $\gamma_\rho = 0$
in the noninteracting case, and $\gamma_\rho$ increases with
increasing interaction strength. Because of spin rotation
invariance, we use $K_\sigma = 1$ and $\gamma_\sigma = 0$.

\section{Results and Discussion}\label{sec:kinkresult}

In order to define a single-hole effective dispersion, we use
momentum distribution curves (MDC's), {\em i.e.}  the single hole
spectral function $A^<(k,\omega_o)$ considered as a function of $k$
at a given value of the frequency $\omega_o$. The effective
dispersion is identified by the position $k_o= k_o(\omega_o)$ of the
maximum of
 $A_{\rm max}^<(k,\omega_o)$ with respect to $k$.
This gives an implicit equation for the effective single hole
dispersion $\omega_o(k_o)$. This method gives a more reliable
definition of the effective dispersion than using energy
distribution curves (EDC's), {\em i.e.} the spectral function
considered as a function of frequency at a given momentum $k_o$,
$A^<(k_o,\omega)$. Whereas EDC's can become quite broad with
increasing interaction strength, MDC's are always sharp due to
kinematic constraints,\cite{frac} so that there is less experimental
uncertainty in identifying the location of a peak in the MDC.

\begin{figure}[!t]
\centering \subfigure[~$A^<(k,\omega)$ at
$\gamma_\rho=0.15$]{\label{fig:intensity-gamma15}
  \includegraphics[width=4in]{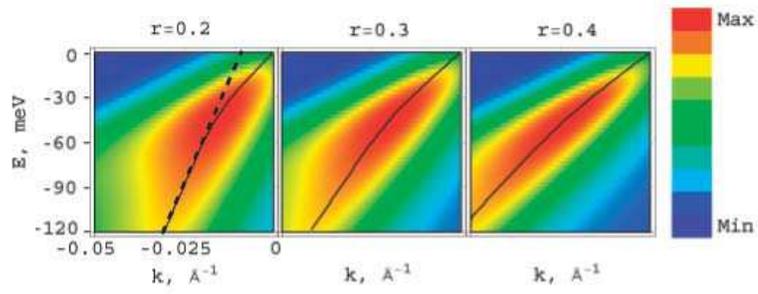}}
  \subfigure[~Effective dispersion at
$\gamma_\rho=0.15$]{\label{fig:dispersion-gamma15}
  \includegraphics[width=4.0in]{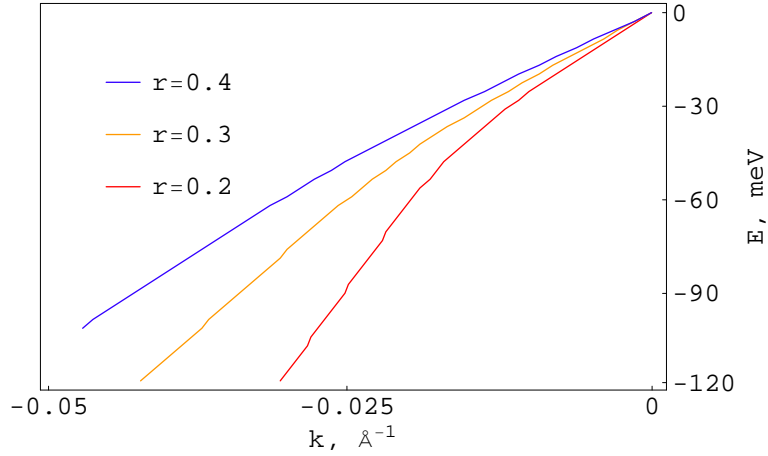}}
\caption{Intensity of the spectral function $A^<(k,\omega)$ and
effective dispersions at an interaction strength $\gamma_\rho =
0.15$. (a) The intensity of $A^<(k,\omega)$ is shown for three
different ratios of the spin to charge velocity, $r=0.2,~0.3,$ and
$0.4$. The black lines are the effective electronic dispersions
derived from MDC peaks, as described in the text. The dashed line in
the first panel shows that the high energy part of the effective
dispersion does not extrapolate back to the Fermi wavevector,
$k_{F}$.  (b) Comparison of the dispersions at different values of
the velocity ratio, $r=0.2,~0.3,$ and $0.4$. In all cases the spin
velocity $v_{\sigma}=1 $eV-$\AA$ and the temperature $k_B
T=14meV$.\label{fig:gamma15}}
\end{figure}
\begin{figure}[!t]
\centering{\includegraphics[width= 5in]{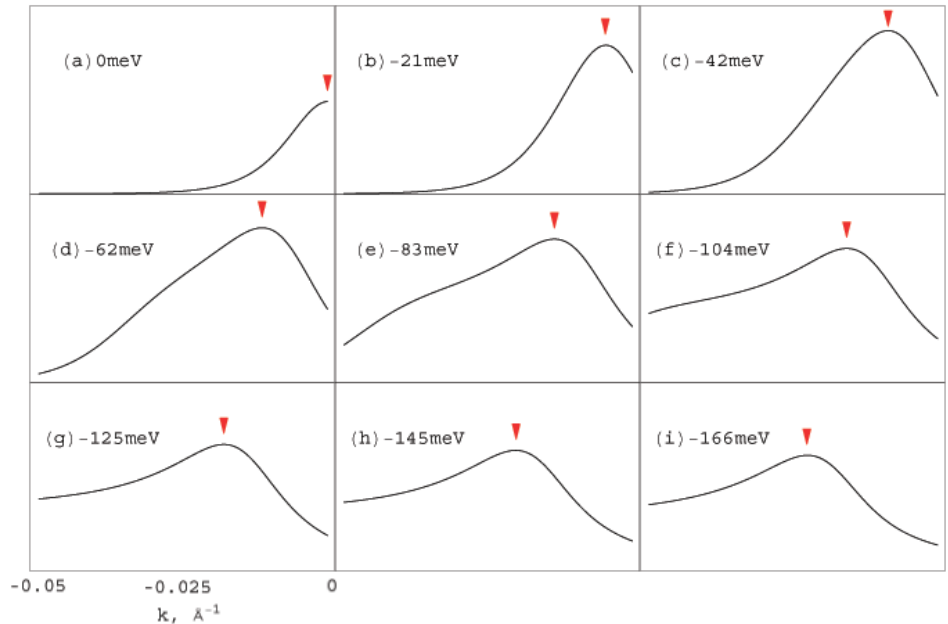} \caption{MDC's for
$\gamma_\rho = 0.15$. The spin velocity $v_{\sigma}=1 $eV-$\AA$ and
the temperature $k_B T=14meV$. The ratio of spin to charge velocity
is $r=0.2$. }\label{fig:MDC-gamma15}}
\end{figure}
Fig.~\ref{fig:intensity-gamma15} shows  representative intensity
plots of the spectral function, $A^<(k,\omega)$. In the figure, we
have used an interaction strength $\gamma_\rho = 0.15$, temperature
$k_B T=14$meV, and velocity ratios $r = 0.2$, $0.3$, and $0.4$.
Fig~\ref{fig:MDC-gamma15} shows the corresponding MDC's for $r=0.2$,
plotted as a function of momentum $k$ at a few representative
energies. The red triangles show the position of the maximum of the
MDC curves. The resulting effective dispersion is denoted by the
solid black lines in Fig.~\ref{fig:intensity-gamma15}. As is evident
from the figure, the effective dispersion is linear as expected at
low energy and also at high energy, but with different velocities.
This gives rise to a ``kink'' in the effective dispersion, {\em
i.e.} a change in the effective velocity. While at zero temperature,
there are two well-defined peaks in the MDC's, one dispersing with
the charge velocity and the other with the spin velocity, when the
temperature is finite, the width of these MDC peaks is thermally
broadened. (See Fig.~\ref{fig:thermalbroadening}.) At low energies
and finite temperatures, the sum of the two broad peaks is itself
one broad peak, as can be seen  in  panels (a)-(c) in
Fig.~\ref{fig:MDC-gamma15}, and the maximum in the MDC will track a
velocity $v_l$ which is between the spin and charge velocities,
$v_\sigma < v_l < v_\rho$. At high enough energies, the temperature
broadened singularities due to the spin and charge part become
sufficiently separated, and the spin peak is sufficiently muted,
that the MDC peak tracks the charge velocity. The separation of the
muted spin peak from the stronger charge peak can be seen in panels
(d) and (e) of Fig.~\ref{fig:MDC-gamma15}. In panels (f)-(i), the
charge and spin peaks have moved sufficiently apart that the peak in
the MDC will track the charge part.
Aside from the presence of a kink in the effective dispersion, the
position of the high energy linear effective dispersion is another
signature of Luttinger liquid behavior.  In
Fig.~\ref{fig:intensity-gamma15}, the dotted line is an
extrapolation of the high energy linear part of the effective
dispersion back to the Fermi energy. As can be seen from the figure,
the dotted line extrapolates to $E = E_F$ at a wavevector $k_{ex} =
k_F + \alpha(r,\gamma_\rho) T$ which is shifted from the Fermi
wavevector by an amount which scales linearly with temperature.

Fig.~\ref{fig:dispersion-gamma15} shows the effective dispersion for
each of the three values of $r$, overlaid for comparison. As one
might expect, as $r \rightarrow 1$, the kink vanishes, since then
the charge and spin pieces disperse with the same velocity. As $r$
is decreased, so that now $v_\rho > v_\sigma$, a kink appears, and
strengthens as $r$ is further decreased. One can see the general
features that the low energy part disperses with a velocity $v_l$
which is between the spin and charge velocities, $v_\sigma <
v_l<v_\rho$, and that the high energy part disperses with the charge
velocity. However, this high energy effective dispersion
extrapolates back to the Fermi energy at a wavevector $k_{ex} \ne
k_F$. At higher interaction strengths, EDC's broaden significantly,
so that $k_{ex}$ is smaller and the kink diminishes in strength.
However, $E_{\rm kink}$ moves to deeper binding energy as the
interaction strength is increased. Additional calculations exploring
the dependence of the kink on various interaction strengths are
shown in appendix~\ref{sec:kinkcalculations}.

\begin{figure}[!t]
\centering
 \subfigure[~$A^<(k,\omega)$ as $U$ varies]{\label{subfig:figdensityplot16815Hubbard}
  \includegraphics [width=4in]{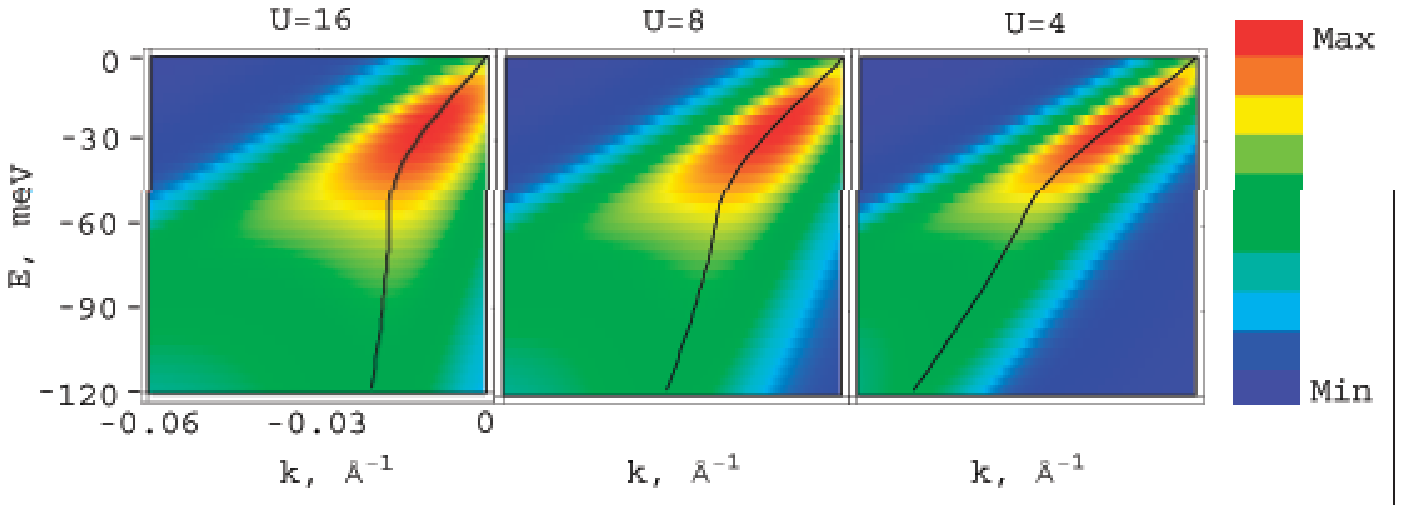}}
\subfigure[~Effective dispersion as $U$
varies]{\label{subfig:figdispersion16815Hubbard}
  \includegraphics[width=4.0in]{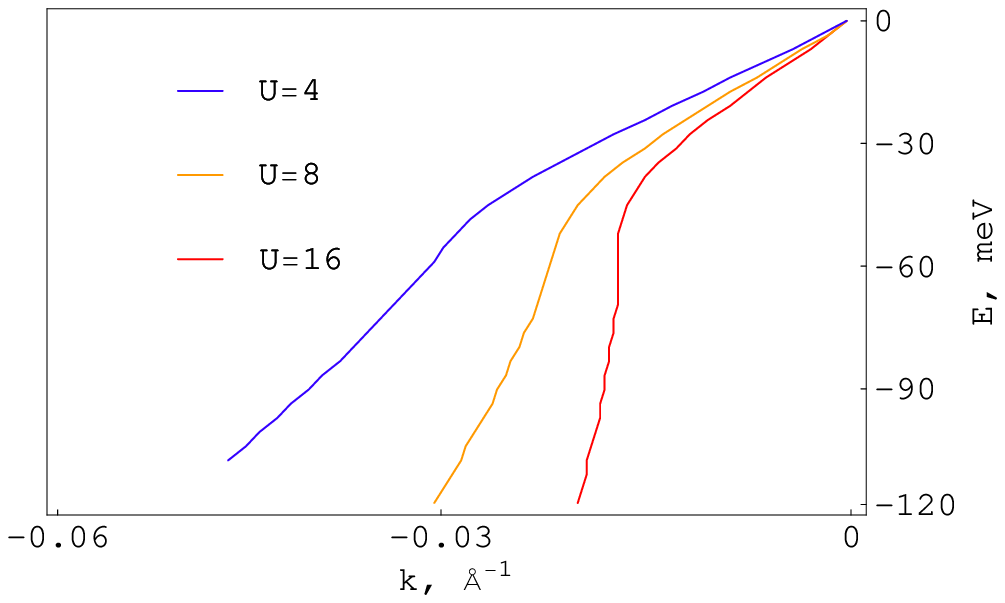}}
 \caption{Intensity of the spectral function $A^<(k,\omega)$ and effective
dispersions for $U = 16, 8,$ and $4$ in units of the hopping
integral $t$. The density $n=0.3$. (a) The intensity of
$A^<(k,\omega)$. The black lines are the effective electronic
dispersions derived from MDC peaks, as described in the text. (b)
Comparison of the dispersions at different values of $U/t$. In all
cases the spin velocity $v_{\sigma}=1 $eV-$\AA$ and the temperature
$k_B T=14meV$. \label{fig:hubbard}}
\end{figure}

In Fig.~\ref{fig:tempvar} we show how the effective dispersion
changes with temperature. Because the Luttinger liquid is quantum
critical, the spectral function has a scaling form, and the only
energy scale in the effective dispersion is the temperature itself.
As a result, the kink energy depends linearly on the temperature,
$E_{\rm kink} \propto T$. As can be seen in the figure, varying only
the temperature merely moves the kink to deeper binding energy,
leaving the low energy velocity $v_l$ (and therefore the strength of
the kink) unchanged. in addition, as temperature is increased, the
high energy part extrapolates back to the Fermi energy at a higher
value of $k_{ex}$.
\begin{figure}[b]
\centering{
\includegraphics[width=5in]{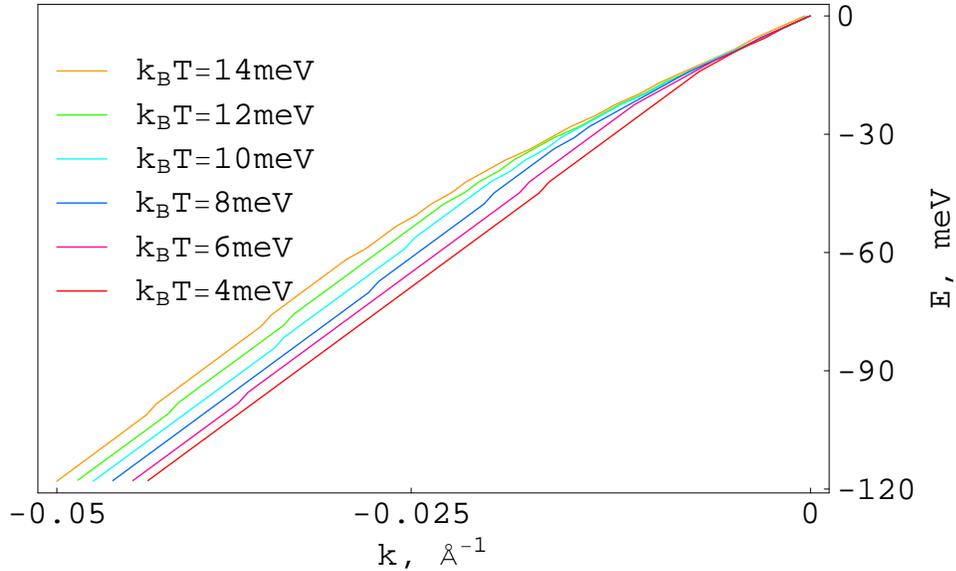}
\caption{Temperature variation of the effective dispersion. The
temperature varies from $k_B T=4$meV to $k_B T=14$meV, starting from
the lower curve and moving to the upper curve. The interaction
strength $\gamma_\rho = 0.15$, the spin velocity $v_\sigma = 1
$eV-$\AA$, and the ratio of spin to charge velocity $r =
0.3$.}\label{fig:tempvar}}
\end{figure}
Up until now, we have studied the Luttinger kink in a
phenomenological manner, allowing $\gamma_\rho$ and $r$ to vary
independently of each other. It is also useful to consider the
systematics of the kink strength and energy within the context of a
microscopic model.  As an example, we show in Fig.~\ref{fig:hubbard}
results for a Luttinger liquid derived from an incommensurate
repulsive 1D Hubbard model. We take the density to be away from half
filling, $n = 0.3$.
For a given value of $U/t$, renormalized values of $\gamma_c^*$ and
$r^*$ are taken from Ref.~\cite{schulz1995}, where Bethe-ansatz was
used to find the renormalized values of $K_\rho^*$, $v_\rho^*$, and
$v_\sigma^*$.
In Fig.~\ref{fig:hubbard}, we show the  intensity plots of the
spectral function along with the effective dispersion, for the
values $U/t = 16, 8,$ and $4$ of the repulsive Hubbard model.
This corresponds to renormalized values of $\gamma_\rho^* = 0.05,
0.04, 0.02$, and $r^* = 0.1, 0.2, 0.4$, respectively. Upon
increasing the Hubbard interaction strength $U$, the strength of the
kink is enhanced due to the change in the renormalized velocity
ratio $r^*$. Notice that in this case, the kink is more pronounced,
and there is a sharper distinction between the low energy and high
energy linear parts.

It is worth noting that behavior reminiscent of this physics was
recently reported in ARPES experiments on the quasi-one-dimensional
Mott-Hubbard insulator SrCuO$_2$.\cite{zx-2006} Being an insulating
material, SrCuO$_2$ is gapped, whereas the Luttinger spectral
functions presented here are not. Nevertheless, the effective
dispersion (measured by EDC's) shows a single peak at energies close
to the gap, which then separates into two peaks at higher binding
energy.
\section{Conclusion}

In conclusion, we have shown the existence of a
temperature-dependent kink in the effective electronic dispersion of
a spin-rotationally invariant Luttinger liquid, due to spin-charge
separation. At low energies, the effective dispersion is linear,
with a velocity between the spin and charge velocities, $v_\sigma <
v_l < v_\rho$. At high energies, the MDC peak disperses with the
charge velocity. Because the Luttinger liquid is quantum critical,
the kink between the high energy and low energy behavior has an
energy set by temperature, $E_{\rm kink} \propto T$.
In addition, the high energy effective dispersion extrapolates back
to the Fermi energy at a wavevector $k_{ex} \ne k_F$ which is
shifted from the Fermi wavevector by an amount which is proportional
to temperature. As interactions are increased, the kink diminishes
in strength, and moves to higher binding energy. In cases where
finite temperature and interactions along with experimental
uncertainties obscure the detection of two separate peaks in the
effective dispersion, the kink analysis presented here can be used
as a signature of spin-charge separation in Luttinger liquids.



\bibliography{thesis}

  \coversheet{APPENDICES}
    \makeatletter
    \@@nonchapter{odd}{APPENDICES}{y}{0pt}
    \makeatother

    \makeatletter
    \renewcommand{\thesection}{\thechapter}
    \renewcommand{\thesubsection}{\thechapter.\@arabic\c@subsection}
    \renewcommand{\thesubsubsection}{\thechapter.\@arabic\c@subsection.\@arabic\c@subsubsection}
    \makeatother

    \renewcommand{\thechapter}{A}
    \refstepcounter{chapter}
    \section*{Appendix A: Luttinger parameters for the quantum wire\label{sec:luttparam}}
    \addcontentsline{toc}{section}{Appendix A: Luttinger parameters for the quantum wire}
The symmetric and the anti-symmetric
velocity ($v^{\pm}_{\rho}$,$v^{\pm}_{\sigma}$) and the Luttinger
parameters, ($K^{\pm}_{\rho}$,$K^{\pm}_{\sigma}$) are expressed in
terms of the original velocity and the Luttinger parameters as
\begin{equation}
v^{+}_{\rho}K^{+}_{\rho}=v_{\rho}K_{\rho} +
\frac{M(2k^{o}_{F})}{2\pi}- \frac{b^{-}}{2\pi}
\end{equation}
\begin{equation}
\frac{v^{+}_{\rho}}{K^{+}_{\rho}}=\frac{v_{\rho}}{K_{\rho}}
+\frac{8M(0)-M(2k^{o}_{F})}{2\pi} - \frac{b^{+}}{2\pi}
\end{equation}
\begin{equation}
v^{-}_{\rho}K^{-}_{\rho}=v_{\rho}K_{\rho}+
\frac{M(2k^{o}_{F})}{2\pi} + \frac{b^{-}}{2\pi}
\end{equation}
\begin{equation}
\frac{v^{-}_{\rho}}{K^{-}_{\rho}}=\frac{v_{\rho}}{K_{\rho}} -
\frac{M(2k^{o}_{F})}{2\pi} + \frac{b^{+}}{2\pi}
\end{equation}
\begin{equation}
v^{+}_{\sigma}K^{+}_{\sigma}=v_{\sigma}K_{\sigma}+
\frac{M(2k^{o}_{F})}{2\pi}- \frac{b^{-}}{2\pi}
\end{equation}
\begin{equation}
\frac{v^{+}_{\sigma}}{K^{+}_{\sigma}}=\frac{v_{\sigma}}{K_{\sigma}}
- \frac{M(2k^{o}_{F})}{2\pi}- \frac{b^{+}}{2\pi}
\end{equation}
\begin{equation}
v^{-}_{\sigma}K^{-}_{\sigma}= v_{\sigma}K_{\sigma} +
\frac{M(2k^{o}_{F})}{2\pi} + \frac{b^{-}}{2\pi}
\end{equation}
\begin{equation}
\frac{v^{-}_{\sigma}}{K^{-}_{\sigma}}=\frac{v_{\sigma}}{K_{\sigma}}
- \frac{M(2k^{o}_{F})}{2\pi} + \frac{b^{+}}{2\pi}
\end{equation}
where the $b^{\pm}$ are defined as
\begin{equation}
b^{\pm}=  \pm\frac{e^{2}}
 {4\pi\epsilon_{o}}\int da \frac{e^{-(a/d)}\cos\bigg[\frac{k_{U}a}{2}\bigg](M(2k^{o}_{F})\pm M(0))}{\sqrt{w^{2}+ a^{2}}}
\end{equation}
and $M(0)=\int da M(a)$ and $M(2k^{o}_{F})=\int da
M(a)\cos[2k^{o}_{F}a]$. The interaction kernel is $M(a) =
\frac{e^{2}}{4\pi\epsilon_{o}}\frac{e^{-(a/d)}}{\sqrt{w^{2}+
a^{2}}}$ (see Eq.~\ref{eqn:intxnkernel}) where $a$ is the distance
along the long direction of the wire, $e$ is the electronic charge,
and $\epsilon_{o}$ the permittivity of free space. The
phenomenological charge and spin Luttinger parameters are $K_{\rho}$
and $K_{\sigma}$ respectively, and the charge and the spin
velocities are $v_{\rho}$ and $v_{\sigma}$. The velocities are taken
to be identical for the two bands since they are degenerate. The
explicit expressions for the bare Luttinger parameters and the bare
velocities are given by
\begin{equation}
K_{\nu}=\sqrt{\frac{2\pi v_{F}+2g_{4,\nu}+g_{\nu}}{2\pi
v_{F}+2g_{4,\nu}-g_{\nu}}}
\end{equation}
and
\begin{equation}
v_{\nu}=\sqrt{\bigg(v_{F}+\frac{g_{4,\nu}}{\pi}\bigg)^{2}-
\bigg(\frac{g_{\nu}}{2\pi}\bigg)^{2}}
\end{equation}
where $g_{\rho}=g_{1}- 2g_{2},g_{\sigma}=g_{1},g_{4,\rho}=g_{4}$ and
$g_{4,\sigma}=0$. In terms of the forward and backscattering
amplitudes the above g-ology coefficients are given by
$g_{1}=M(2k^{o}_{F}),g_{2}=M(0)-M(2k^{o}_{F})$ and $g_{4}=M(0)$. The
wavevectors $k_{U}$ and $k_{F}^{o}$, respectively, are the umklapp
wavevector and the magnitude of the Fermi wavevector measured from
the bottom of each band (see Fig.~\ref{fig:2}).
    \renewcommand{\thechapter}{B}
    \refstepcounter{chapter}
    \section*{Appendix B: Renormalization group equations\label{sec:rg}}
    \addcontentsline{toc}{section}{Appendix B: Renormalization group equations}
The renormalization group equation for the coupling constants (see
Eq.~\ref{interactionsI}) of the problem are
\begin{equation}
\frac{dt_{1}}{dL}= \bigg(2 -K^{+}_{\sigma}-K^{-}_{\sigma}\bigg)t_{1}
- \frac{t_{2}t_{3}}{2\pi}- \frac{t_{5}t_{6}}{2\pi}-
\frac{t_{9}t_{10}}{2\pi}
\end{equation}

\begin{equation}
\frac{dt_{2}}{dL}= \bigg(2 - K^{-}_{\rho} -K^{+}_{\sigma}\bigg)t_{2}
- \frac{t_{1}t_{3}}{2\pi}- \frac{t_{4}t_{6}}{2\pi}-
\frac{t_{9}t_{11}}{2\pi}
\end{equation}

\begin{equation}
\frac{dt_{3}}{dL}= \bigg(2 - K^{-}_{\rho}-K^{-}_{\sigma}\bigg)t_{3}
- \frac{t_{1}t_{2}}{2\pi}- \frac{t_{4}t_{5}}{2\pi}-
\frac{t_{10}t_{11}}{2\pi}
\end{equation}

\begin{equation}
\frac{dt_{4}}{dL}= \bigg(2 -
\frac{1}{K^{-}_{\sigma}}-K^{-}_{\rho}\bigg)t_{4} -
\frac{t_{2}t_{6}}{2\pi} - \frac{t_{3}t_{5}}{2\pi}-
\frac{t_{8}t_{11}}{2\pi}
\end{equation}

\begin{equation}
\frac{dt_{5}}{dL}= \bigg(2 -
\frac{1}{K^{-}_{\sigma}}-K^{-}_{\sigma}\bigg)t_{5} -
\frac{t_{1}t_{6}}{2\pi} - \frac{t_{3}t_{4}}{2\pi}-
\frac{t_{8}t_{10}}{2\pi}
\end{equation}

\begin{equation}
\frac{dt_{6}}{dL}= \bigg(2 -
\frac{1}{K^{-}_{\sigma}}-K^{+}_{\sigma}\bigg)t_{6} -
\frac{t_{1}t_{5}}{2\pi} - \frac{t_{2}t_{4}}{2\pi} -
\frac{t_{8}t_{9}}{2\pi}
\end{equation}

\begin{equation}
\frac{dt_{7}}{dL}= \bigg(2 -
\frac{1}{K^{-}_{\sigma}}-K^{+}_{\sigma}\bigg)t_{7}
\end{equation}

\begin{equation}
\frac{dt_{8}}{dL}= \bigg(2 -
\frac{1}{K^{-}_{\rho}}-\frac{1}{K^{-}_{\sigma}}\bigg)t_{8} -
\frac{t_{4}t_{11}}{2\pi} - \frac{t_{5}t_{10}}{2\pi}-
\frac{t_{6}t_{9}}{2\pi}
\end{equation}

\begin{equation}
\frac{dt_{9}}{dL}= \bigg(2 - \frac{1}{K^{-}_{\rho}} -
{K^{+}_{\sigma}}\bigg)t_{9} - \frac{t_{1}t_{10}}{2\pi} -
\frac{t_{2}t_{11}}{2\pi}- \frac{t_{6}t_{8}}{2\pi}
\end{equation}

\begin{equation}
\frac{dt_{10}}{dL}= \bigg(2 - K^{-}_{\sigma}-
\frac{1}{K^{-}_{\rho}}\bigg)t_{10} - \frac{t_{1}t_{9}}{2\pi} -
\frac{t_{3}t_{11}}{2\pi} - \frac{t_{5}t_{8}}{2\pi}
\end{equation}

\begin{equation}
\frac{dt_{11}}{dL}= \bigg(2 - K^{-}_{\rho}-
\frac{1}{K^{-}_{\rho}}\bigg)t_{11} - \frac{t_{2}t_{9}}{2\pi} -
\frac{t_{3}t_{10}}{2\pi} - \frac{t_{4}t_{8}}{2\pi}
\end{equation}

 \begin{equation}
\frac{d\ln K^{-}_{\rho}}{dL}=
\frac{1}{8\pi^{2}}\bigg[-K^{-}_{\rho}\bigg(t^{2}_{2} + t^{2}_{3}
+t^{2}_{4}\bigg)+ \frac{1}{K^{-}_{\rho}}\bigg(t^{2}_{8}+
t^{2}_{9}+t^{2}_{10}+t^{2}_{11}\bigg)\bigg]
 \end{equation}

    \begin{equation}
\frac{d\ln K^{+}_{\sigma}}{dL}=
-\frac{K^{+}_{\sigma}}{8\pi^{2}}\bigg(t^{2}_{1} +
t^{2}_{2}+t^{2}_{6}+t^{2}_{7}+t^{2}_{9}\bigg)
 \end{equation}

\begin{eqnarray}
\frac{d\ln K^{-}_{\sigma}}{dL}=
\frac{1}{8\pi^{2}}\bigg[-K^{-}_{\sigma}\bigg(t^{2}_{1} +
t^{2}_{3}+t^{2}_{10}+t^{2}_{5}\bigg)+
\frac{1}{K^{-}_{\sigma}}\bigg(t^{2}_{4}+t^{2}_{5}+
t^{2}_{6}+t^{2}_{7}+t^{2}_{8}\bigg)\bigg]
 \end{eqnarray}
where $L=\ln(\Lambda / \alpha)$, with length scale $\Lambda$. The
initial conditions of the problem are
\begin{eqnarray}
\label{initconditionsappendix}
 &t_{1}(0)=t_{2}(0)=2\int da M(a)\cos[2k^{o}_{F}a] \nonumber \\
 &t_{3}(0)= 2\int da M(a)\bigg(\cos[2k^{o}_{F}a]-\cos\bigg[\frac{k_{U}a}{2}\bigg]\bigg) \nonumber \\
 &t_{4}(0)= t_{5}(0)= -t_{10}(0)= -t_{11}(0)= -2\int M(a)\cos\bigg[\frac{k_{U}a}{2}\bigg] \nonumber \\
 &t_{6}(0)= -t_{9}(0)= -2\int da M(a)\cos[2k^{o}_{F}a]\cos\bigg[\frac{k_{U}a}{2}\bigg]\nonumber \\
 &t_{7}(0)=  -2\int da M(a)\sin[2k^{o}_{F}a]\sin \bigg[\frac{k_{U}a}{2}\bigg]\nonumber \\
 &t_{8}(0)= 2\int da M(a)\bigg(1- \cos[2k^{o}_{F}a]\bigg)\cos \bigg[\frac{k_{U}a}{2}\bigg]
 \end{eqnarray}
In the above equations the interaction kernel is $M(a) =
\frac{e^{2}}{4\pi\epsilon_{o}}\frac{e^{-(a/d)}}{\sqrt{w^{2}+
a^{2}}}$ (see Eq.~\ref{eqn:intxnkernel}) where $a$ is the distance
along the long direction of the wire. The coupling constants of the
interaction terms (see Eq.~\ref{interactionsI}) are denoted by
$t_{i}$, where $i=1,\dots,11$. They are related to the fourier
components of the interaction kernel, $M(a)$. The wavevectors
$k_{U}$ and $k_{F}^{o}$, respectively, are the umklapp wavevector
and the magnitude of the Fermi wavevector measured from the bottom
of each band (see Fig.~\ref{fig:2}). The symmetric and
anti-symmetric velocity ($v^{\pm}_{\rho}$,$v^{\pm}_{\sigma}$) and
Luttinger parameters, ($K^{\pm}_{\rho}$,$K^{\pm}_{\sigma}$) are
given by the expressions quoted in appendix~\ref{sec:luttparam}.
\renewcommand{\thechapter}{C}
    \refstepcounter{chapter}
    \section*{Appendix C: Correlation functions for the quantum wire\label{sec:correlation}}
    \addcontentsline{toc}{section}{Appendix C: Correlation functions for the quantum wire}
    The fermionic definition together with
the bosonized version of the correlation functions are stated below.
The correlation functions have been classified into two categories -
intraband and interband. In the expressions the band index is
$n=A,B$. The spin index is $s=\{\uparrow,\downarrow$\}, and the
Fermi momenta are defined by $ k^{A\pm}_{F}=-\frac{k_{U}}{4}\pm
k^{o}_{F}$ and $k^{B\pm}_{F}= \frac{k_{U}}{4}\pm k^{o}_{F}$ where
$k_U$ is the umklapp vector, and $k_{F}^{o}$ is the magnitude of the
Fermi wavevector measured from the bottom of each band, as shown in
Fig.~\ref{subfig:bandstructure}. The correlation functions are
expressed in terms of the right- and left- moving excitations,
$R_{ns}(x)$ and ${L}_{ns}(x)$, respectively, residing around the
Fermi points of the two bands (indicated by the black dots in
Fig.~\ref{subfig:bandstructure}). The fields in the correlation
function are the symmetric and the anti-symmetric fields defined as
$\phi^{\pm}_{\nu} =\frac{1}{\sqrt{2}}(\phi_{A\nu} \pm \phi_{B\nu})$
and its dual $\theta^{\pm}_{\nu} = \frac{1}{\sqrt{2}}(\theta_{A\nu}
\pm \theta_{B\nu})$ where $\nu = \rho,\sigma$ are the charge and
spin modes. The Klein factors are denoted by $\eta_{Rns}$ and
$\eta_{Lns}$ for the right- and left- moving fermions, respectively,
in band $n$ with spin $s$.
\subsection{Charge density wave (CDW) correlation functions}
\subsubsection{Intraband CDW (even combination), $2k_{F}^{o}$ wavevector}
\begin{eqnarray}
\hat{O}^{intra,even}_{CDW,2k_{F}^{o}}(x)=
[e^{i2k_{F}^{o}x}L^{\dag}_{As}R_{As}+e^{-i2k_{F}^{o}x}R^{\dag}_{As}L_{As}]+
(A \rightarrow B) \\
=\frac{2}{\pi\alpha}\sin[\sqrt{\pi}(\phi^{+}_{\rho}+\phi^{-}_{\rho})-
2k_{F}^{o}x]\cos[\sqrt{\pi}(\phi^{+}_{\sigma}+\phi^{-}_{\sigma})]
\nonumber\\+\frac{2}{\pi\alpha}\sin[\sqrt{\pi}(\phi^{+}_{\rho}-\phi^{-}_{\rho})-
2k_{F}^{o}x]\cos[\sqrt{\pi}(\phi^{+}_{\sigma}-\phi^{-}_{\sigma})]
\end{eqnarray}
\subsubsection{Intraband CDW (odd combination), $2k_{F}^{o}$ wavevector}
\begin{eqnarray}
\hat{O}^{intra,odd}_{CDW,2k_{F}^{o}}(x)=[e^{i2k_{F}^{o}x}L^{\dag}_{As}R_{As}+e^{-i2k_{F}^{o}x}R^{\dag}_{As}L_{As}]
- (A \rightarrow B) \\
=\frac{2}{\pi\alpha}\sin[\sqrt{\pi}(\phi^{+}_{\rho}+\phi^{-}_{\rho})-
2k_{F}^{o}x]\cos[\sqrt{\pi}(\phi^{+}_{\sigma}+\phi^{-}_{\sigma})]
\nonumber
\\-\frac{2}{\pi\alpha}\sin[\sqrt{\pi}(\phi^{+}_{\rho}-\phi^{-}_{\rho})-
2k_{F}^{o}x]\cos[\sqrt{\pi}(\phi^{+}_{\sigma}-\phi^{-}_{\sigma})]
\end{eqnarray}
\subsection{Spin density wave (SDW) correlation functions}
\subsubsection{Intraband SDW (even combination), $2k_{F}^{o}$ wavevector}
\begin{eqnarray}
&\hat{\vec{O}}^{intra,even}_{SDW,2k_{F}^{o}}(x)=[e^{i2k_{F}^{o}x}L^{\dag}_{A\alpha}\bigg(\frac{\vec{\sigma}}{2}\bigg)_{\alpha,\beta}R_{A\beta}
+e^{-i2k_{F}^{o}x}R^{\dag}_{A\alpha}\bigg(\frac{\vec{\sigma}}{2}\bigg)^{\dag}_{\alpha,\beta}L_{A\beta}]
+(A \rightarrow B)\\
&[\hat{O}^{intra,even}_{SDW,2k_{F}^{o}}(x)]_{z}=
-\frac{1}{\pi\alpha}\cos[\sqrt{\pi}(\phi^{+}_{\rho}+\phi^{-}_{\rho})-
2k_{F}^{o}x]\sin[\sqrt{\pi}(\phi^{+}_{\sigma}+\phi^{-}_{\sigma})] \nonumber \\
&-\frac{1}{\pi\alpha}\cos[\sqrt{\pi}(\phi^{+}_{\rho}-\phi^{-}_{\rho})-
2k_{F}^{o}x]\sin[\sqrt{\pi}(\phi^{+}_{\sigma}-\phi^{-}_{\sigma})]\\
&[\hat{O}^{intra,even}_{SDW,2k_{F}^{o}}(x)]_{y}=q
\frac{1}{\pi\alpha}\cos[\sqrt{\pi}(\phi^{+}_{\rho}+\phi^{-}_{\rho})-
2k_{F}^{o}x]\cos[\sqrt{\pi}(\theta^{+}_{\sigma}+\theta^{-}_{\sigma})] \nonumber \\
&+\frac{1}{\pi\alpha}\cos[\sqrt{\pi}(\phi^{+}_{\rho}-\phi^{-}_{\rho})-
2k_{F}^{o}x]\cos[\sqrt{\pi}(\theta^{+}_{\sigma}-\theta^{-}_{\sigma})]\\
&[\hat{O}^{intra,even}_{SDW,2k_{F}^{o}}(x)]_{x}=
\frac{1}{\pi\alpha}\cos[\sqrt{\pi}(\phi^{+}_{\rho}+\phi^{-}_{\rho})-
2k_{F}^{o}x]\sin[\sqrt{\pi}(\theta^{+}_{\sigma}+\theta^{-}_{\sigma})] \nonumber \\
&+\frac{1}{\pi\alpha}\cos[\sqrt{\pi}(\phi^{+}_{\rho}-\phi^{-}_{\rho})-
2k_{F}^{o}x]\sin[\sqrt{\pi}(\theta^{+}_{\sigma}-\theta^{-}_{\sigma})]
\end{eqnarray}
\subsubsection{Intraband SDW (odd combination), $2k_{F}^{o}$ wavevector}
\begin{eqnarray}
&\hat{\vec{O}}^{intra,odd}_{SDW,2k_{F}^{o}}(x)=[e^{i2k_{F}^{o}x}L^{\dag}_{A\alpha}\bigg(\frac{\vec{\sigma}}{2}\bigg)_{\alpha,\beta}R_{A\beta}
+e^{-i2k_{F}^{o}x}R^{\dag}_{A\alpha}\bigg(\frac{\vec{\sigma}}{2}\bigg)^{\dag}_{\alpha,\beta}L_{A\beta}]
- (A \rightarrow B)\\
&[\hat{O}^{intra,odd}_{SDW,2k_{F}^{o}}(x)]_{z}=-\frac{1}{\pi\alpha}\cos[\sqrt{\pi}(\phi^{+}_{\rho}+\phi^{-}_{\rho})-
2k_{F}^{o}x]\sin[\sqrt{\pi}(\phi^{+}_{\sigma}+\phi^{-}_{\sigma})]\nonumber \\
&+\frac{1}{\pi\alpha}\cos[\sqrt{\pi}(\phi^{+}_{\rho}-\phi^{-}_{\rho})-
2k_{F}^{o}x]\sin[\sqrt{\pi}(\phi^{+}_{\sigma}-\phi^{-}_{\sigma})]
\\
&[\hat{O}^{intra,odd}_{SDW,2k_{F}^{o}}(x)]_{y}=
\frac{1}{\pi\alpha}\cos[\sqrt{\pi}(\phi^{+}_{\rho}+\phi^{-}_{\rho})-
2k_{F}^{o}x]\cos[\sqrt{\pi}(\theta^{+}_{\sigma}+\theta^{-}_{\sigma})] \nonumber \\
&-\frac{1}{\pi\alpha}\cos[\sqrt{\pi}(\phi^{+}_{\rho}-\phi^{-}_{\rho})-
2k_{F}^{o}x]\cos[\sqrt{\pi}(\theta^{+}_{\sigma}-\theta^{-}_{\sigma})]
\\
&[\hat{O}^{intra,odd}_{SDW,2k_{F}^{o}}(x)]_{x}=
\frac{1}{\pi\alpha}\cos[\sqrt{\pi}(\phi^{+}_{\rho}+\phi^{-}_{\rho})-
2k_{F}^{o}x]\sin[\sqrt{\pi}(\theta^{+}_{\sigma}+\theta^{-}_{\sigma})] \nonumber \\
&-\frac{1}{\pi\alpha}\cos[\sqrt{\pi}(\phi^{+}_{\rho}-\phi^{-}_{\rho})-
2k_{F}^{o}x]\sin[\sqrt{\pi}(\theta^{+}_{\sigma}-\theta^{-}_{\sigma})]
\end{eqnarray}
\subsection{Interband CDW }
\subsubsection{Interband CDW (even combination), $\frac{k_{U}}{2}$ wavevector}
\begin{eqnarray}
&\hat{O}^{inter,even}_{CDW,\frac{k_{U}}{2}}(x)={e^{i\frac{k_{U}}{2}x}(R^{\dag}_{As}R_{Bs}+L^{\dag}_{As}L_{Bs})}
+ e^{-i\frac{k_{U}}{2}x}(A \rightarrow B) \\
&=-\frac{i\eta_{RA\uparrow}\eta_{RB\uparrow}}{\pi\alpha}\sin[\sqrt{\pi}(\theta_{\rho}^{-}+\theta_{\sigma}^{-}
-\phi^{-}_{\rho}-\phi^{-}_{\sigma})
-\frac{k_{U}}{2}x]\nonumber\\&-\frac{i\eta_{RA\downarrow}\eta_{RB\downarrow}}{\pi\alpha}\sin[\sqrt{\pi}(\theta_{\rho}^{-}-\theta_{\sigma}^{-}
-\phi^{-}_{\rho}+\phi^{-}_{\sigma})-\frac{k_{U}}{2}x]+(R \rightarrow
L)
\end{eqnarray}
\subsubsection{Interband CDW (odd combination), $\frac{k_{U}}{2}$ wavevector }
\begin{eqnarray}
&\hat{O}^{inter,odd}_{CDW,\frac{k_{U}}{2}}(x)={(e^{i\frac{k_{U}}{2}x})(R^{\dag}_{As}R_{Bs}+L^{\dag}_{As}L_{Bs})}
- e^{-i\frac{k_{U}}{2}x}(A \rightarrow B) \\
&=-\frac{\eta_{RA\uparrow}\eta_{RB\uparrow}}{\pi\alpha}\cos[\sqrt{\pi}(\theta_{\rho}^{-}+\theta_{\sigma}^{-}
-\phi^{-}_{\rho}-\phi^{-}_{\sigma})-\frac{k_{U}}{2}x]\nonumber\\&-\frac{\eta_{RA\downarrow}\eta_{RB\downarrow}}{\pi\alpha}\cos[\sqrt{\pi}(\theta_{\rho}^{-}-\theta_{\sigma}^{-}
-\phi^{-}_{\rho}+\phi^{-}_{\sigma})-\frac{k_{U}}{2}x]+(R \rightarrow
L)
\end{eqnarray}
\subsubsection{Interband CDW (even combination), $\frac{k_{U}}{2}+2k_{F}^{o}$ wavevector}
\begin{eqnarray}
&\hat{O}^{inter,even}_{CDW,\frac{k_{U}}{2}+2k_{F}^{o}}(x)={e^{i(\frac{k_{U}}{2}+2k_{F}^{o})x}}L^{\dag}_{As}R_{Bs}
+{e^{-i(\frac{k_{U}}{2}+2k_{F}^{o})x}}R^{\dag}_{Bs}L_{As} \\
&=-\frac{i\eta_{LA\uparrow}\eta_{RB\uparrow}}{\pi\alpha}\sin[\sqrt{\pi}(\theta_{\rho}^{-}+\theta_{\sigma}^{-}
+\phi^{+}_{\rho}+\phi^{+}_{\sigma})-(\frac{k_{U}}{2}+ 2k_{F}^{o})x]
\nonumber\\&-\frac{i\eta_{LA\downarrow}\eta_{RB\downarrow}}{\pi\alpha}\sin[\sqrt{\pi}(\theta_{\rho}^{-}-\theta_{\sigma}^{-}
+\phi^{+}_{\rho}-\phi^{+}_{\sigma})-(\frac{k_{U}}{2}+2k_{F}^{o})x]
\end{eqnarray}
\subsubsection{Interband CDW (odd combination), $\frac{k_{U}}{2}+2k_{F}^{o}$ wavevector}
\begin{eqnarray}
&\hat{O}^{inter,odd}_{CDW,\frac{k_{U}}{2}+2k_{F}^{o}}(x)={e^{i(\frac{k_{U}}{2}+2k_{F}^{o})x}}L^{\dag}_{As}R_{Bs}
-{e^{-i(\frac{k_{U}}{2}+2k_{F}^{o})x}}R^{\dag}_{Bs}L_{As}\\
&=\frac{\eta_{LA\uparrow}\eta_{RB\uparrow}}{\pi\alpha}\cos[\sqrt{\pi}(\theta_{\rho}^{-}+\theta_{\sigma}^{-}
+\phi^{+}_{\rho}+\phi^{+}_{\sigma})-(\frac{k_{U}}{2}+ 2k_{F}^{o})x]
\nonumber\\&+\frac{\eta_{LA\downarrow}\eta_{RB\downarrow}}{\pi\alpha}\cos[\sqrt{\pi}(\theta_{\rho}^{-}-\theta_{\sigma}^{-}
+\phi^{+}_{\rho}-\phi^{+}_{\sigma})-(\frac{k_{U}}{2}+2k_{F}^{o})x]
\end{eqnarray}
\subsubsection{Interband CDW (even combination), $\frac{k_{U}}{2}-2k_{F}^{o}$ wavevector}
\begin{eqnarray}
&\hat{O}^{inter,even}_{CDW,\frac{k_{U}}{2}-2k_{F}^{o}}(x)={e^{i(\frac{k_{U}}{2}-2k_{F}^{o}
)x}}R^{\dag}_{As}L_{Bs} +{e^{-i(\frac{k_{U}}{2}-2k_{F}^{o} )x}}L^{\dag}_{Bs}R_{As}\\
&=\frac{-i\eta_{RA\uparrow}\eta_{LB\uparrow}}{\pi\alpha}\sin[\sqrt{\pi}(\theta_{\rho}^{-}+\theta_{\sigma}^{-}
-\phi^{+}_{\rho}-\phi^{+}_{\sigma})-(\frac{k_{U}}{2}-
2k_{F}^{o})x]\nonumber\\&-\frac{i\eta_{RA\downarrow}\eta_{LB\downarrow}}{\pi\alpha}\sin[\sqrt{\pi}(\theta_{\rho}^{-}-\theta_{\sigma}^{-}
-\phi^{+}_{\rho}+\phi^{+}_{\sigma})-(\frac{k_{U}}{2}-2k_{F}^{o})x]
\end{eqnarray}
\subsubsection{Interband CDW (odd combination), $\frac{k_{U}}{2}-2k_{F}^{o}$ wavevector}
\begin{eqnarray}
&\hat{O}^{inter,odd}_{CDW,\frac{k_{U}}{2}-2k_{F}^{o}}(x)={e^{i(\frac{k_{U}}{2}-2k_{F}^{o}
)x}}R^{\dag}_{As}L_{Bs} -
{e^{-i(\frac{k_{U}}{2}-2k_{F}^{o} )x}}L^{\dag}_{Bs}R_{As}\\
&=\frac{\eta_{RA\uparrow}\eta_{LB\uparrow}}{\pi\alpha}\cos[\sqrt{\pi}(\theta_{\rho}^{-}+\theta_{\sigma}^{-}
-\phi^{+}_{\rho}-\phi^{+}_{\sigma})-(\frac{k_{U}}{2}-
2k_{F}^{o})x]\nonumber\\&+\frac{\eta_{RA\downarrow}\eta_{LB\downarrow}}{\pi\alpha}\cos[\sqrt{\pi}(\theta_{\rho}^{-}-\theta_{\sigma}^{-}
-\phi^{+}_{\rho}+\phi^{+}_{\sigma})-(\frac{k_{U}}{2}-2k_{F}^{o})x]
\end{eqnarray}
\subsection{Interband SDW}

\subsubsection{Interband SDW (even combination), $\frac{k_{U}}{2}$ wavevector}
\begin{eqnarray}
&\hat{\vec{O}}^{inter,even}_{SDW,\frac{k_{U}}{2}}(x)={e^{i\frac{k_{U}}{2}x}(R^{\dag}_{A\alpha}
\bigg(\frac{\vec{\sigma}}{2}\bigg)_{\alpha,\beta}R_{B\beta}+L^{\dag}_{A\alpha}\bigg(\frac{\vec{\sigma}}{2}\bigg)_{\alpha,\beta}L_{B\beta})}
\nonumber\\&+e^{-i\frac{k_{U}}{2}x}(R^{\dag}_{A\alpha}\bigg(\frac{\vec{\sigma}}{2}\bigg)^{\dag}_{\alpha,\beta}R_{B\beta}+L^{\dag}_{A\alpha}\bigg(\frac{\vec{\sigma}}{2}\bigg)^{\dag}_{\alpha,\beta}L_{B\beta})
\\
&[\hat{O}^{inter,even}_{SDW,\frac{k_{U}}{2}}(x)]_{z}=-\frac{i\eta_{RA\uparrow}\eta_{RB\uparrow}}{2\pi\alpha}\sin[\sqrt{\pi}(\theta_{\rho}^{-}+\theta_{\sigma}^{-}
-\phi^{-}_{\rho}-\phi^{-}_{\sigma}) -\frac{k_{U}}{2}x]\nonumber \\
&+\frac{i\eta_{RA\downarrow}\eta_{RB\downarrow}}{2\pi\alpha}\sin[\sqrt{\pi}(\theta_{\rho}^{-}-\theta_{\sigma}^{-}
-\phi^{-}_{\rho}+\phi^{-}_{\sigma})-\frac{k_{U}}{2}x]+(R \rightarrow
L)\\
&[\hat{O}^{inter,even}_{SDW,\frac{k_{U}}{2}}(x)]_{y}=-\frac{i\eta_{RA\uparrow}\eta_{RB\downarrow}}{2\pi\alpha}\cos[\sqrt{\pi}(\theta_{\rho}^{-}+\theta_{\sigma}^{+}
-\phi^{-}_{\rho}-\phi^{+}_{\sigma}) -\frac{k_{U}}{2}x]\nonumber \\
&+\frac{i\eta_{RA\downarrow}\eta_{RB\uparrow}}{2\pi\alpha}\cos[\sqrt{\pi}(\theta_{\rho}^{-}-\theta_{\sigma}^{+}
-\phi^{-}_{\rho}+\phi^{+}_{\sigma})-\frac{k_{U}}{2}x]+(R \rightarrow
L)\\
&[\hat{O}^{inter,even}_{SDW,\frac{k_{U}}{2}}(x)]_{x}=-\frac{i\eta_{RA\uparrow}\eta_{RB\downarrow}}{2\pi\alpha}\sin[\sqrt{\pi}(\theta_{\rho}^{-}+\theta_{\sigma}^{+}
-\phi^{-}_{\rho}-\phi^{+}_{\sigma}) -\frac{k_{U}}{2}x]\nonumber \\
&+\frac{i\eta_{RA\downarrow}\eta_{RB\uparrow}}{2\pi\alpha}\sin[\sqrt{\pi}(\theta_{\rho}^{-}-\theta_{\sigma}^{+}
-\phi^{-}_{\rho}+\phi^{+}_{\sigma})-\frac{k_{U}}{2}x]+(R \rightarrow
L)
\end{eqnarray}
\subsubsection{Interband SDW (odd combination), $\frac{k_{U}}{2}$ wavevector}
\begin{eqnarray}
&\hat{\vec{O}}^{inter,odd}_{SDW,\frac{k_{U}}{2}}(x)={e^{i\frac{k_{U}x}{2}}(R^{\dag}_{A\alpha}\bigg(\frac{\vec{\sigma}}{2}\bigg)_{\alpha,\beta}R_{B\beta}+L^{\dag}_{A\alpha}\bigg(\frac{\vec{\sigma}}{2}\bigg)_{\alpha,\beta}L_{B\beta})}
\nonumber\\&-e^{-i\frac{k_{U}x}{2}}(R^{\dag}_{A\alpha}\bigg(\frac{\vec{\sigma}}{2}\bigg)^{\dag}_{\alpha,\beta}R_{B\beta}+L^{\dag}_{A\alpha}\bigg(\frac{\vec{\sigma}}{2}\bigg)^{\dag}_{\alpha,\beta}L_{B\beta})
\\
&[\hat{O}^{inter,odd}_{SDW,\frac{k_{U}}{2}}(x)]_{z}=\frac{\eta_{RA\uparrow}\eta_{RB\uparrow}}{2\pi\alpha}\cos[\sqrt{\pi}(\theta_{\rho}^{-}+\theta_{\sigma}^{-}
-\phi^{-}_{\rho}-\phi^{-}_{\sigma}) -\frac{k_{U}}{2}x]\nonumber \\
&-\frac{\eta_{RA\downarrow}\eta_{RB\downarrow}}{2\pi\alpha}\cos[\sqrt{\pi}(\theta_{\rho}^{-}-\theta_{\sigma}^{-}
-\phi^{-}_{\rho}+\phi^{-}_{\sigma})-\frac{k_{U}}{2}x]+(R \rightarrow
L) \\
&[\hat{O}^{inter,odd}_{SDW,\frac{k_{U}}{2}}(x)]_{y}=-\frac{\eta_{RA\uparrow}\eta_{RB\downarrow}}{2\pi\alpha}\sin[\sqrt{\pi}(\theta_{\rho}^{-}+\theta_{\sigma}^{+}
-\phi^{-}_{\rho}-\phi^{+}_{\sigma}) -\frac{k_{U}}{2}x]\nonumber \\
&+\frac{\eta_{RA\downarrow}\eta_{RB\uparrow}}{2\pi\alpha}\sin[\sqrt{\pi}(\theta_{\rho}^{-}-\theta_{\sigma}^{+}
-\phi^{-}_{\rho}+\phi^{+}_{\sigma})-\frac{k_{U}}{2}x]+(R \rightarrow
L)\\
&[\hat{O}^{inter,odd}_{SDW,\frac{k_{U}}{2}}(x)]_{x}=\frac{\eta_{RA\uparrow}\eta_{RB\downarrow}}{2\pi\alpha}\sin[\sqrt{\pi}(\theta_{\rho}^{-}+\theta_{\sigma}^{+}
-\phi^{-}_{\rho}-\phi^{+}_{\sigma}) -\frac{k_{U}}{2}x]\nonumber \\
&+\frac{\eta_{RA\downarrow}\eta_{RB\uparrow}}{2\pi\alpha}\sin[\sqrt{\pi}(\theta_{\rho}^{-}-\theta_{\sigma}^{+}
-\phi^{-}_{\rho}+\phi^{+}_{\sigma})-\frac{k_{U}}{2}x]+(R \rightarrow
L)
\end{eqnarray}
\subsubsection{Interband SDW (even combination), $\frac{k_{U}}{2}+2k_{F}^{o}$ wavevector}
\begin{eqnarray}
&\hat{\vec{O}}^{inter,even}_{SDW,\frac{k_{U}}{2}+2k_{F}^{o}}(x)={e^{i(\frac{k_{U}}{2}+2k_{F}^{o})x}}L^{\dag}_{A\alpha}\bigg(\frac{\vec{\sigma}}{2}\bigg)_{\alpha,\beta}R_{B\beta}
+ {e^{-i(\frac{k_{U}}{2}+2k_{F}^{o}
)x}}R^{\dag}_{B\alpha}\bigg(\frac{\vec{\sigma}}{2}\bigg)^{\dag}_{\alpha,\beta}L_{A\beta}
\nonumber \\
\\&[\hat{O}^{inter,even}_{SDW,\frac{k_{U}}{2}+2k_{F}^{o}}(x)]_{z}=-\frac{i\eta_{LA\uparrow}\eta_{RB\uparrow}}{2\pi\alpha}
\sin[\sqrt{\pi}(\theta_{\rho}^{-}+\theta_{\sigma}^{-}+\phi^{+}_{\rho}+\phi^{+}_{\sigma}) -(\frac{k_{U}}{2}+2k_{F}^{o})x]\nonumber \\
&+\frac{i\eta_{LA\downarrow}\eta_{RB\downarrow}}{2\pi\alpha}\sin[\sqrt{\pi}(\theta_{\rho}^{-}-\theta_{\sigma}^{-}
+\phi^{+}_{\rho}-\phi^{+}_{\sigma})-(\frac{k_{U}}{2}+2k_{F}^{o})x]\\
&[\hat{O}^{inter,even}_{SDW,\frac{k_{U}}{2}+2k_{F}^{o}}(x)]_{y}=-\frac{i\eta_{LA\uparrow}\eta_{RB\downarrow}}{2\pi\alpha}
\cos[\sqrt{\pi}(\theta_{\rho}^{-}+\theta_{\sigma}^{+}+\phi^{+}_{\rho}+\phi^{-}_{\sigma}) -(\frac{k_{U}}{2}+2k_{F}^{o})x]\nonumber \\
&+\frac{i\eta_{LA\downarrow}\eta_{RB\uparrow}}{2\pi\alpha}\cos[\sqrt{\pi}(\theta_{\rho}^{-}-\theta_{\sigma}^{+}
+\phi^{+}_{\rho}-\phi^{-}_{\sigma})-(\frac{k_{U}}{2}+2k_{F}^{o})x]\\
&[\hat{O}^{inter,even}_{SDW,\frac{k_{U}}{2}+2k_{F}^{o}}(x)]_{x}=-\frac{i\eta_{LA\uparrow}\eta_{RB\downarrow}}{2\pi\alpha}
\sin[\sqrt{\pi}(\theta_{\rho}^{-}+\theta_{\sigma}^{+}+\phi^{+}_{\rho}+\phi^{-}_{\sigma}) -(\frac{k_{U}}{2}+2k_{F}^{o})x]\nonumber \\
&-\frac{i\eta_{LA\downarrow}\eta_{RB\uparrow}}{2\pi\alpha}\sin[\sqrt{\pi}(\theta_{\rho}^{-}-\theta_{\sigma}^{+}
+\phi^{+}_{\rho}-\phi^{-}_{\sigma})-(\frac{k_{U}}{2}+2k_{F}^{o})x]
\end{eqnarray}
\subsubsection{Interband SDW (odd combination), $\frac{k_{U}}{2}+2k_{F}^{o}$ wavevector}
\begin{eqnarray}
&\hat{\vec{O}}^{inter,odd}_{SDW,\frac{k_{U}}{2}+2k_{F}^{o}}(x)={e^{i(\frac{k_{U}}{2}+2k_{F}^{o}
)x}}L^{\dag}_{A\alpha}\bigg(\frac{\vec{\sigma}}{2}\bigg)_{\alpha,\beta}R_{B\beta}
-{e^{-i(\frac{k_{U}}{2}+2k_{F}^{o}
)x}}R^{\dag}_{B\alpha}\bigg(\frac{\vec{\sigma}}{2}\bigg)^{\dag}_{\alpha,\beta}L_{A\beta}
\nonumber \\ \\
&[\hat{O}^{inter,odd}_{SDW,\frac{k_{U}}{2}+2k_{F}^{o}}(x)]_{z}=\frac{\eta_{LA\uparrow}\eta_{RB\uparrow}}{2\pi\alpha}
\cos[\sqrt{\pi}(\theta_{\rho}^{-}+\theta_{\sigma}^{-}+\phi^{+}_{\rho}+\phi^{+}_{\sigma}) -(\frac{k_{U}}{2}+2k_{F}^{o})x]\nonumber \\
&-\frac{\eta_{LA\downarrow}\eta_{RB\downarrow}}{2\pi\alpha}\cos[\sqrt{\pi}(\theta_{\rho}^{-}-\theta_{\sigma}^{-}
+\phi^{+}_{\rho}-\phi^{+}_{\sigma})-(\frac{k_{U}}{2}+2k_{F}^{o})x] \\
&[\hat{O}^{inter,odd}_{SDW,\frac{k_{U}}{2}+2k_{F}^{o}}(x)]_{y}=-\frac{\eta_{LA\uparrow}\eta_{RB\downarrow}}{2\pi\alpha}
\sin[\sqrt{\pi}(\theta_{\rho}^{-}+\theta_{\sigma}^{+}+\phi^{+}_{\rho}+\phi^{-}_{\sigma}) -(\frac{k_{U}}{2}+2k_{F}^{o})x]\nonumber \\
&+\frac{\eta_{LA\downarrow}\eta_{RB\uparrow}}{2\pi\alpha}\sin[\sqrt{\pi}(\theta_{\rho}^{-}-\theta_{\sigma}^{+}
+\phi^{+}_{\rho}-\phi^{-}_{\sigma})-(\frac{k_{U}}{2}+2k_{F}^{o})x] \\
&[\hat{O}^{inter,odd}_{SDW,\frac{k_{U}}{2}+2k_{F}^{o}}(x)]_{x}=\frac{\eta_{LA\uparrow}\eta_{RB\downarrow}}{2\pi\alpha}
\cos[\sqrt{\pi}(\theta_{\rho}^{-}+\theta_{\sigma}^{+}+\phi^{+}_{\rho}+\phi^{-}_{\sigma}) -(\frac{k_{U}}{2}+2k_{F}^{o})x]\nonumber \\
&+\frac{\eta_{LA\downarrow}\eta_{RB\uparrow}}{2\pi\alpha}\cos[\sqrt{\pi}(\theta_{\rho}^{-}-\theta_{\sigma}^{+}
+\phi^{+}_{\rho}-\phi^{-}_{\sigma})-(\frac{k_{U}}{2}+2k_{F}^{o})x]
\end{eqnarray}
\subsubsection{Interband SDW (even combination), $\frac{k_{U}}{2}-2k_{F}^{o}$ wavevector}
\begin{eqnarray}
&\hat{\vec{O}}^{inter,even}_{SDW,\frac{k_{U}}{2}-2k_{F}^{o}}(x)={e^{i(\frac{k_{U}}{2}-2k_{F}^{o})x}}R^{\dag}_{A\alpha}\bigg(\frac{\vec{\sigma}}{2}\bigg)_{\alpha,\beta}L_{B\beta}+
{e^{-i(\frac{k_{U}}{2}-2k_{F}^{o})x}}L^{\dag}_{B\alpha}\bigg(\frac{\vec{\sigma}}{2}\bigg)^{\dag}_{\alpha,\beta}R_{A\beta}
\nonumber \\ \\
&[\hat{O}^{inter,even}_{SDW,\frac{k_{U}}{2}-2k_{F}^{o}}(x)]_{z}=-\frac{i\eta_{RA\uparrow}\eta_{LB\uparrow}}{2\pi\alpha}
\sin[\sqrt{\pi}(\theta_{\rho}^{-}+\theta_{\sigma}^{-}-\phi^{+}_{\rho}-\phi^{+}_{\sigma}) -(\frac{k_{U}}{2}-2k_{F}^{o})x]\nonumber \\
&+\frac{i\eta_{RA\downarrow}\eta_{LB\downarrow}}{2\pi\alpha}\sin[\sqrt{\pi}(\theta_{\rho}^{-}-\theta_{\sigma}^{-}
-\phi^{+}_{\rho}+\phi^{+}_{\sigma})-(\frac{k_{U}}{2}-2k_{F}^{o})x]\\
&[\hat{O}^{inter,even}_{SDW,\frac{k_{U}}{2}-2k_{F}^{o}}(x)]_{y}=-\frac{i\eta_{RA\uparrow}\eta_{LB\downarrow}}{2\pi\alpha}
\cos[\sqrt{\pi}(\theta_{\rho}^{-}+\theta_{\sigma}^{+}-\phi^{+}_{\rho}-\phi^{-}_{\sigma}) -(\frac{k_{U}}{2}+2k_{F}^{o})x]\nonumber \\
&+\frac{i\eta_{RA\downarrow}\eta_{LB\uparrow}}{2\pi\alpha}\cos[\sqrt{\pi}(\theta_{\rho}^{-}-\theta_{\sigma}^{+}
-\phi^{+}_{\rho}+\phi^{-}_{\sigma})-(\frac{k_{U}}{2}-2k_{F}^{o})x]\\
&[\hat{O}^{inter,even}_{SDW,\frac{k_{U}}{2}-2k_{F}^{o}}(x)]_{x}=-\frac{i\eta_{RA\uparrow}\eta_{LB\downarrow}}{2\pi\alpha}
\sin[\sqrt{\pi}(\theta_{\rho}^{-}+\theta_{\sigma}^{+}-\phi^{+}_{\rho}-\phi^{-}_{\sigma}) -(\frac{k_{U}}{2}-2k_{F}^{o})x]\nonumber \\
&-\frac{i\eta_{RA\downarrow}\eta_{LB\uparrow}}{2\pi\alpha}\sin[\sqrt{\pi}(\theta_{\rho}^{-}-\theta_{\sigma}^{+}
-\phi^{+}_{\rho}+\phi^{-}_{\sigma})-(\frac{k_{U}}{2}-2k_{F}^{o})x]
\end{eqnarray}
\subsubsection{Interband SDW (odd combination), $\frac{k_{U}}{2}-2k_{F}^{o}$ wavevector}
\begin{eqnarray}
&\hat{\vec{O}}^{inter,odd}_{SDW,\frac{k_{U}}{2}-2k_{F}^{o}}(x)={e^{i(\frac{k_{U}}{2}-2k_{F}^{o})x}}R^{\dag}_{A\alpha}\bigg(\frac{\vec{\sigma}}{2}\bigg)_{\alpha,\beta}L_{B\beta}
-{e^{-i(\frac{k_{U}}{2}-2k_{F}^{o})x}}L^{\dag}_{B\alpha}\bigg(\frac{\vec{\sigma}}{2}\bigg)^{\dag}_{\alpha,\beta}R_{A\beta}\nonumber \\ \\
&[\hat{O}^{inter,odd}_{SDW,\frac{k_{U}}{2}-2k_{F}^{o}}(x)]_{z}=\frac{\eta_{RA\uparrow}\eta_{LB\uparrow}}{2\pi\alpha}
\cos[\sqrt{\pi}(\theta_{\rho}^{-}+\theta_{\sigma}^{-}-\phi^{+}_{\rho}-\phi^{+}_{\sigma}) -(\frac{k_{U}}{2}-2k_{F}^{o})x]\nonumber \\
&-\frac{\eta_{RA\downarrow}\eta_{LB\downarrow}}{2\pi\alpha}\cos[\sqrt{\pi}(\theta_{\rho}^{-}-\theta_{\sigma}^{-}
-\phi^{+}_{\rho}+\phi^{+}_{\sigma})-(\frac{k_{U}}{2}-2k_{F}^{o})x]\\
&[\hat{O}^{inter,odd}_{SDW,\frac{k_{U}}{2}-2k_{F}^{o}}(x)]_{y}=-\frac{\eta_{RA\uparrow}\eta_{LB\downarrow}}{2\pi\alpha}
\sin[\sqrt{\pi}(\theta_{\rho}^{-}+\theta_{\sigma}^{+}-\phi^{+}_{\rho}-\phi^{-}_{\sigma}) -(\frac{k_{U}}{2}-2k_{F}^{o})x]\nonumber \\
&+\frac{\eta_{RA\downarrow}\eta_{LB\uparrow}}{2\pi\alpha}\sin[\sqrt{\pi}(\theta_{\rho}^{-}-\theta_{\sigma}^{+}
-\phi^{+}_{\rho}+\phi^{-}_{\sigma})-(\frac{k_{U}}{2}-2k_{F}^{o})x]\\
&[\hat{O}^{inter,odd}_{SDW,\frac{k_{U}}{2}-2k_{F}^{o}}(x)]_{x}=\frac{\eta_{RA\uparrow}\eta_{LB\downarrow}}{2\pi\alpha}
\cos[\sqrt{\pi}(\theta_{\rho}^{-}+\theta_{\sigma}^{+}-\phi^{+}_{\rho}-\phi^{-}_{\sigma}) -(\frac{k_{U}}{2}-2k_{F}^{o})x]\nonumber \\
&+\frac{\eta_{RA\downarrow}\eta_{LB\uparrow}}{2\pi\alpha}\cos[\sqrt{\pi}(\theta_{\rho}^{-}-\theta_{\sigma}^{+}
-\phi^{+}_{\rho}+\phi^{-}_{\sigma})-(\frac{k_{U}}{2}-2k_{F}^{o})x]
\end{eqnarray}

\subsection{Superconducting correlation functions}
\subsubsection{Intraband singlet}
\begin{eqnarray}
&[\Delta^{intra,singlet}]_{s}(x)=(L_{A\uparrow}R_{A\downarrow}-
L_{A\downarrow}R_{A\uparrow})e^{-i\frac{k_{U}}{2}x}
+(A \rightarrow B)e^{i\frac{k_{U}}{2}x}\\
&=\frac{2i}{\pi\alpha}e^{i\sqrt{\pi}\theta^{+}_{\rho}}\cos[\sqrt{\pi}\theta^{-}_{\rho}-\frac{k_{U}}{2}x]\cos[\sqrt{\pi}\phi^{+}_{\sigma}]
\cos[\sqrt{\pi}\phi^{-}_{\sigma}]\nonumber\\&+\frac{2}{\pi\alpha}e^{i\sqrt{\pi}\theta^{+}_{\rho}}\sin[\sqrt{\pi}\theta^{-}_{\rho}-\frac{k_{U}}{2}x]\sin[\sqrt{\pi}\phi^{+}_{\sigma}]
\sin[\sqrt{\pi}\phi^{-}_{\sigma}]\\
&[\Delta^{intra,singlet}]_{d}(x)=(L_{A\uparrow}R_{A\downarrow}-
L_{A\downarrow}R_{A\uparrow})e^{-i\frac{k_{U}}{2}x}-(A \rightarrow B)e^{i\frac{k_{U}}{2}x}\\
&=-\frac{2}{\pi\alpha}e^{i\sqrt{\pi}\theta^{+}_{\rho}}\sin[\sqrt{\pi}\theta^{-}_{\rho}-\frac{k_{U}}{2}x]\cos[\sqrt{\pi}\phi^{+}_{\sigma}]
\cos[\sqrt{\pi}\phi^{-}_{\sigma}]
\nonumber\\&-\frac{2i}{\pi\alpha}e^{i\sqrt{\pi}\theta^{+}_{\rho}}\cos[\sqrt{\pi}\theta^{-}_{\rho}-\frac{k_{U}}{2}x]\sin[\sqrt{\pi}\phi^{+}_{\sigma}]
\sin[\sqrt{\pi}\phi^{-}_{\sigma}]
\end{eqnarray}
\subsubsection{Intraband $\eta$-pairing operators at $\frac{k_{U}}{2} \pm 2k_{F}^{o}$ wavevectors}
\begin{eqnarray}
&\Delta^{intra}_{\eta, \frac{k_{U}}{2}-
2k^{o}_{F}}(x)=R_{A\uparrow}R_{A\downarrow}e^{-i(\frac{k_{U}}{2}-
2k^{o}_{F})x} \pm
L_{B\uparrow}L_{B\downarrow}e^{i(\frac{k_{U}}{2}-\xi 2k^{o}_{F})x}
\\
&=\frac{i}{\pi\alpha}e^{i\sqrt{\pi}(\theta^{+}_{\rho}-
\phi^{-}_{\rho})}\left\{
\begin{tabular}{l}
$\cos[\sqrt{\pi}(\theta^{-}_{\rho}- \phi^{+}_{\rho})-(\frac{k_{U}}{2}- 2k^{o}_{F})x]$\\
$i\sin[\sqrt{\pi}(\theta^{-}_{\rho}-
\phi^{+}_{\rho})-(\frac{k_{U}}{2}- 2k^{o}_{F})x]$
\end{tabular}\right\}
\end{eqnarray}
\begin{eqnarray}
&\Delta^{intra}_{\eta,
\frac{k_{U}}{2}+2k^{o}_{F}}(x)=L_{A\uparrow}L_{A\downarrow}e^{-i(\frac{k_{U}}{2}+2k^{o}_{F})x}
\pm R_{B\uparrow}R_{B\downarrow}e^{i(\frac{k_{U}}{2}+2k^{o}_{F})x}\\
&=\frac{i}{\pi\alpha}e^{i\sqrt{\pi}(\theta^{+}_{\rho}+\phi^{-}_{\rho})}\left\{
\begin{tabular}{l}
$\cos[\sqrt{\pi}(\theta^{-}_{\rho}+\phi^{+}_{\rho})-(\frac{k_{U}}{2}+2k^{o}_{F})x]$\\
$i\sin[\sqrt{\pi}(\theta^{-}_{\rho}+\phi^{+}_{\rho})-(\frac{k_{U}}{2}+2k^{o}_{F})x]$
\end{tabular}\right\}
\end{eqnarray}
where the upper row refers to the even combination and the lower to
the odd.
\subsubsection{Intraband Triplet}
\begin{eqnarray}
&\Delta^{intra,triplet}_{s}(x)=(L_{A\uparrow}R_{A\downarrow}+
L_{A\downarrow}R_{A\uparrow})e^{-i\frac{k_{U}}{2}x} + (A \rightarrow
B)e^{i\frac{k_{U}}{2}x} \\
&= -\frac{2}{\pi\alpha}e^{i\sqrt{\pi}\theta^{+}_{\rho}}
\cos[(\sqrt{\pi}\theta^{-}_{\rho}-\frac{k_{U}}{2}x)]\sin[\sqrt{\pi}\phi^{+}_{\sigma}]\cos[\sqrt{\pi}\phi^{-}_{\sigma}]
\nonumber\\&-\frac{2i}{\pi\alpha}e^{i\sqrt{\pi}\theta^{+}_{\rho}}
\sin[(\sqrt{\pi}\theta^{-}_{\rho}-\frac{k_{U}}{2}x)]\cos[\sqrt{\pi}\phi^{+}_{\sigma}]\sin[\sqrt{\pi}\phi^{-}_{\sigma}]
\end{eqnarray}
\begin{eqnarray}
&\Delta^{intra,triplet}_{d}(x)=(L_{A\uparrow}R_{A\downarrow}+
L_{A\downarrow}R_{A\uparrow})e^{-i\frac{k_{U}}{2}x} - (A \rightarrow
B)e^{i\frac{k_{U}}{2}x} \\
&= -\frac{2i}{\pi\alpha}e^{i\sqrt{\pi}\theta^{+}_{\rho}}
\sin[(\sqrt{\pi}\theta^{-}_{\rho}-\frac{k_{U}}{2}x)]\sin[\sqrt{\pi}\phi^{+}_{\sigma}]\cos[\sqrt{\pi}\phi^{-}_{\sigma}]
\nonumber\\&-\frac{2}{\pi\alpha}e^{i\sqrt{\pi}\theta^{+}_{\rho}}
\cos[(\sqrt{\pi}\theta^{-}_{\rho}-\frac{k_{U}}{2}x)]\cos[\sqrt{\pi}\phi^{+}_{\sigma}]\sin[\sqrt{\pi}\phi^{-}_{\sigma}]
\end{eqnarray}

\begin{eqnarray}
&\Delta^{intra,triplet}_{\uparrow\uparrow}(x)=
R_{A\uparrow}L_{A\uparrow}e^{-i\frac{k_{U}}{2}x} \pm (A \rightarrow
B)e^{i\frac{k_{U}}{2}x} \\
&=\frac{i}{\pi\alpha}e^{i\sqrt{\pi}(\theta^{+}_{\rho}+\theta^{+}_{\sigma})}\left\{
\begin{tabular}{l}
$\cos[\sqrt{\pi}(\theta^{-}_{\rho}+\theta^{-}_{\sigma})-\frac{k_{U}}{2}x]$\\
$i\sin[\sqrt{\pi}(\theta^{-}_{\rho}+\theta^{-}_{\sigma})-\frac{k_{U}}{2}x]$
\end{tabular}\right\}
\end{eqnarray}
\begin{eqnarray}
&\Delta^{intra,triplet}_{\downarrow\downarrow}(x)=R_{A\downarrow}L_{A\downarrow}e^{-i\frac{k_{U}}{2}x}
\pm (A \rightarrow B)e^{i\frac{k_{U}}{2}x}\\
&=\frac{i}{\pi\alpha}e^{i\sqrt{\pi}(\theta^{+}_{\rho}-\theta^{+}_{\sigma})}\left\{
\begin{tabular}{l}
$\cos[\sqrt{\pi}(\theta^{-}_{\rho}-\theta^{-}_{\sigma})-\frac{k_{U}}{2}x]$\\
$i\sin[\sqrt{\pi}(\theta^{-}_{\rho}-\theta^{-}_{\sigma})-\frac{k_{U}}{2}x]$
\end{tabular}\right\}
\end{eqnarray}
where the upper row refers to the even combination and the lower to
the odd.
\subsubsection{Interband singlet at $k=0$ wavevector}
\begin{eqnarray}
&\Delta^{inter,singlet}(x)=(L_{A\uparrow}R_{B\downarrow}-
L_{A\downarrow}R_{B\uparrow})\pm (A \rightarrow B) \\
&=\frac{\eta_{LA\uparrow}\eta_{RB\downarrow}}{2\pi\alpha}e^{i\sqrt{2\pi}(\theta_{\rho}^{+}+\theta_{\sigma}^{-}
+\phi^{-}_{\rho}+\phi^{+}_{\sigma})} \pm
\frac{\eta_{LA\downarrow}\eta_{RB\uparrow}}{2\pi\alpha}e^{i\sqrt{2\pi}(\theta_{\rho}^{+}-\theta_{\sigma}^{-}
+\phi^{-}_{\rho}-\phi^{+}_{\sigma})}+(A \rightarrow B)
\end{eqnarray}
\subsubsection{Interband singlet at $2k_{F}^{o}$ wavevector}
\begin{eqnarray}
&\Delta^{inter,singlet}_{2k^{o}_{F} }(x)=
[R_{A\uparrow}R_{B\downarrow} - (\uparrow \rightarrow
\downarrow)]e^{i2k^{o}_{F}x} \pm [L_{A\uparrow}L_{B\downarrow} -
(\uparrow
\rightarrow \downarrow)]e^{-i2k^{o}_{F}x} \\
&= \left [ \frac{\eta_{RA\uparrow}
\eta_{RB\downarrow}}{2\pi\alpha}e^{\sqrt{\pi}(\theta^{+}_{\rho}+\theta^{-}_{\sigma}-\phi^{+}_{\rho}-\phi^{-}_{\sigma})
e^{i2k^{o}_{F}x}} - \frac{\eta_{RA\downarrow}
\eta_{RB\uparrow}}{2\pi\alpha}
e^{\sqrt{\pi}(\theta^{+}_{\rho}-\theta^{-}_{\sigma}-\phi^{+}_{\rho}+\phi^{-}_{\sigma})e^{i2k^{o}_{F}x}}
\right ] \nonumber \\
&\pm \left [ \frac{\eta_{LA\uparrow}
\eta_{LB\downarrow}}{2\pi\alpha}e^{\sqrt{\pi}(\theta^{+}_{\rho}+\theta^{-}_{\sigma}+\phi^{+}_{\rho}+\phi^{-}_{\sigma})
e^{-i2k^{o}_{F}x}} - \frac{\eta_{LA\downarrow}
\eta_{LB\uparrow}}{2\pi\alpha}
e^{\sqrt{\pi}(\theta^{+}_{\rho}-\theta^{-}_{\sigma}+\phi^{+}_{\rho}-\phi^{-}_{\sigma})e^{-i2k^{o}_{F}x}}
\right ]
\end{eqnarray}


\subsubsection{Interband triplet at $k=0$ wavevector}

\begin{eqnarray}
&\Delta^{inter,triplet}(x)=(L_{A\uparrow}R_{B\downarrow}+
L_{A\downarrow}R_{B\uparrow})\pm (A \rightarrow B) \\
&=\frac{\eta_{LA\uparrow}\eta_{RB\downarrow}}{2\pi\alpha}e^{i\sqrt{2\pi}(\theta_{\rho}^{+}+\theta_{\sigma}^{-}
+\phi^{-}_{\rho}+\phi^{+}_{\sigma})}
+\frac{\eta_{LA\downarrow}\eta_{RB\uparrow}}{2\pi\alpha}e^{i\sqrt{2\pi}(\theta_{\rho}^{+}-\theta_{\sigma}^{-}
+\phi^{-}_{\rho}-\phi^{+}_{\sigma})}\pm (A \rightarrow B)
\end{eqnarray}

\begin{eqnarray}
&\Delta^{inter,triplet}_{\uparrow\uparrow}(x) =
R_{A\uparrow}L_{B\uparrow} \pm
L_{A\uparrow}R_{B\uparrow}\\
&=\frac{\eta_{RA\uparrow}\eta_{LB\uparrow}}{2\pi\alpha}e^{i\sqrt{\pi}(\theta^{+}_{\rho}+\theta^{+}_{\sigma}
-\phi^{-}_{\rho}-\phi^{-}_{\sigma})}\pm
\frac{\eta_{LA\uparrow}\eta_{RB\uparrow}}{2\pi\alpha}e^{i\sqrt{\pi}(\theta^{+}_{\rho}+\theta^{+}_{\sigma}
+\phi^{-}_{\rho}+\phi^{-}_{\sigma})}
\end{eqnarray}

\begin{eqnarray}
&\Delta^{inter,triplet}_{\downarrow\downarrow}(x)=
R_{A\downarrow}L_{B\downarrow} \pm
L_{A\downarrow}R_{B\downarrow}\\
&=\frac{\eta_{RA\downarrow}\eta_{LB\downarrow}}{2\pi\alpha}e^{i\sqrt{\pi}(\theta^{+}_{\rho}-\theta^{+}_{\sigma}
-\phi^{-}_{\rho}+\phi^{-}_{\sigma})}\pm
\frac{\eta_{LA\downarrow}\eta_{RB\downarrow}}{2\pi\alpha}e^{i\sqrt{\pi}(\theta^{+}_{\rho}-\theta^{+}_{\sigma}
+\phi^{-}_{\rho}-\phi^{-}_{\sigma})}
\end{eqnarray}

\subsubsection{Inter-band triplet at $2k_{F}^{o}$ wavevector}
\begin{eqnarray}
&\Delta^{inter,triplet}_{2k^{o}_{F}}(x)=
[R_{A\uparrow}R_{B\downarrow} + (\uparrow \rightarrow
\downarrow)]e^{i2k^{o}_{F}x} \pm [L_{A\uparrow}L_{B\downarrow} +
(\uparrow
\rightarrow \downarrow)]e^{-i2k^{o}_{F}x} \\
&= \left [ \frac{\eta_{RA\uparrow}
\eta_{RB\downarrow}}{2\pi\alpha}e^{\sqrt{\pi}(\theta^{+}_{\rho}+\theta^{-}_{\sigma}-\phi^{+}_{\rho}-\phi^{-}_{\sigma})
e^{i2k^{o}_{F}x}} + \frac{\eta_{RA\downarrow}
\eta_{RB\uparrow}}{2\pi\alpha}
e^{\sqrt{\pi}(\theta^{+}_{\rho}-\theta^{-}_{\sigma}-\phi^{+}_{\rho}+\phi^{-}_{\sigma})e^{i2k^{o}_{F}x}}
\right ] \nonumber \\
&\pm \left [ \frac{\eta_{LA\uparrow}
\eta_{LB\downarrow}}{2\pi\alpha}e^{\sqrt{\pi}(\theta^{+}_{\rho}+\theta^{-}_{\sigma}+\phi^{+}_{\rho}+\phi^{-}_{\sigma})
e^{-i2k^{o}_{F}x}} + \frac{\eta_{LA\downarrow}
\eta_{LB\uparrow}}{2\pi\alpha}
e^{\sqrt{\pi}(\theta^{+}_{\rho}-\theta^{-}_{\sigma}+\phi^{+}_{\rho}-\phi^{-}_{\sigma})e^{-i2k^{o}_{F}x}}
\right ]
\end{eqnarray}

\begin{eqnarray}
&\Delta^{inter,triplet}_{\uparrow\uparrow,2k^{o}_{F}}(x) =
R_{A\uparrow}R_{B\uparrow}e^{2ik^{o}_{F}x} \pm
L_{A\uparrow}L_{B\uparrow}e^{-2ik^{o}_{F}x}\\
&=\frac{\eta_{RA\uparrow}\eta_{RB\uparrow}}{2\pi\alpha}e^{i\sqrt{\pi}(\theta^{+}_{\rho}+\theta^{+}_{\sigma}
-\phi^{+}_{\rho}-\phi^{+}_{\sigma})}e^{i2k^{o}_{F}x}\pm
\frac{\eta_{LA\uparrow}\eta_{LB\uparrow}}{2\pi\alpha}e^{i\sqrt{\pi}(\theta^{+}_{\rho}+\theta^{+}_{\sigma}
+\phi^{+}_{\rho}+\phi^{+}_{\sigma})}e^{-i2k^{o}_{F}x}
\end{eqnarray}

\begin{eqnarray}
&\Delta^{inter,triplet}_{\downarrow\downarrow, 2k^{o}_{F}}(x) =
R_{A\downarrow}R_{B\downarrow}e^{2ik^{o}_{F}x} \pm
L_{A\downarrow}L_{B\downarrow}e^{-2ik^{o}_{F}x}\\
&=\frac{\eta_{RA\downarrow}\eta_{RB\downarrow}}{2\pi\alpha}e^{i\sqrt{\pi}(\theta^{+}_{\rho}-\theta^{+}_{\sigma}
-\phi^{+}_{\rho}+\phi^{+}_{\sigma})}e^{i2k^{o}_{F}x}\pm
\frac{\eta_{LA\downarrow}\eta_{LB\downarrow}}{2\pi\alpha}e^{i\sqrt{\pi}(\theta^{+}_{\rho}-\theta^{+}_{\sigma}
+\phi^{+}_{\rho}-\phi^{+}_{\sigma})}e^{-i2k^{o}_{F}x}
\end{eqnarray}
    \renewcommand{\thechapter}{D}
    \refstepcounter{chapter}
    \section*{Appendix D: Variation of Luttinger liquid kink with interaction strength\label{sec:kinkcalculations}}
    \addcontentsline{toc}{section}{Appendix D: Variation of Luttinger liquid kink with interaction strength}
\begin{figure}[h]
\centering
 \subfigure[~$A^<(k,\omega)$ at
$\gamma_\rho=0.20$]{\label{subfig:figdensityplot0720}
  \includegraphics [width=4in]{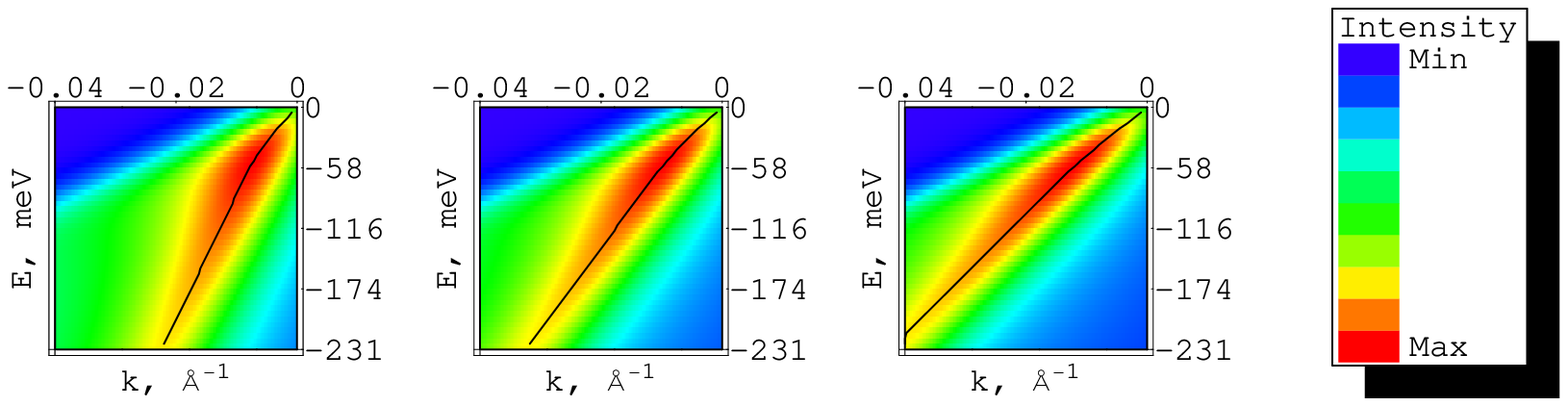}}
\subfigure[~Effective dispersion at
$\gamma_\rho=0.20$]{\label{subfig:figdispersion0720}
  \includegraphics[width=4.0in]{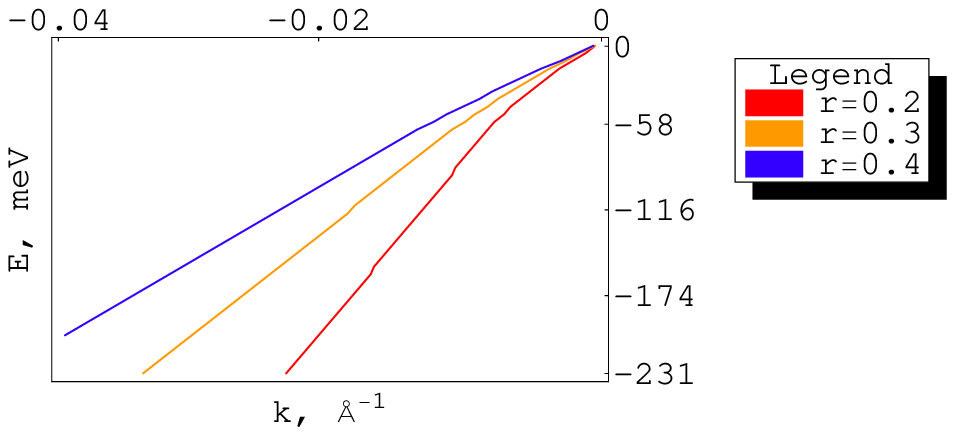}}
  \caption{Intensity of the spectral function $A^<(k,\omega)$ and
effective dispersions at an interaction strength $\gamma_\rho =
0.20$. (a) The intensity of $A^<(k,\omega)$ is shown for three
different ratios of the spin to charge velocity, $r=0.2,~0.3,$ and
$0.4$. The black lines are the effective electronic dispersions
derived from MDC peaks, as described in the text. The dashed line in
the first panel shows that the high energy part of the effective
dispersion does not extrapolate back to the Fermi wavevector,
$k_{F}$.  (b) Comparison of the dispersions at different values of
the velocity ratio, $r=0.2,~0.3,$ and $0.4$. In all cases the spin
velocity $v_{\sigma}=0.7 $eV-$\AA$ and the temperature $k_B
T=14meV$.\label{fig:gamma15}}
\end{figure}
\begin{figure}[!t]
\centering
 \subfigure[~$A^<(k,\omega)$ at
$\gamma_\rho=0.25$]{\label{subfig:figdensityplot0725}
  \includegraphics [width=4in]{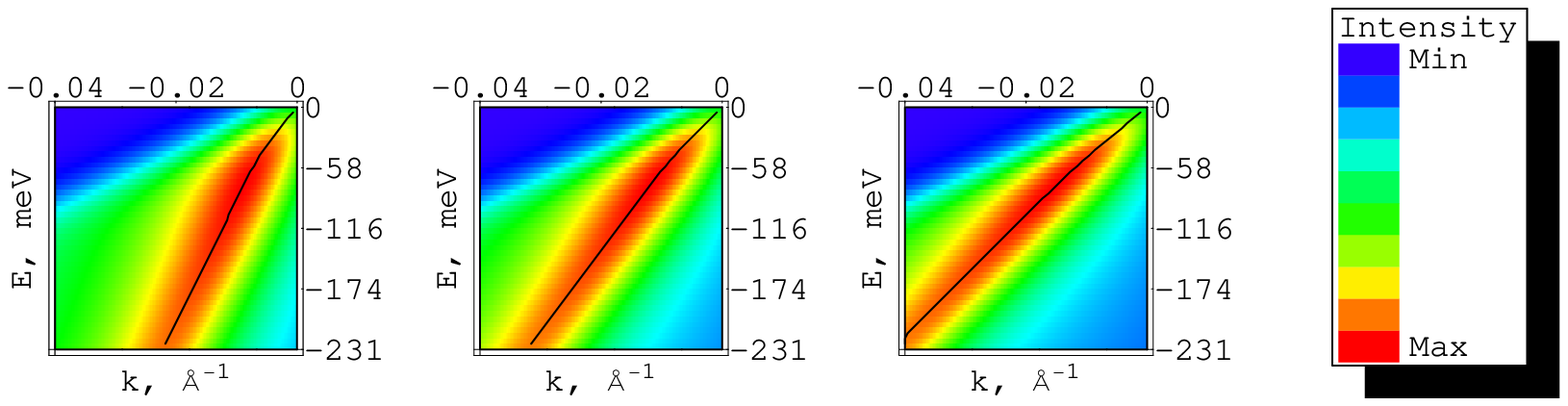}}
\subfigure[~Effective dispersion at
$\gamma_\rho=0.25$]{\label{subfig:figdispersion0725}
  \includegraphics[width=4.0in]{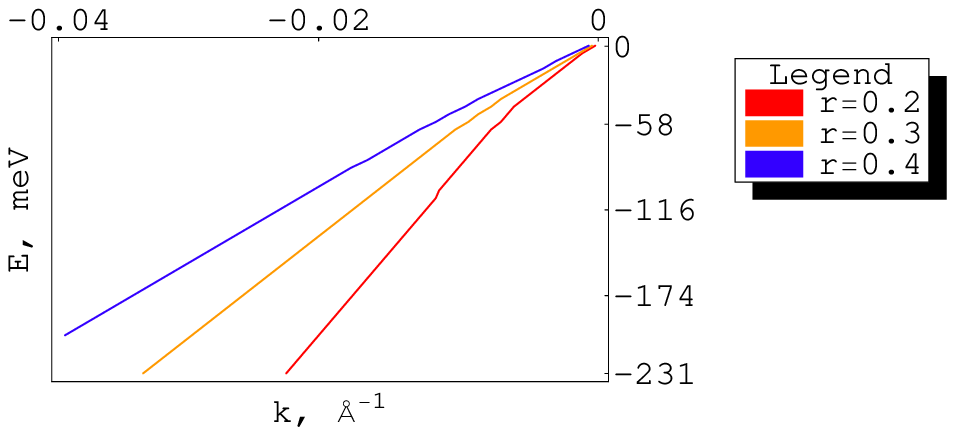}}
  \caption{Intensity of the spectral function $A^<(k,\omega)$ and
effective dispersions at an interaction strength $\gamma_\rho =
0.25$. (a) The intensity of $A^<(k,\omega)$ is shown for three
different ratios of the spin to charge velocity, $r=0.2,~0.3,$ and
$0.4$. The black lines are the effective electronic dispersions
derived from MDC peaks, as described in the text. The dashed line in
the first panel shows that the high energy part of the effective
dispersion does not extrapolate back to the Fermi wavevector,
$k_{F}$.  (b) Comparison of the dispersions at different values of
the velocity ratio, $r=0.2,~0.3,$ and $0.4$. In all cases the spin
velocity $v_{\sigma}=0.7 $eV-$\AA$ and the temperature $k_B
T=14meV$.\label{fig:gamma15}}
\end{figure}
\begin{figure}[!t]
\centering
 \subfigure[~$A^<(k,\omega)$ at
$\gamma_\rho=0.30$]{\label{subfig:figdensityplot0730}
  \includegraphics [width=4in]{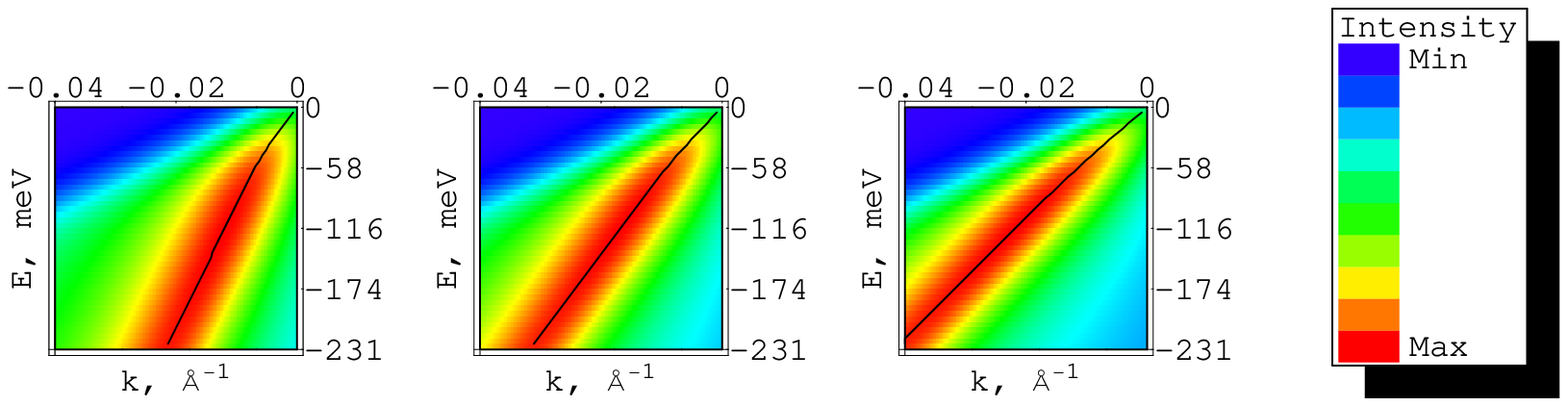}}
\subfigure[~Effective dispersion at
$\gamma_\rho=0.30$]{\label{subfig:figdispersion0730}
  \includegraphics[width=4.0in]{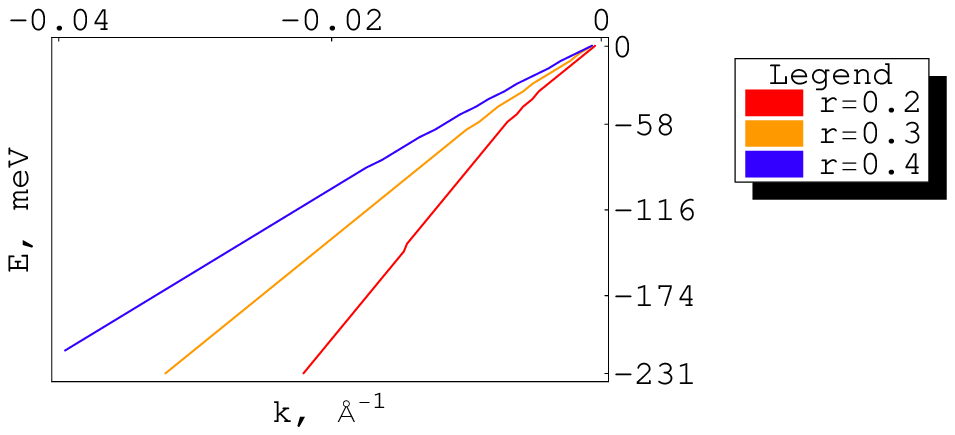}}
  \caption{Intensity of the spectral function $A^<(k,\omega)$ and
effective dispersions at an interaction strength $\gamma_\rho =
0.30$. (a) The intensity of $A^<(k,\omega)$ is shown for three
different ratios of the spin to charge velocity, $r=0.2,~0.3,$ and
$0.4$. The black lines are the effective electronic dispersions
derived from MDC peaks, as described in the text. The dashed line in
the first panel shows that the high energy part of the effective
dispersion does not extrapolate back to the Fermi wavevector,
$k_{F}$.  (b) Comparison of the dispersions at different values of
the velocity ratio, $r=0.2,~0.3,$ and $0.4$. In all cases the spin
velocity $v_{\sigma}=0.7 $eV-$\AA$ and the temperature $k_B
T=14meV$.\label{fig:gamma15}}
\end{figure}
\begin{figure}[!t]
\centering
 \subfigure[~$A^<(k,\omega)$ at
$\gamma_\rho=0.35$]{\label{subfig:figdensityplot0735}
  \includegraphics [width=4in]{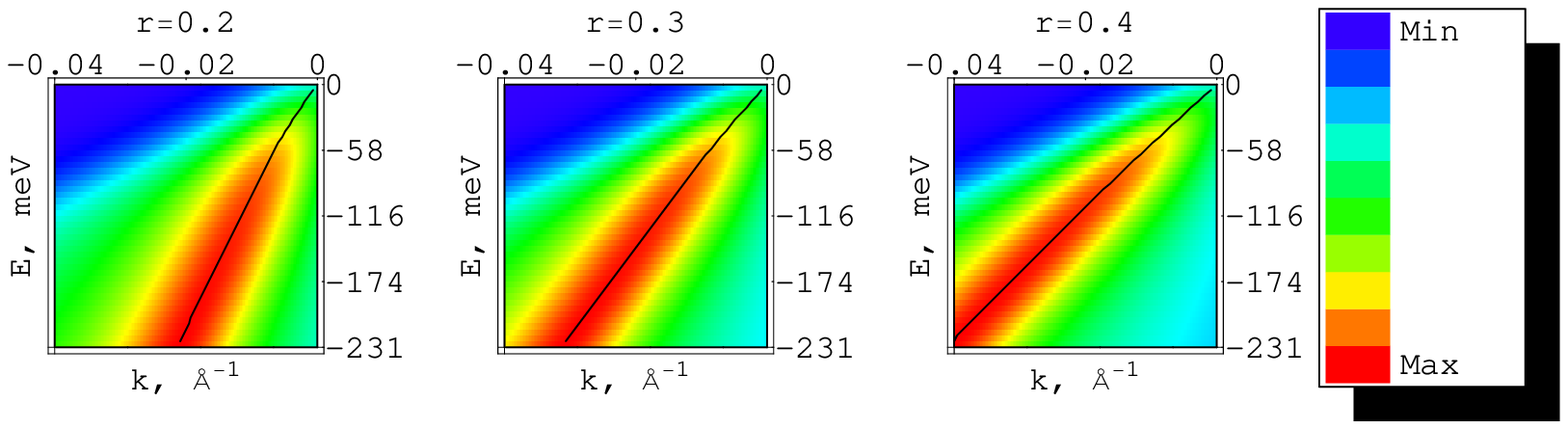}}
\subfigure[~Effective dispersion at
$\gamma_\rho=0.35$]{\label{subfig:figdispersion0735}
  \includegraphics[width=4.0in]{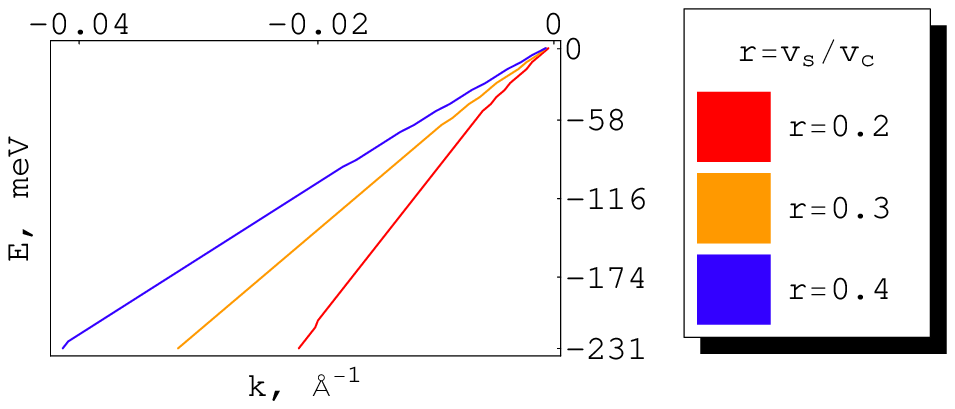}}
  \caption{Intensity of the spectral function $A^<(k,\omega)$ and
effective dispersions at an interaction strength $\gamma_\rho =
0.35$. (a) The intensity of $A^<(k,\omega)$ is shown for three
different ratios of the spin to charge velocity, $r=0.2,~0.3,$ and
$0.4$. The black lines are the effective electronic dispersions
derived from MDC peaks, as described in the text. The dashed line in
the first panel shows that the high energy part of the effective
dispersion does not extrapolate back to the Fermi wavevector,
$k_{F}$.  (b) Comparison of the dispersions at different values of
the velocity ratio, $r=0.2,~0.3,$ and $0.4$. In all cases the spin
velocity $v_{\sigma}=0.7 $eV-$\AA$ and the temperature $k_B
T=14meV$.\label{fig:gamma15}}
\end{figure}


%
%
%
%

\begin{vita}
Trinanjan Datta was born in Calcutta (presently Kolkata), West
Bengal, India. He studied at St. Xavier's College, Calcutta and
obtained his Bachelor of Science degree from the University of
Calcutta in 1999 and then his Master of Science degree from the
Indian Institute of Technology, Kanpur in 2001. He joined Purdue
University as a doctoral candidate in the Fall of 2001. During his
stay at Purdue University he held teaching assistantship positions
for the initial few years. He was then awarded the Purdue Research
Foundation Grant and also the Bilsland Dissertation Fellowship to
complete the doctoral degree. He received his Ph.D from Purdue
University in August, 2007.
\end{vita}

\end{document}